\pdfoutput=1

\documentclass[AMA,STIX1COL,sort&compress,hidelinks]{WileyNJD-v2}

\RequirePackage{amsfonts}
\RequirePackage{graphicx}
\RequirePackage{threeparttable,adjustbox,longtable,booktabs,multirow,tabularx,caption,subcaption,dcolumn,bbm}

\usepackage{tabularx}
\newcolumntype{Y}{>{\raggedleft\arraybackslash}X}

\usepackage{colortbl}
\definecolor{Gray}{gray}{0.9}

\usepackage{xr}

\makeatletter
\newcommand*{\addFileDependency}[1]{
  \typeout{(#1)}
  \@addtofilelist{#1}
  \IfFileExists{#1}{}{\typeout{No file #1.}}
}
\makeatother

\newcommand*{\myexternaldocument}[1]{%
    \externaldocument{#1}%
    \addFileDependency{#1.tex}%
    \addFileDependency{#1.aux}%
}

\myexternaldocument{supporting_information_R2}

\newcommand{\indep}{\perp \!\!\! \perp}

\newcommand{\bfepsilon}{\mbox{\begin{boldmath}$\epsilon$\end{boldmath}}}

\def\bSig\mathbf{\Sigma}

\usepackage{mathtools,tikz}
\usepackage[justification=centering]{caption}
\DeclareRobustCommand\sampleline[1]{%
	\tikz\draw[#1] (0,0) (0,\the\dimexpr\fontdimen22\textfont2\relax)
	-- (2em,\the\dimexpr\fontdimen22\textfont2\relax);%
}

\usetikzlibrary{plotmarks}
\definecolor{Rred}{HTML}{F8766D}
\definecolor{Rgold}{HTML}{A3A500}
\definecolor{Rgreen}{HTML}{00BF7D}
\definecolor{Rblue}{HTML}{00B0F6}
\definecolor{Rpink}{HTML}{E76BF3}

\definecolor{darkblue}{rgb}{0.0, 0.0, 0.55}

\articletype{Article Type}%

\received{26 April 2016}
\revised{6 June 2016}
\accepted{6 June 2016}

\begin{document}

\title{Targeted learning in observational studies with multi-valued treatments: An evaluation of antipsychotic drug treatment safety}

\author[1]{Jason Poulos*}

\author[2]{Marcela Horvitz-Lennon}

\author[1]{Katya Zelevinsky}

\author[3]{Tudor Cristea-Platon}
\author[3]{Thomas Huijskens}
\author[3]{Pooja Tyagi}
\author[3]{Jiaju Yan}
\author[3]{Jordi Diaz}

\author[1,4]{Sharon-Lise Normand}

\authormark{POULOS \textsc{et al.}}

\address[1]{\orgdiv{Department of Health Care Policy}, \orgname{Harvard Medical School}, \orgaddress{\state{Boston, Massachusetts}, \country{USA}}}

\address[2]{\orgname{RAND Corporation}, \orgaddress{\state{Boston, Massachusetts}, \country{USA}}}

\address[3]{\orgname{QuantumBlack}, \orgaddress{\state{London}, \country{UK}}}

\address[4]{\orgdiv{Department of Biostatistics}, \orgname{Harvard Chan School of Public Health}, \orgaddress{\state{Boston, Massachusetts}, \country{USA}}}

\corres{*Jason Poulos, Division of Endocrinology, Brigham and Women's Hospital, 221 Longwood Avenue, Boston, MA, USA. \email{jvpoulos@bwh.harvard.edu}}

\fundingInfo{National Institute of Mental Health, Grant/Award Number: R01-MH106682; QuantumBlack-McKinsey and Company, Grant/Award Number: A42960}


\abstract[Summary]{We investigate estimation of causal effects of multiple competing (multi-valued) treatments in the absence of randomization. Our work is motivated by an intention-to-treat study of the relative cardiometabolic risk of assignment to one of six commonly prescribed antipsychotic drugs in a cohort of nearly 39,000 adults with serious mental illnesses. Doubly-robust estimators, such as targeted minimum loss-based estimation (TMLE), require correct specification of either the treatment model or outcome model to ensure consistent estimation; however, common TMLE implementations estimate treatment probabilities using multiple binomial regressions rather than multinomial regression. We implement a TMLE estimator that uses multinomial treatment assignment and ensemble machine learning to estimate average treatment effects. Our multinomial implementation improves coverage, but does not necessarily reduce bias, relative to the binomial implementation in simulation experiments with varying treatment propensity overlap and event rates. Evaluating the causal effects of the antipsychotics on three-year diabetes risk or death, we find a safety benefit of moving from a second-generation drug considered among the safest of the second-generation drugs to an infrequently prescribed first-generation drug known for having low cardiometabolic risk.} 

\jnlcitation{\cname{%
\author{Poulos J.},
\author{M. Horvitz-Lennon},
\author{K. Zelevinsky},
\author{T. Cristea-Platon}, 
\author{T. Huijskens}, 
\author{P. Tyagi}, 
\author{J. Yan},
\author{J. DIaz}, and
\author{S.L. Normand}},
\ctitle{Targeted learning in observational studies with multi-valued treatments: An evaluation of antipsychotic drug treatment safety}, \cjournal{Statistics in Medicine}. \cyear{2023}; \cvol{submitted}.}

\maketitle


\section{Introduction}\label{sec1}

Antipsychotic drugs effectively control some of the most disturbing symptoms of schizophrenia, and no other treatments have comparable effectiveness.\citep{keepers2020american} These drug are also valuable for the treatment of bipolar I disorder \citep{carvalho2020bipolar} and treatment-resistant major depressive disorder (MDD). \citep{zhou2015atypical} While more than 20 antipsychotic drugs are available in the U.S., the most widely used are the subset of second generation antipsychotics (SGAs). SGAs are generally as effective as first generation antipsychotics (FGAs) and avoid some common FGA side effects, but there is evidence that some frequently used SGAs carry a higher risk for cardiometabolic morbidity (which includes diabetes) relative to FGAs.\citep{correll2015effects} Diabetes is a serious condition, and its rising prevalence in the general population is a major target of efforts to improve the health of the public,\citep{pandya2013more,virani2020heart} and it is at least twice as prevalent among people with serious mental illnesses (SMI) than their peers.\citep{holt2015diabetes} Compared to the general population, people with SMI have a higher risk for cardiometabolic morbidity in general,\citep{de2011physical} which accounts for a large fraction of their reduced life expectancy.\citep{olfson2015premature}

The existing evidence on the relative safety of different antipsychotics is limited. Randomized controlled trials (RCTs) have focused on the drugs' effects on risk factors for diabetes and other cardiometabolic morbidity\citep{pillinger2020comparative} rather than the occurrence of these morbidities, and mortality trials have been conducted in samples of elderly adults with dementia.\citep{reus2016american} Otherwise, most evidence comes from observational studies. Some studies compare (i.) recent initiators of antipsychotic drugs to a control group not receiving an antipsychotic drug;\citep{gianfrancesco2006influence,taipale202020} (ii.) FGA users to SGA users, not differentiating specific antipsychotics;\citep{guo2007risk} (iii.) individuals receiving a specific SGA to those receiving any other SGA;\citep{yood2009incidence} or (iv.), individuals receiving one versus two or more antipsychotic drugs.\citep{katona2014real} These studies have some important limitations. Given the effect heterogeneity for cardiometabolic morbidity for specific SGAs, pooled analyses of all SGAs may mask risks for specific drugs, potentially incorrectly implicating all SGAs. Studies that compare SGA outcomes to individuals receiving no SGA are of little help for those with SMI because an antipsychotic is required. Moreover, these studies cannot draw conclusions on the relative safety of antipsychotics based on one-to-one comparisons because they do not balance covariates across the drug groups. An exception is the study of Gianfrancesco et al.,\cite{gianfrancesco2006influence} which models diabetes risk using a single logistic regression controlling for patient characteristics. 

Our paper is motivated by an intention-to-treat study of the relative cardiometabolic risk of non-random assignment to one of six commonly prescribed antipsychotic drugs in a cohort of adults with SMI. We compare four SGAs and one FGA to a reference drug (a SGA) thought to have a relatively lower risk for cardiometabolic morbidity and mortality compared to other SGAs. The clinical relevance of these comparisons is bolstered by evidence that switching to safer drugs can improve some metabolic indices without causing significant psychiatric deterioration.\citep{mukundan2010antipsychotic}

Several estimators have been proposed for estimating causal effects in observational data settings with multiple competing (multi-valued) treatments,\citep{linden2016estimating, lopez2017estimation} although their applications  have been limited to a small number of treatment levels and a focus on continuous outcomes. The propensity score, the probability of receiving a treatment given the observed covariates,\citep{rosenbaum1983central} has played a central role in causal inference. For instance, inverse probability of treatment weighted (IPTW) estimators, which weight the outcomes of units in each treatment group by the inverse of the propensity score with the goal of matching the covariate distribution of a target population, is a common strategy for binary treatments.\citep{frolich2004finite,li2013weighting,mccaffrey2013inverse,austin2015moving} Two decades ago, Imbens\cite{imbens2000role} and Imai and van Dyk\cite{imai2004causal} generalized the propensity score framework from the binary treatment setting to the setting of multi-valued treatments. Generalized propensity score (GPS) methods have since been proposed for the case of a single continuous treatment,\cite{hirano2005propensity,kreif2015evaluation,fong2018covariate,wu2022matching} and multiple continuous\cite{egger2013generalized,kallus2018policy,huffman2018covariate} or multi-valued\cite{feng2012generalized,nguyen2019use,hu2020estimation} treatments. Yang et al.\cite{yang2016propensity} proposed subclassification or matching on the GPS to estimate pairwise average causal effects, and Li and Li\cite{li2019propensity} introduce generalized overlap weights for pairwise comparisons that focus on the target population with the most covariate overlap across multiple levels of treatment. Similar to other propensity score methods, these generalized approaches depend on the correct specification of the treatment model and do not eliminate bias from unmeasured confounding. A different approach proposed by Bennett et al.\cite{bennett2020building} does not require estimation of a GPS. Rather, the authors directly match on the covariates using mixed integer programming methods to balance each treatment group to a representative sample drawn from the target population. This approach has the advantage of directly balancing covariates without the need to specify a statistical model.

Doubly-robust estimators require that either the treatment model or outcome model is correctly specified to ensure consistent estimation.\citep{bang2005doubly,kang2007demystifying,benkeser2017doubly,rose2019double} Targeted minimum loss-based estimation (TMLE) is a widely-used doubly-robust estimator that permits data-adaptive estimation strategies to improve specification of models for causal inference.\citep{van2006targeted,van2011targeted,schuler2017targeted} The TMLE, which is doubly robust for both consistency and asymptotic linearity, reweights an initial estimator with a function of the estimated GPS. Augmented IPTW (A-IPTW), which adds an augmentation term to the IPTW estimator, is another doubly-robust method which aims to solve an estimating equation in candidate values of the causal parameter.\cite{robins1994estimation,robins2000marginal} 

TMLE's flexibility in model specification and its focus on minimizing bias through a log-likelihood loss function offer advantages in terms of estimator efficiency and robustness. Unlike A-IPTW, TMLE does not aim to solve an estimating equation, but rather uses a log-likelihood loss function to minimize the bias of the causal parameter. This allows TMLE to leverage nonparametric methods for estimating the outcome and treatment models, thereby reducing the likelihood of model misspecification, especially in high-dimensional settings.\citep{lee2010improving,austin2012using,pirracchio2015improving} Moreover, TMLE has been shown to outperform A-IPTW in finite samples.\cite{porter2011relative}

Few researchers have used TMLE for causal inference in multi-valued treatments settings. Cattaneo\cite{cattaneo2010efficient} focuses on estimation of multi-valued treatment effects using a generalized method of moments approach that is also doubly-robust and semiparametrically efficient. Wang et al.\cite{wang2020estimating} adapt TMLE to estimate a global treatment importance metric for numerous studies that make comparisons between multiple concurrent treatments, where the availability of treatments may differ across studies. Similarly, Liu et al.\cite{liu2022modeling} adapts TMLE to measure effect heterogeneity in a meta-analysis of numerous studies with multiple concurrent treatments. While the goal of Wang et al. and Liu et al. is to obtain a global measure of effect or effect modification, respectively, in a meta-analysis, the focus of the present paper is to estimate pairwise average causal effects between multiple competing treatments in a single study. Siddique et al.\cite{siddique2019causal} focus on the TMLE of multiple concurrent treatments, resulting in a potentially large number of possible treatment combinations which may not be observed in the data. Our study is the first to our knowledge to evaluate TMLE in the multi-valued treatment setting in simulations. In the simulation studies of Siddique et al., Liu et al., and Wang et al., multiple treatments are assigned using binomial logistic models, whereas in our simulations, multi-valued treatment is assigned using a single multinomial logistic model.

Our paper contributes to the causal inference literature along several dimensions. First, we add to the sparse literature on inference for multi-valued treatments with a focus on implementation. We review the assumptions required to make causal inference in the multi-valued treatment setting, define several key causal parameters, and describe approaches for assessing the validity of the common support assumption. Second, we extend TMLE to the multi-level treatment setting through accurate estimation of a multinomial treatment model. One of the key advantages of using a multinomial treatment model in the TMLE framework is its ability to jointly model the multiple treatment categories. This joint modeling allows for a more efficient use of the available data, as it can account for the correlations between different treatment levels. This advantage becomes particularly evident when using the IPTW estimator. In TMLE, the outcome model tends to moderate the influence of the weights, making the advantages of a well-specified treatment model less pronounced. However, in IPTW, a poorly specified treatment model can lead to unstable estimates, primarily due to errors in the weighting mechanism.

Surprisingly, all peer-reviewed implementations of TMLE for multi-valued treatments use a series of binomial treatment assignments. McCaffrey et al.\cite{mccaffrey2013tutorial} propose using a gradient boosting algorithm to estimate the probabilities for multiple treatments, and suggest a binomial modeling approach for computational ease, noting that while the sum of the estimated probabilities across all treatment levels may not equal one, this poses no problem for weighted estimators because only the estimated probability of the treatment actually received for each individual is used. However, this approach will result in a loss in efficiency because the wrong treatment model is estimated and complicates the assessment of common support because estimates for all treatment levels for each unit are required. Another computational reason for the binomial assignment approach is that the software implementation of the super learner\cite{van2007super,polley2010super,polley2011super,gruber2012tmle,ottoboni2020estimating,phillips2023practical} used to estimate the treatment model does not support multinomial outcomes. The super learner is an ensemble method that uses cross-validated log-likelihood to select the optimal weighted average of estimators from a pre-selected library of nonparametric classification algorithms.

Third, we evaluate the comparative performance of the current implementations and our approach of TMLE through numerical studies using data adaptive approaches. While theory dictates that TMLE estimators will be unbiased if only one model is misspecified, efficiency will suffer. Simulations demonstrate that our multinomial implementation improves coverage, but does not always minimize bias, compared to the binomial implementation. 

\section{Doubly-robust estimators for multi-valued treatments}\label{estimation}

\subsection{Notation and setup}\label{notation}

We observe a sample of size $n$ in which each subject $i$ has been assigned to one of $J$ treatment levels. In our application, we focus on monotherapy users of one of six drugs; i.e., those who use a single drug for treatment. The observed treatment level is denoted $a_i \in \mathcal{A}$, with $\bf{a}$ the length-$n$ vector of treatment assignments, and $\mathcal{A} = \{ j = 1, 2, \cdots, J\}$ the collection of possible treatment levels. The sample size for each treatment level $j$ is denoted $n_j$, with $\sum_{j=1}^{J} n_j = n$. We also observe a $p \times 1$ vector of covariates measured prior to treatment initiation, ${\bf x}_i$, with ${\bf x} ~ \in ~\mathbb{X}$.

The observed outcome is $y_i = \sum_{j=1}^J \mathbbm{1} (a_{i}=j) y_i(j)$, with $\mathbbm{1}(\cdot)$ denoting the indicator function. For each subject, we observe $o_i = (y_i, a_i,{\bf x}_i)$ arising from some probability distribution $\mathbb{P}$. Under the potential outcomes framework, subject $i$'s potential outcome under treatment level $j$, $y_i(j)$, depends only on the treatment the subject receives and not by treatments received by other subjects.
\noindent
\begin{assumption}
{\em Stable unit treatment value assumption (SUTVA). (i.)  The potential outcome for any subject does not vary with treatments assigned to other subjects; and (ii.), a single version of each treatment level exists:} $y_i(a_1, a_2, \ldots, a_n) = y_i(a_i),~ \forall~ a_i \in \mathcal{A}. $
\end{assumption}
\noindent
The no interference assumption (i.) is plausible for the treatment examined in this paper --- a subject's diabetes status cannot be caused by another subject's antipsychotic treatment assignment. The assumption of a single version of each treatment level (ii.), which ensures that each subject has the same number of potential outcomes, may be violated if treatment levels are loosely defined. In our example, we include both oral and injectable versions of the same drug. While the biological effects of the oral and injectable versions may be similar, variations in adherence could lead to meaningful variations in a treatment level and potentially introduce bias in the estimated treatment effects in an intent-to-treat study.

We denote the conditional probability subject $i$ is assigned treatment level $j$, $\mbox{Pr}(a_i=j \mid {\bf x}_i)$, by ${p}_j({\bf x}_i)$ such that $\sum_{j=1}^J{p_j}({\bf x}_i) = 1$. For causal inference in the multi-valued treatment setting, Imbens\cite{imbens2000role} refers to $p_j({\bf x}_i)$ as the GPS. We let  $\mu_{j} = \mathbb{E}\{y_i(j)\}$ denote the marginal mean outcome and $e_j({\bf x}_i, \mu_j) = \mathbb{E}( y_i(j) \mid {\bf x}_i)$ denote the conditional mean outcome. We make the following two assumptions, which are explicitly made in the work of Imbens. 
\noindent
\begin{assumption}
{\em Weak Unconfoundedness. The distribution of the potential outcomes is independent of treatment assignment, conditional on the observed covariates:} $ y_i(j) \indep \mathbbm{1} (a_{i}=j)  \mid {\bf x}_i,~ \forall~ {\bf x}_i \in \mathbb{X} ~\mbox{and}~a_i \in \mathcal{A}.$
\end{assumption}
\noindent
The assumption is weak because the conditional independence is assumed at each level of treatment rather than joint independence of all the potential outcomes. This assumption is not testable and typically justified on substantive grounds. Bolstering its validity requires conditioning on many covariates, making the dimensionality of ${\bf x}$ large. In our setting, six-month medical history information prior to the index antipsychotic fill is available, including all drugs filled by the subject and billable medical services utilized. Demographic information that includes place of residence is known. All subjects have the same health insurer, although how the benefits are managed may differ across states. Nonetheless, treatment preferences, results of diagnostic tests, and some information known only to physicians, such as the subjects' body mass index, are unknown.
\begin{assumption}
{\em Positivity. There is a positive probability that someone with covariates ${\bf x}_i$ could be assigned to each $j$: } $\mbox{Pr}(a_{i}=j \mid {\bf x}_i) > 0,~ \forall~ {\bf x}_i \in \mathbb{X} ~\mbox{and}~a_i \in \mathcal{A}.$
\end{assumption}
\noindent
The positivity assumption is required to avoid extrapolating treatment effects for covariate patterns where there are no observations for some treatments. Structural violations occur if subjects with specific covariate patterns cannot receive one of the treatment levels, due to, in our case, absolute contraindications. However, practical violations of the positivity assumption could occur due to finite sample sizes. While the positivity assumption is testable in high dimensions, detecting violations is challenging.\citep{petersen2012diagnosing}

The number of treatments levels complicates inferences in the observational setting. First, meeting the unconfoundedness assumption requires the availability of a large number of covariates to differentiate among the treatment choices. Second, several target populations exist and our clinical problem requires a population of individuals eligible for any of the six drugs. Finding individuals from all treatment groups in subsets determined by the covariate space becomes increasingly difficult as the number of treatment choices increases. Regression, some machine-learning algorithms, propensity-score based approaches, and matching methods often extrapolate over areas of the covariate space with no common support (referred to as areas of ``non-overlap''). To circumvent non-overlap, a common strategy is to winsorize extreme probabilities to a threshold.\citep{cole2008constructing,lee2011weight} Third, the probability of assignment varies considerably across treatment levels which will impact the precision of estimates, and with more treatment levels, the observed number of individuals in any treatment arm may be small.

\subsection{Positivity and common support}\label{ESS}
We use the effective sample size (ESS) associated with each treatment level as a diagnostic for assessing common support. Comparison of this metric among different estimators provides a rough measure of the amount of information in the sample used to estimate the marginal mean outcome. 
\begin{definition}
{\em Effective Sample Size (ESS). The ESS is a measure of the weighted sample size for treatment level $j$ defined as} 
\begin{eqnarray*}
ESS_j = \frac{\left( \sum_i^n \mathbbm{1} (a_{i}=j) w_j({\bf x}_i) \right)^2}{\sum_i^n \mathbbm{1} (a_{i}=j) w_j({\bf x}_i)^2}, \quad ~~\mbox{with}~~\sum_j \widehat{p}_{j}({\bf x}_i) = 1,~~\mbox{and}~~w_j({\bf x}_i) = 1/\widehat{p}_{j}({\bf x}_i). \end{eqnarray*}
\end{definition}
McCaffrey et al.\cite{mccaffrey2013tutorial} suggests the ratio $ESS_j/n_j$ as a measure of the loss of precision due to weighting, with relatively small values of the ratio indicating weak overlap among the treatment groups.

\subsection{Causal parameter}\label{causal-parameter}

Our inferential goal is the estimation of the difference in the average outcome if everyone was treated with any other treatment $j_{\text{Alt}}$ and the average outcome if everyone was treated with a reference treatment $j_{\text{Ref}}$.
\begin{definition}
{\em Average Treatment Effect (ATE). The average effect caused by any other treatment $j_{\text{Alt}}$ over the reference treatment $j_{\text{Ref}}$ in the sample.} %
\begin{eqnarray*}
\text{ATE}_{j_{\text{Ref}}, \, j_{\text{Alt}}} = \mathbb{E}(y_i(j_{\text{Alt}}) - y_i(j_{\text{Ref}})) = \mu_{j_{\text{Alt}}} - \mu_{j_{\text{Ref}}}; ~~ j_{\text{Alt}} \neq j_{\text{Ref}}.
\end{eqnarray*} 
\end{definition}
\noindent
The ATE represents the causal effect of moving from one treatment to another for all units in the sample. Identification of the ATE in the context of multi-valued treatments is provided in Cattaneo.\cite{cattaneo2010efficient} In the application, we want to understand how patients treated with any antipsychotic other than the Reference drug would fare in terms of diabetes or mortality risk if they were instead treated with the Reference, which is purported to have a more favorable cardiometabolic risk profile. 

\subsection{Targeted minimum loss-based estimation (TMLE)} \label{TMLE}
TMLE updates an initial estimate of a parameter with a correction determined by optimizing the bias-variance trade-off using a loss function for the causal parameter. The estimator is asymptotically linear with the influence curve equal to the canonical gradient. We focus on estimation of the marginal mean outcome for each treatment level $j$, $\mu_j$, and collect estimators into a vector. This strategy is useful for making joint inferences between multiple treatment levels, as demonstrated by Cattaneo.\cite{cattaneo2010efficient} Let $\widehat{e}^0(\cdot)$ denote the initial estimate of $\mathbb{E}(y_i(j) \mid {\bf x}_i)$, also called the G-computation estimate,\cite{robins1986new} and $\widehat{e}^1(\cdot)$ denote the adjusted estimate. The TMLE estimator for $\mu_j$ is 
\begin{eqnarray} \label{eq:TMLE}
\widehat{\mu}_{j, \, \text{TMLE}} = \frac{1}{n} \sum_{i=1}^n \widehat{e}^1({\bf x}_i, \mu_j) = \frac{1}{n}\sum_{i=1}^n h^{-1}\left(h\left(\widehat{e}^0({\bf x}_i, \mu_j)\right) + \frac{\widehat{\epsilon}_j \mathbbm{1} (a_{i}=j)}{\widehat{p}_{j}({\bf x}_i)} \right),
\end{eqnarray}
\noindent
where $h$ is a link function and $\left(\widehat{\epsilon}_j \mathbbm{1} (a_{i}=j)\right)/\widehat{p}_{j}(x_i)$ is a correction that targets the unknown parameter $\mu_j$. In comparison, the IPTW estimator for $\mu_j$ instead reweights the observed outcomes with the inverse of the GPS
\begin{eqnarray*} 
\widehat{\mu}_{j, \, \text{IPTW}} = \frac{1}{n}\sum_{i=1}^n \frac{ \mathbbm{1} (a_{i}=j)}{\widehat{p}_{j}({\bf x}_i)} y_i.
\end{eqnarray*}

The two-step TMLE estimation procedure is as follows. First, super learner estimates for $e_j({\bf x}_i, \mu_j)$ and $p_j({\bf x}_i)$ are substituted into Equation (\ref{eq:TMLE}). Second, the term $\widehat{\epsilon}_j$ is obtained by estimating a parametric regression model 
\begin{eqnarray} \label{eq:epsilon}
h\left(\mathbb{E}(y_i=1\mid a_i, {\bf x}_i, \bfepsilon)\right) = 
h\left(\widehat{e}^0({\bf x}_i, \mu_{a_i})\right) + \sum_{j=1}^J \epsilon_j \frac{\mathbbm{1} (a_{i}=j)}{\widehat{p}_{j}({\bf x}_i)},
\end{eqnarray} 
and fixing the coefficient of $h\left(\widehat{e}^0({\bf x}_i, \mu_{a_i})\right)$ at one. The correction is determined using a log-likelihood loss function to minimize the bias of $\mu_j$. 
\noindent
When Assumptions (1) -- (3) are met, van der Laan and Rubin\cite{van2006targeted} demonstrate that the efficient influence curve for $\text{ATE}_{j_{\text{Ref}}, \, j_{\text{Alt}}}$ is 
\begin{eqnarray*} 
\text{IC}_{j_{\text{Ref}}, \, j_{\text{Alt}}}(o_i) = \left(\frac{\mathbbm{1} (a_i=j_{\text{Alt}})}{{p}_{j_{\text{Alt}}}({\bf x}_i)} - \frac{\mathbbm{1} (a_{i}=j_{\text{Ref}})}{{p}_{j_{\text{Ref}}}({\bf x}_i)} \right) \left(y_i- {e}({\bf x}_i,\mu_{a_i})\right) + {e}_{j_{\text{Alt}}}({\bf x}_i,\mu_{j_{\text{Alt}}}) - {e}_{j_{\text{Ref}}}({\bf x}_i,\mu_{j_{\text{Ref}}}) - \widehat{\text{ATE}}_{j_{\text{Ref}}, \, j_{\text{Alt}}}.
\end{eqnarray*} %
\noindent
The influence curve tells us how much an estimate will change if the input changes, and is used to estimate the variance and standard error $\sigma$ of the ATE
\begin{eqnarray}\label{influence}
\text{V}\left(\text{ATE}_{j_{\text{Ref}}, \, j_{\text{Alt}}}\right) = \frac{1}{n} \sum_{i=1}^n \widehat{\text{IC}}_{j_{\text{Ref}}, \, j_{\text{Alt}}}^2(o_i) \quad ~~\mbox{and}~~ \quad \sigma_{j_{\text{Ref}}, \, j_{\text{Alt}}} = \sqrt{\text{V}\left(\text{ATE}_{j_{\text{Ref}}, \, j_{\text{Alt}}}\right)/n}.
\end{eqnarray}
The standard error is used to construct a 95\% Wald-type confidence interval, $\text{ATE}_{j_{\text{Ref}}, \, j_{\text{Alt}}} \pm 1.96 \sigma_{j_{\text{Ref}}, \, j_{\text{Alt}}}$.

\subsubsection{Binomial treatment model}\label{tmle-bin}
In the software implementation of TMLE, each $p_j({\bf x}_i)$ is modelled separately as a Bernouilli random variable, estimating the probability of receiving treatment level $j$ relative to all other treatment levels. Thus, Equation (\ref{eq:epsilon}) is replaced by 
\begin{eqnarray*}\label{eq:epsilon-bin}
h\left(\mathbb{E}(y_i=1\mid a_i, {\bf x}_i, \bfepsilon)\right) = 
h\left(\widehat{e}^0({\bf x}_i, \mu_{a_i})\right) + \epsilon_j \frac{\mathbbm{1} (a_{i}=j)}{\widehat{p}_j({\bf x}_i)} + \epsilon_{-j} \frac{\mathbbm{1} (a_i = -j)}{\widehat{p}_{-j}({\bf x}_i)},
\end{eqnarray*}
where the subscript $-j$ refers to all treatments except $j$. In this strategy, there is no guarantee that $\sum_j p_j({\bf x}_i) = 1$ and the estimate of $\epsilon_j$ may differ from those obtained using Equation (\ref{eq:epsilon}). In both our simulation results (Section \ref{sim-results}) and the application findings (Section \ref{results}), we observed that the binomial estimates of $p_j({\bf x}_i)$ generally assume more extreme values and exhibit greater variability compared to the multinomial estimates, resulting in narrower confidence intervals.

Comparisons of the use of repeated binomial models with a multinomial model for nominal response options have been previously studied. Agresti\cite{agresti2013categorical} indicated that the standard errors of maximum likelihood estimates of regression parameters when fitting separate binary regression models are larger relative to those obtained when fitting a single multinomial model. In earlier work, Becg and Gray\cite{becg1984calculation} demonstrated that when using the same reference group via a logit link, the multinomial and repeated binomial models are parametrically similar, and the maximum likelihood estimators of regression coefficients from both models are asymptotically normal and unbiased but have different covariance matrices. Using simulation studies, the authors demonstrated that while the relative asymptotic efficiencies of the regression parameters obtained via the repeated binomial modeling approach were sufficient, the efficiencies were lower for predicted probabilities and declined as (i.) the number of covariates, (ii.) the number of treatment groups, and (iii.) the differences in magnitude of the regression coefficients across response options increase. Thus, the use of repeated binomial models for the treatment assignment mechanism rather than a multinomial model in the TMLE when the outcome model is correctly specified should result in differences in coverage or confidence interval widths for marginal outcome estimates, but bias should not differ. 

While the aforementioned work has shown that repeated binomial models and multinomial models yield asymptotically normal and unbiased maximum likelihood estimators of regression coefficients with different covariance matrices, it is important to note that coverage is not solely determined by these factors. Coverage also depends on the bias of the estimator, the bias in the estimated standard errors, and the degree to which the distribution of the estimator approximates normality.

\subsubsection{Implementation details}\label{implementation}
We rely on the \texttt{sl3} package in \textsf{R} for constructing the super learner (hereafter, ``SL'') for the treatment and outcome models, since this package supports multinomial classification algorithms and a multinomial loss function for the SL.\citep{coyle2021sl3-rpkg} When estimating a multinomial treatment model, the SL combines the predictions from multiple classification algorithms by multinomial linear regression. For binomial treatment or outcome models, the SL combines algorithmic predictions by binomial logistic regression. The SL weights are optimized by minimizing a negative log-likelihood loss function that is cross-validated with 5 folds, each consisting of a validation set and a training set. The routine for optimizing the SL weights, given the loss function and combination function, is nonlinear optimization using Lagrange multipliers.\citep{ghalanos2012rsolnp}  

We use a variety of flexible and nonparametric classification algorithms for the treatment and outcome model ensembles. These algorithms include gradient boosting;\citep{chen2016xgboost} random forests with varying forest sizes;\citep{wright2017ranger} $\ell_1$-penalized lasso regression; and elastic net regressions, weighting the $\ell_1$ penalty at $\alpha \in \{0.25,0.50,0.75\}$ and the $\ell_2$ penalty at $1-\alpha$.\citep{friedman2010regularization} The lasso and elastic net regressions internally perform 5-fold cross-validation to select the optimal regularization strength. Web Table \ref{tab:ensemble} provides additional details on the candidate algorithms used in the treatment and outcome model ensembles.

\section{Numerical studies} \label{sec:simulations}

We conduct numerical studies to assess the operating characteristics of various estimators in the finite sample-size setting, following the simulation design of Yang et al., who examined multi-valued treatments but focused on continuous outcomes. Li and Li also used this design to assess the comparative performance of matching, weighting, and subclassification estimators using the GPS. Specifically, we generate potential outcomes under each of $J=6$ treatments assigned to $n=10000$ individuals and estimate ATEs for each of the 15 pairwise comparisons, denoted $\boldsymbol{\lambda}_{j_{\text{Ref}}, \,j_{\text{Alt}}}$. We iterate this process $H=1000$ times, and for each simulation run $h$, calculate mean absolute bias, coverage probability of 95\% confidence intervals, and confidence interval widths (defined in Web Appendix A). We use the influence curve for each estimator to estimate standard errors. 

Our focus on the ATE for the sample rather than for a population has specific implications for our simulation design. While this choice complicates inference relative to the simulation, we opted for a standard simulation approach that generates a new sample with each run. This method allows us to assess the robustness of our estimator under varying conditions, providing a more comprehensive evaluation of its performance. Although one could argue for an alternative approach --- treating the sample as fixed and repeatedly sampling treatment assignments using propensity scores as sample probabilities --- we believe our chosen method offers valuable insights into how the estimator behaves across different samples.

We evaluate two different TMLE implementations for the treatment model, both estimated with SL: TMLE using multinomial treatment probabilities (hereafter, TMLE-multinomial) and TMLE using binomial treatment probabilities (TMLE-binomial), both estimated with SL. The outcome model uses binomial outcome probabilities estimated with SL. We also include three non-doubly-robust estimators for comparison, each also estimated with SL: IPTW using multinomial treatment probabilities (IPTW-multinomial) or binomial treatment probabilities (IPTW-binomial), and G-computation. TMLE-multinomial is the approach we use in the application because it is doubly-robust and reflects the multinomial stochastic structure of the treatment probabilities. TMLE-binomial closely aligns with the software implementation of TMLE, which incorrectly assumes binomial treatment probabilities. The IPTW and G-computation estimators are included to demonstrate the doubly-robust property of TMLE.

\subsection{Multinomial treatment assignment}\label{treatment-gen}

We assign treatments according to six covariates: $x_{1i}$, $x_{2i}$, and $x_{3i}$ are generated from a multivariate normal distribution with means zero, variances of (2,1,1) and covariances of (1, -1, -0.5). The latter three covariates are generated as follows: $x_{4i} \sim \text { Uniform }[-3,3]$, $x_{5i} \sim \chi_{1}^{2}$, and $x_{6i} \sim \operatorname{Bern}(0.5)$, with the covariate vector $\mathbf{x}_i^{\top} = (\mathbf{1}, x_{1i}, x_{2i}, \ldots, x_{6i})$. The treatment model follows the multinomial logistic model, $\left(\mathbbm{1} (a_{i}=1), \ldots, \mathbbm{1} (a_{i}=J)\right) \mid \mathbf{x}_{i} \sim \operatorname{Multinom}\left(p_{1}\left(\mathbf{x}_{i}\right), \ldots, p_{J}\left(\mathbf{x}_{i}\right)\right)$, with $p_{j}\left(\boldsymbol{x}_{i}\right) = \frac{\exp \left(\mathbf{x}_{i}^{\top} \beta_{j}\right)}{\sum_{k}^{J} \exp \left(\mathbf{x}_{i}^{\top} \beta_{k}\right)}$, where $\beta_{1}^{\top} = (0,0,0,0,0,0,0)$, $\beta_{2}^{\top} = \kappa_{2} \, \times(0, 1, 1, 2, 1, 1, 1)$, $\beta_{3}^{\top} = \kappa_{3} \times(0, 1, 1, 1, 1, 1, -5)$, $\beta_{4}^{\top} = \kappa_{4} \, \times(0, 1, 1, 1, 1, 1, 5)$, $\beta_{5}^{\top} = \kappa_{5} \times(0, 1, 1, 1, -2, 1, 1)$, and $\beta_{6}^{\top} = \kappa_{6} \, \times(0, 1, 1, 1, -2, -1, 1)$. Different values of $\kappa$ are selected to vary the amount of overlap, or similarity in the distributions of the propensity scores across treatment levels, and thus produce three treatment model settings. Following Li and Li, we use $\left(\kappa_{2}, \ldots, \kappa_{6}\right)=(0.1, 0.15, 0.2, 0.25, 0.3)$ to simulate experiments with ``adequate overlap''; i.e., similarity in the distributions of propensity scores across treatment groups. Treatment probabilities range from 8.7\% to 25.6\% in this setting (Web Figure \ref{simulation_A}). In a different setting, we set $\left(\kappa_{2}, \ldots, \kappa_{6}\right)=(0.4, 0.6, 0.8, 1.0, 1.2)$, which are the same values used in Yang et al., to simulate an ``inadequate overlap'' scenario with strong propensity tails; simulated treatment probabilities range from 3.9\% to 33.9\% in this setting. We examine a third  setting that is reflective of a RCT. In the RCT setting, $\left(\kappa_{2}, \ldots, \kappa_{6}\right)=(0, 0, 0, 0, 0)$ so that the covariates have no influence in assignment treatment; i.e., there's a $1/6$ probability of treatment, on average.

\subsection{Outcome generation}\label{outcome-gen}
In each simulation run, we generate potential outcomes using the Bernoulli model, $y_{i}(j) \sim \operatorname{Bern}\left(\frac{\exp \left(\mathbf{x}_{i}^{\top} \gamma_{j} + \mathbbm{1} (a_{i}=j)\right)}{1+\exp \left(\mathbf{x}_{i}^{\top} \gamma_{j} + \mathbbm{1} (a_{i}=j)\right)}\right)$. We simulate three different settings to vary the event rate, or the probability an outcome is observed under each treatment level. In a ``low event rate'' setting, we generate outcome event rates using $\gamma_{1}^{\top} = (-4, 1, -2, -1, 1, 1, 1)$, $\gamma_{2}^{\top} = (-6, 1, -2, -1, 1, 1, 1)$, $\gamma_{3}^{\top} = (-2, 1, -1, -1, -1, -1, -4)$, $\gamma_{4}^{\top} = (1, 2, 1, 2, -1, -1, -3)$, $\gamma_{5}^{\top} = (-2, 2, -1, 1, -2, -1, -3)$, and $\gamma_{6}^{\top} = (-3, 3, -1, 1, -2, -1, -2)$. This setting generates event rates that range from 3.5\% to 55.4\% (Web Figure \ref{simulation_Y}). In a ``moderate event rate'' setting, we use the same $\gamma_{j}$ values in Yang et al.: $\gamma_{1}^{\top} = (-1.5, 1, 1, 1, 1, 1, 1)$, $\gamma_{2}^{\top} = (-3, 2, 3, 1, 2, 2, 2)$, $\gamma_{3}^{\top} = (3, 3, 1, 2, -1, -1, -4)$, $\gamma_{4}^{\top} = (2.5, 4, 1, 2, -1, -1, -3)$, $\gamma_{5}^{\top} = (2, 5, 1, 2, -1, -1, -2)$, and $\gamma_{6}^{\top} = (1.5, 6, 1, 2, -1, -1, -1)$. Outcomes generated using these parameters range from 21.1\% to 99.6\%. Lastly, we specify $\gamma_{1}^{\top}, \ldots, \gamma_{6}^{\top} = (0, 0, 0, 0, 0, 0, 0)$ to study a setting where there is no treatment effect. In this setting, the outcome model does not use covariate information and the event rates are simulated at 73.1\%, on average.

\subsection{Results}\label{sim-results}

\begin{figure}[htbp]
	\centering
	\includegraphics[width=0.8\columnwidth]{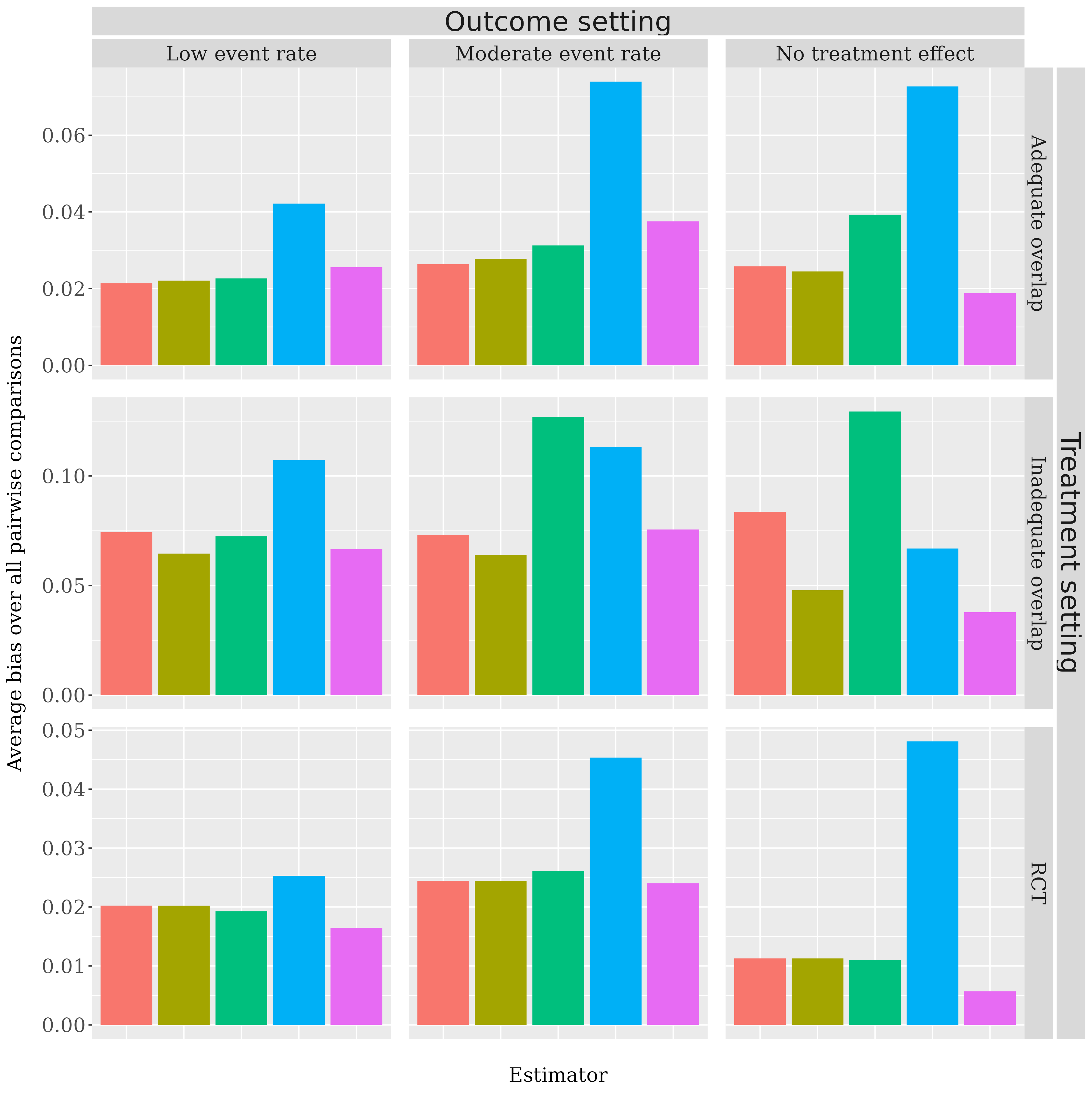}
	\caption{Average bias for the ATE over all 15 pairwise comparisons and 1000 simulated datasets. Estimator: \hspace{1mm}
		{\protect\tikz \protect\draw[color={Rred}] (0,0) -- plot[mark=square*, mark options={scale=2}] (0,0) -- (0,0);}\, TMLE-multi. (SL); \hspace{1mm}
	{\protect\tikz \protect\draw[color={Rgold}] (0,0) -- plot[mark=square*, mark options={scale=2}] (0,0) -- (0,0);}\, TMLE-bin. (SL); \hspace{1mm}
	{\protect\tikz \protect\draw[color={Rgreen}] (0,0) -- plot[mark=square*, mark options={scale=2}] (0,0) -- (0,0);}\, IPTW-multi. (SL); \hspace{1mm}
 		{\protect\tikz \protect\draw[color={Rblue}] (0,0) -- plot[mark=square*, mark options={scale=2}] (0,0) -- (0,0);}\, IPTW-bin. (SL); \hspace{1mm}
   		{\protect\tikz \protect\draw[color={Rpink}] (0,0) -- plot[mark=square*, mark options={scale=2}] (0,0) -- (0,0);}\, G-comp. (SL).} \label{bias_average_6}
\end{figure}

Misspecification of only the treatment model is expected to impact efficiency rather than consistency. Therefore, it is not unexpected that the performance of TMLE-multinomial in terms of bias is mixed. Our simulations reveal that, in scenarios with adequate overlap and low or moderate event rates, the TMLE-multinomial estimator exhibits less average bias than its binomial counterpart (Figure \ref{bias_average_6}). Conversely, under the condition of no treatment effect with adequate overlap, as well as across all scenarios of inadequate overlap, the TMLE-binomial estimator shows lower average bias, indicating more accurate estimation in these specific settings. In the RCT settings, average bias is comparable across the estimators. Interestingly, g-computation yields limited bias even without weighting, having lower bias than the IPTW estimators in most settings and the TMLE estimators in five of the nine settings.

\begin{figure}[htbp]
	\centering
	\includegraphics[width=0.8\columnwidth]{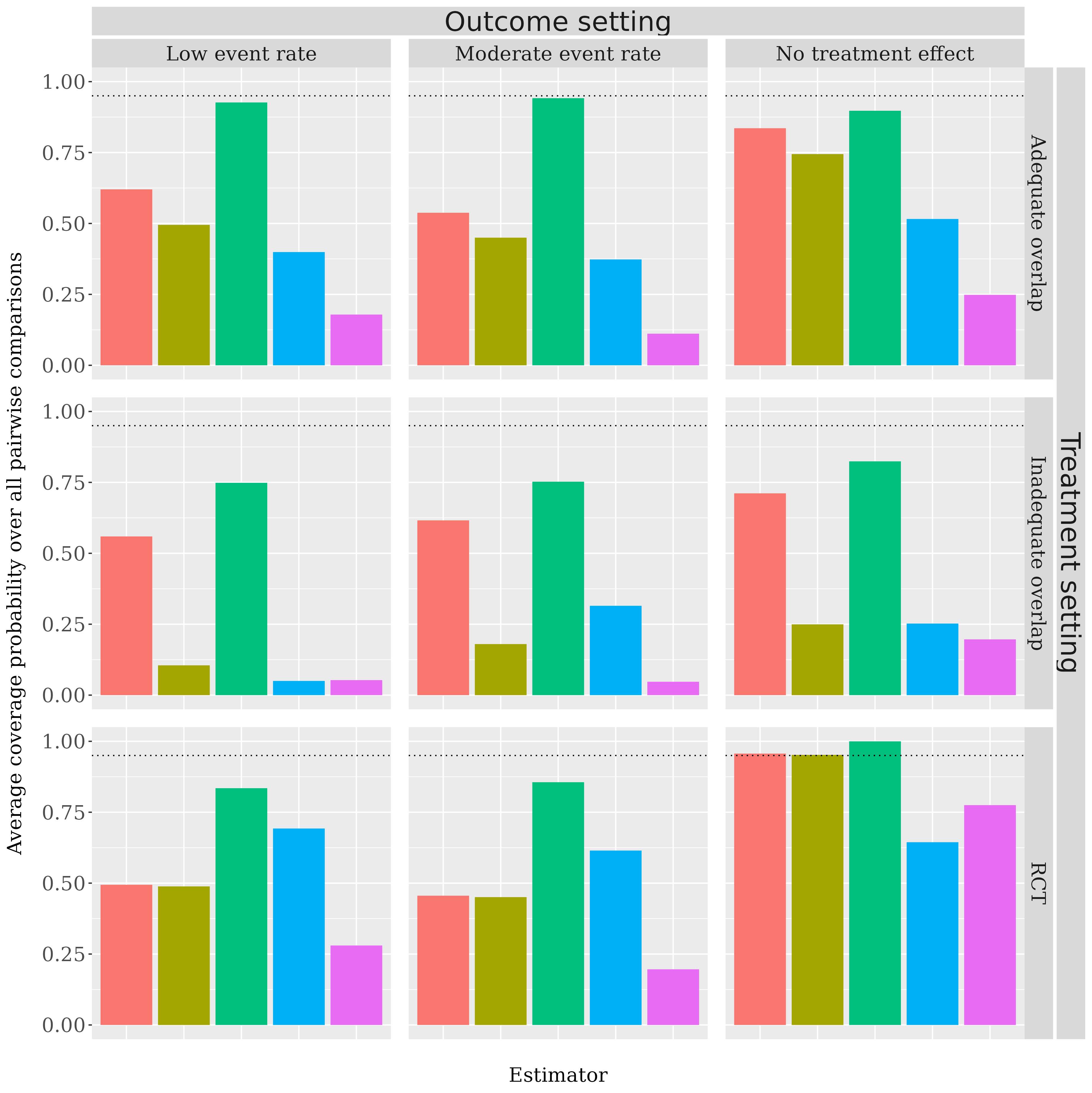}
	\caption{Average coverage probability for the ATE over all 15 pairwise comparisons and 1000 simulated datasets. Estimator:\\ \hspace{1mm}
		{\protect\tikz \protect\draw[color={Rred}] (0,0) -- plot[mark=square*, mark options={scale=2}] (0,0) -- (0,0);}\, TMLE-multi. (SL); \hspace{1mm}
	{\protect\tikz \protect\draw[color={Rgold}] (0,0) -- plot[mark=square*, mark options={scale=2}] (0,0) -- (0,0);}\, TMLE-bin. (SL); \hspace{1mm}
	{\protect\tikz \protect\draw[color={Rgreen}] (0,0) -- plot[mark=square*, mark options={scale=2}] (0,0) -- (0,0);}\, IPTW-multi. (SL); \hspace{1mm}
 		{\protect\tikz \protect\draw[color={Rblue}] (0,0) -- plot[mark=square*, mark options={scale=2}] (0,0) -- (0,0);}\, IPTW-bin. (SL); \hspace{1mm}
   		{\protect\tikz \protect\draw[color={Rpink}] (0,0) -- plot[mark=square*, mark options={scale=2}] (0,0) -- (0,0);}\, G-comp. (SL).} \label{cp_average_6}
\end{figure}

Figure \ref{cp_average_6}, which plots the average coverage probability for the ATE over all 15 pairwise comparisons, shows that TMLE-multinomial achieves superior coverage compared to TMLE-binomial. The exception involves the RCT setting where there is no treatment effect (the ninth simulation setting) where both TMLE implementations achieve the nominal coverage of 95\% represented by the dotted horizontal line. In this particular setting, IPTW-multinomial overcovers, and IPTW-binomial and G-computation undercover. When there is adequate overlap and no treatment effect (the third simulation setting), the coverage probability for both TMLE-multinomial and IPTW-multinomial exceeds 75\%, whereas their binomial counterparts do not achieve the expected coverage probability. All five estimators struggle in the inadequate overlap settings, which feature treatment probabilities that are close to zero. Our preferred implementation, TMLE-multinomial estimated using SL, has an average coverage rate between 46\% and 96\%.

Web Figure \ref{cp_average_6_GLM} provides coverage results for TMLE using multinomial treatment probabilities estimated using a multinomial Generalized Linear Model (GLM). This parametric implementation is a useful benchmark in the simulations because it uses the correct parametric treatment model, except in the RCT setting, where covariates have no role in the treatment model. TMLE-multinomial estimated using GLM has an average coverage rate between 60\% and 96\%. As expected, the GLM estimator performs well when it matches the data-generating model. However, this performance should not overshadow the advantages in choosing more flexible methods like SL-based TMLE and IPTW. These methods offer insurance against model misspecification at the cost of potentially being less efficient when the model is correctly specified.

Web Figure \ref{ciw_average_6}, which plots the average confidence interval widths over all comparisons, shows that in the adequate and inadequate overlap settings, TMLE-multinomial has appropriately wide confidence intervals, reflecting the variability of the treatment model estimator, while TMLE-binomial underestimates the true variability, yielding intervals that are too narrow. The difference between binomial and multinomial implementations is not as large when the TMLE is estimated using GLMs, as shown in Web Figure \ref{ciw_average_6_GLM}. In addition to having narrower confidence intervals, the average relative precision of the binomial approach exceeds the multinomial implementation in the adequate and inadequate overlap settings, and is comparable in the RCT settings (Web Figure \ref{rel_precision_6}). Relative precision is calculated as the variance of TMLE-multinomial estimated using GLM --- which correctly models the multinomial treatment assignment --- divided by the variance of each comparison estimator.

The estimated propensity scores play an important role in the influence curve, and consequently, for the confidence intervals for the estimated ATE. Web Figure \ref{simulation_est_A_diff}, which plots the absolute difference between the estimated and true treatment probabilities, provides insight into how well the methods estimate the treatment probabilities. The multinomial treatment model estimated via SL produces no detectable bias in the estimated treatment probabilities across treatment model settings. In contrast, the binomial treatment model estimated via SL exhibits more bias and greater variability in terms of the estimated treatment probabilities, and slightly underestimates the true treatment probabilities in each of the three treatment settings (Web Figures \ref{simulation_A} and \ref{simulation_est_A}). 

While the binomial approach estimated via SL shows more bias and greater variance in the estimated treatment probabilities, it also generally exhibits higher ratios of $ESS_j/n_j$ in both adequate and inadequate overlap settings (Web Figure \ref{simulation_est_ESS_ratio}). In the RCT setting, both treatment model implementations yield ESS ratios of one, indicating perfect overlap among the treatment groups. It is important to note that these ratios are not influenced by the average treatment probabilities but rather depend on the variance of the weights. Therefore, the observed differences in $ESS_j/n_j$ ratios are not arbitrary but are a reflection of the variance in the weights.

\subsubsection{$J=3$ treatments} Web Appendix C details the data-generating process (DGP) and simulation results for the case of $J=3$ treatments, which is the focus of the simulation studies of Yang et al. Similar to the case of $J=6$, the TMLE-multinomial estimator displays average biases that are comparable with those observed for the TMLE-binomial estimator (Web Figure \ref{bias_average_3}). The TMLE-multinomial estimator shows better coverage averaged across the three pairwise comparisons compared to TMLE-binomial (Web Figure \ref{cp_average_3}), except for the RCT setting with no treatment effect, where both estimators slightly overcover. 

\subsubsection{High-dimensional covariate space}\label{high-dimensional}
Our simulations incorporated a modest number of covariates (six) to assess estimator performance. To explore the impact of high-dimensional settings, we expanded our simulations to include 40 or 100 covariates. Detailed descriptions of the DGP for these scenarios, along with their respective simulation outcomes, can be found in Web Appendix D.

Our findings indicate an improvement in the performance of the TMLE-multinomial estimator, particularly in terms of bias and variance reduction. With an increase in the covariate count to 40, our estimator demonstrated parity in average bias with the TMLE-binomial estimator across all simulation scenarios (Web Figure \ref{bias_average_6_40}). Notably, the TMLE-multinomial estimator's advantage in average coverage was not apparent with 40 covariates (Web Figure \ref{cp_average_6_40}). While the confidence interval widths for all estimators narrowed as the number of covariates increased, the TMLE-multinomial estimator's confidence interval widths were equal to or more narrow than those observed in the binomial implementation (Web Figure \ref{ciw_average_6_40}). These patterns held when the number of covariates were further increased to 100 (Web Figure \ref{bias_cp_ciw_average_6_100}).

\subsubsection{Misspecified models\label{misspecification}}

To provide a comprehensive evaluation of the estimators under varying conditions of model specification, we include three additional scenarios that involve the omission of a single covariate, $x_{6i}$, from the estimation process. The scenarios are as follows: (i.) omitting $x_{6i}$ from the outcome model, resulting in its misspecification with the treatment model being correctly specified; (ii.) omitting $x_{6i}$ when estimating the treatment model, leading to its misspecification while the outcome model remains correctly specified; and (iii.), omitting $x_{6i}$ from both the outcome and treatment models, thereby introducing misspecification into both models.

In scenarios where the outcome or treatment model are misspecified (Web Figures \ref{bias_cp_ciw_average_6_misOut} and \ref{bias_cp_ciw_average_6_misTreat}), TMLE-multinomial yields lower or the same average bias compared to its binomial counterpart in all simulation settings, except when there is no treatment effect and adequate or inadequate overlap (the third and sixth settings). TMLE-multinomial also maintained its advantage in average coverage probability compared to its binomial counterpart in all settings except the RCT setting with no treatment effect (the ninth setting), where both estimators achieve nominal coverage. When both the outcome and treatment models are misspecified (Web Figure \ref{bias_cp_ciw_average_6_misBoth}), all estimators face an inevitable increase in average bias, and TMLE-multinomial maintains its advantage in average coverage probability compared to its binomial counterpart. This highlights its comparative robustness in the presence of double misspecification.

\section{Safety effects of antipsychotic drug treatments} \label{sec:application}

We utilize patient-level data collected by the Centers for Medicare \& Medicaid Services (CMS) from California, Georgia, Iowa, Mississippi, Oklahoma, South Dakota, and West Virginia.\cite{poulos2023antipsychotics} These states are selected for their racial diversity and lower rates of managed care penetration. Our cohort includes Medicare and dual Medicaid--Medicare beneficiaries aged 18-64 years who resided in one of the seven states; i.e., all patients in the cohort have the same public health insurer. We include patients who were diagnosed with schizophrenia, bipolar I disorder, or severe MDD, initiated one of six antipsychotic drugs between 2008 and 2010, and who were relatively new monotherapy users. The latter requirement restricts the cohort to patients who have not used any antipsychotic drugs within the six months prior to treatment assignment. Restricting the initial cohort of size $n=64120$ to patients who complete the three-year follow-up or died before the three-year follow-up yields a final cohort of size $n = 38762$. As in the numerical studies, inferences are made conditional on the study population; thus, we do not assume that our study population is a sample from a larger population. 

The study design is intention-to-treat: patients are non-randomly assigned to the first drug filled regardless of initial dose or duration, except that they must remain on the assigned drug for the first three months for those who are alive during this period. Each of the six antipsychotic drug treatments are available to each patient. We focus on four commonly-used SGAs, denoted drugs ``B'', ``C'', ``D'', and ``E'', a Reference SGA thought to have lower cardiometabolic risk relative to the other SGAs, and a FGA known for having low cardiometabolic risk (denoted drug ``A''). Table \ref{tb:outcomes} summarizes the observed three-year safety outcomes by antipsychotic drug. There is a wide range of treatment assignment rates, with drug A initiated in only 6\% of the cohort and drug C initiated in 26.5\%. The Reference drug was initiated in less than 1 in 5 patients. Across all treatment arms, incident diabetes is 9.3\% and all-cause death 5\%. While the Reference drug is associated with the lowest risk of mortality (3.4\%), drug B is associated with the lowest observed risk of diabetes (6.7\%).

The CMS data include person-level demographic, diagnostic, and pharmacy, behavioral health, physical health, laboratory tests, and other service use information measured six months prior to drug initiation. Tables \ref{tab:binary} and \ref{tab:continous} in the Appendix summarize the baseline covariates included in the outcome and treatment models by treatment drug. Selection into treatment is apparent with 42.8\% of Reference drug initiators having schizophrenia compared to 87\% of drug A initiators, and 11.6\% of drug A initiators having a psychiatric comorbidity compared to 21.9\% of drug C initiators.

\subsection{Censoring}
Our study has an all-cause censoring rate of about 27\%, as shown in the second column of Web Table \ref{tab:censoring}, which is non-negligible. While the mean days to the end of follow-up may not vary substantially across treatment levels for the initial cohort of size $n=64120$, this does not negate the potential for bias due to censoring. The types of censoring events include the conclusion of the study period (19.3\%), loss of insurance coverage (6.3\%), and turning 65 years (1.8\%). To assess patient retention across treatment groups, Kaplan-Meier curves were plotted for the initial cohort over 36 months in Web Appendix E. The Reference drug showed the highest retention rates, followed closely by drug A. Drugs B, C, D, and E displayed similar patterns of retention decline, with drug E showing a noticeable drop towards the end of the study period.

To further investigate the impact of treatment on survival time, we examined the Restricted Mean Survival Time (RMST) differences between the Reference drug and each comparator drug (Web Table \ref{rmst}). The RMST estimates, both with and without covariates, revealed varying effects across different antipsychotics. For example, drug E had a positive RMST difference in both models, suggesting better survival time compared to the Reference drug. Conversely, drug B had a negative RMST difference when covariates were included, indicating worse survival time. We also fit Cox proportional hazards models to assess the impact of treatment assignment on the hazard of being censored (Web Table \ref{cox_ph}). Both models with and without covariates yielded statistically significant results for the Wald test, indicating that the antipsychotics collectively have a significant influence on the hazard of being censored. The test for the assumption of proportional hazards indicated that both models violated this assumption, suggesting that the hazard ratios are not constant over time and that censoring could introduce bias into our estimates.

\subsection{Causal estimates} \label{results} 
We compare the ATE estimates using our preferred estimator, TMLE-multinomial, with the binomial treatment model version (TMLE-binomial), and non-doubly-robust estimators (IPTW and G-computation). Similar to the numerical studies, we estimate standard errors for the ATE using the influence curve in all estimators. The outcome and treatment models are both estimated by SL and each model relies on the same set of baseline covariates: binary indicators for state, race and ethnicity, and health status (Table \ref{tab:binary}), and count variables of health service utilization such as ER visits (Table \ref{tab:continous}) that are centered and scaled when fitting the models.  In the SL ensembles for the binomial outcome model and multinomial treatment model (Web Table \ref{tab:ensemble}), the gradient boosting classifier is favored, while random forests, elastic net regression, and lasso regression also receive positive weights.

Compared to the binomial implementation, TMLE-multinomial does better in terms of overlap, ensuring the treatment probabilities sum to one. Table \ref{tb:ESS} summarizes the estimated treatment probabilities for the multinomial and binomial implementations estimated with SL, along with the ESS and the ratio $ESS_j/n_j$ for each drug. The predicted probabilities of TMLE-binomial typically assume more extreme values and are more variable compared to TMLE-multinomial. The estimated values of the ratio are all above 0.8, except for the TMLE-binomial estimated ratio with respect to drug A, and the values of the TMLE-multinomial estimated ratio are equal or greater to those of TMLE-binomial, except for the drug C comparison. Values of the ratio $ESS_j/n_j$ that are close to one indicate a similarity in estimated propensity scores and adequate overlap among drug groups.

\begin{figure}[htbp] 
	\centering
	\includegraphics[width=0.8\columnwidth]{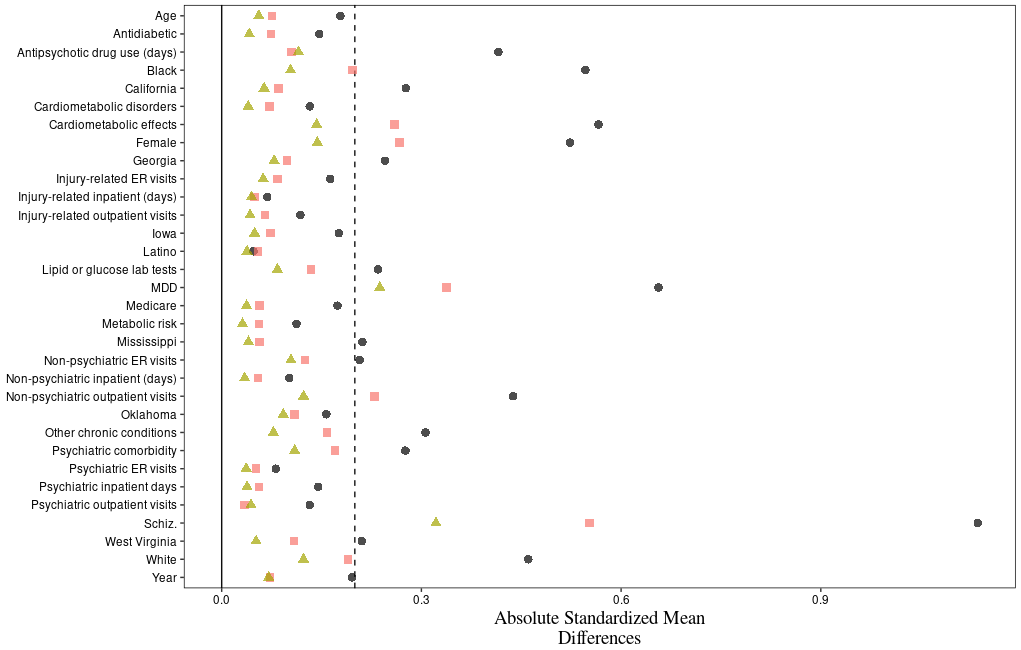}
	\caption{Covariate balance in terms of the maximum absolute standardized mean difference across treatment pairs. The dotted vertical line corresponds to the balance threshold of 0.2.\\ Treatment model: \hspace{1mm}
 		{\protect\tikz \protect\draw[color={black}] (0,0) -- plot[mark=*, mark options={scale=2}] (0,0) -- (0,0);}\, Unadjusted; \hspace{1mm}
		{\protect\tikz \protect\draw[color={Rred}] (0,0) -- plot[mark=square*, mark options={scale=2}] (0,0) -- (0,0);}\, Multinomial (SL); \hspace{1mm}
	{\protect\tikz \protect\draw[color={Rgold}] (0,0) -- plot[mark=triangle*, mark options={scale=2}] (0,0) -- (0,0);}\, Binomial (SL).} \label{covariate_balance}
\end{figure}

The high ratios of ESS to the sample size suggests there is overlap prior to weighting, but does not guarantee that the covariates are balanced. Figure \ref{covariate_balance} plots the balance of the covariates after weighting, measured in terms of the maximum absolute pairwise bias at each covariate, standardized by the pooled standard error of the covariate.\cite{lopez2017estimation,cobalt2023} The plot shows that balance was improved on all 32 covariates after adjustment, bringing all but five below the threshold of 0.2 for absolute mean differences, as suggested by McCaffrey et al.\cite{mccaffrey2013tutorial} It is worth noting that Austin\cite{austin2007performance} recommends a smaller cutoff of 0.1 for assessing balance on covariates. While both binomial and multinomial approaches improved balance, the binomial approach provided better balance for most covariates.

\begin{figure}
     \centering
     \begin{subfigure}[b]{0.49\columnwidth}
         \centering
         \includegraphics[width=\columnwidth]{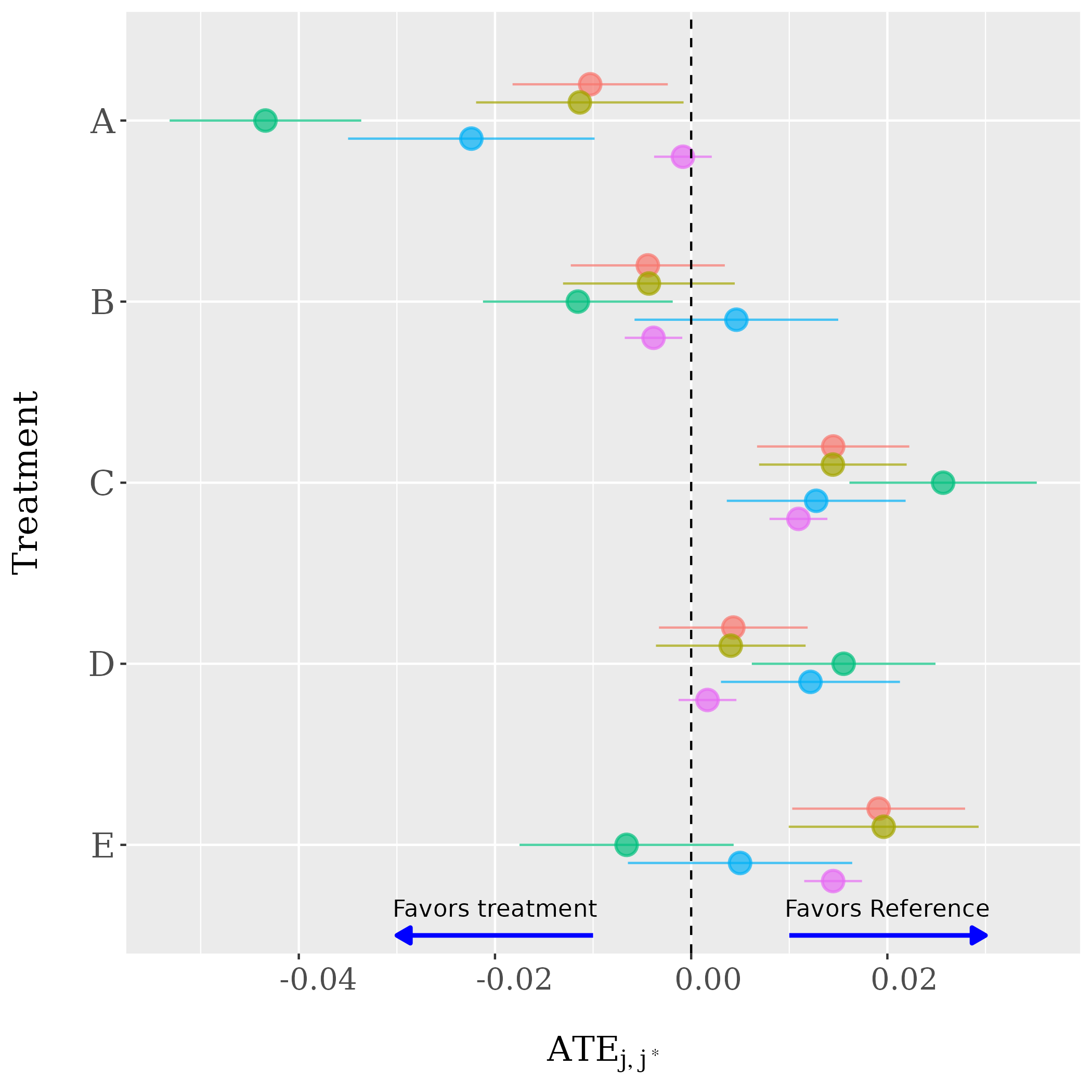}
         \caption{Diabetes diagnosis or death within 36 months.}
         \label{fig:combined}
     \end{subfigure}
     \hfill
     \begin{subfigure}[b]{0.49\columnwidth}
         \centering
         \includegraphics[width=\columnwidth]{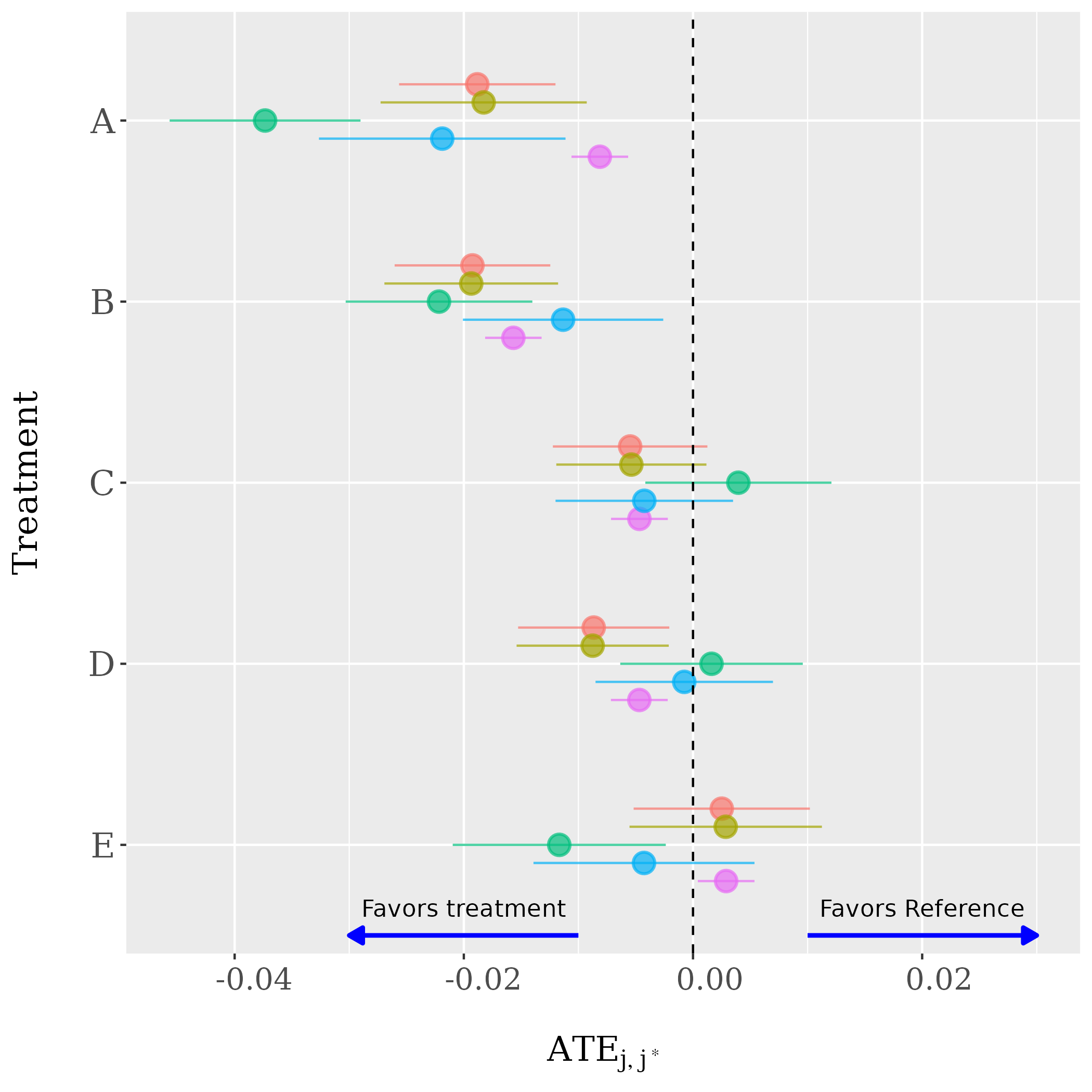}
         \caption{Diabetes diagnosis within 36 months.}
         \label{fig:diabetes}
     \end{subfigure}
     \hfill
     \begin{subfigure}[b]{0.49\columnwidth}
         \centering
         \includegraphics[width=\columnwidth]{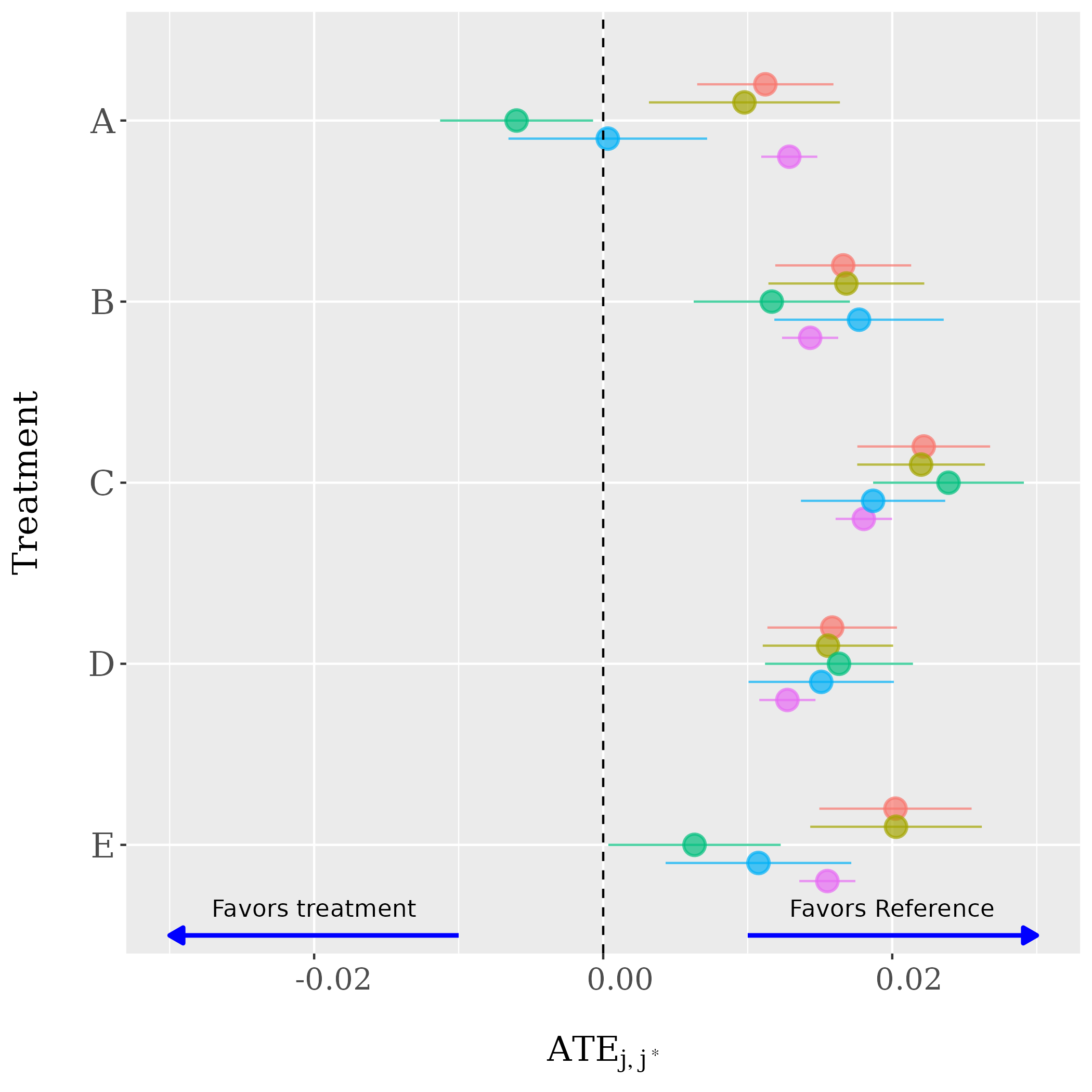}
         \caption{Death within 36 months.}
         \label{fig:death}
     \end{subfigure}
          \hfill
     \begin{subfigure}[b]{0.49\columnwidth}
         \centering
         \includegraphics[width=\columnwidth]{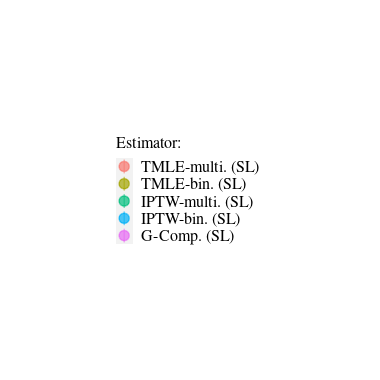}
     \end{subfigure}
        \caption{ATE estimates for each pairwise comparison (relative to Reference drug). Horizontal ranges are 95\% confidence intervals calculated using standard errors estimated from the influence function.}
        \label{ATEs}
\end{figure}

Figure \ref{ATEs} presents the estimated ATE for each treatment drug relative to the Reference drug on the combined outcome of diabetes diagnosis or death within 36 months, or each outcome separately. The TMLE-multinomial estimate indicates moving patients from the Reference to drug A yields a 1.0 [0.2, 1.8] percentage point reduction in diabetes incidence or death (Figure \ref{fig:combined}). Relative to the unadjusted risk of diabetes or death among those treated with the Reference (13.3\%), the point estimate of this ATE represents a 7.5 percentage point reduction. Moving patients from the Reference to Drugs C or E yields a 1.4 [0.7, 2.2] or 1.9 [1.0, 2.8] percentage point increase in the risk of diabetes or death, respectively. For the remaining two pairwise comparisons, the confidence intervals cover zero. The ATEs estimated using TMLE-binomial are similar in magnitude compared to those from TMLE-multinomial, and the interpretation of these results do not depend on the treatment model distribution used for the TMLE. However, the interpretation of the results do vary in certain comparisons if a non-doubly-robust estimator is used rather than TMLE. 

The finding that drug A is favorable to the Reference in terms of death or diabetes can be explained by a reduction in diabetes risk rather than mortality: moving patients from the Reference to drug A confers a 1.9 [1.2, 2.6] percentage point reduction in diabetes risk (Figure \ref{fig:diabetes}). Relative to the unadjusted rate of diabetes among those treated with Reference (10.2\%), this point estimate represents a 18.5 percentage point reduction. There is an equivalent size reduction in diabetes risk favoring drug B over the Reference, 1.9 [1.2, 2.6], and a smaller treatment effect favoring drug D over the Reference, 0.9 [0.2, 1.5]. 

The Reference drug is the safest in terms of mortality risk: moving patients from the Reference to the treatment drugs would increase the risk of death, with percentage point increases ranging from 1.1 [0.7, 1.6] to 2.2 [1.8, 2.7] corresponding to drugs A and C, respectively (Figure \ref{fig:death}). Relative to the unadjusted rate of mortality in the Reference group (3.4\%), these point estimates represent a 32.4 to 64.7 percentage point reduction in mortality.

\section{Discussion}\label{discussion}

Our research focused on estimating the pairwise ATEs of antipsychotic drugs using TMLE, a doubly-robust estimator, implemented with a multinomial treatment model. The study offers valuable insights into the comparative effectiveness of these treatments, particularly for individuals with SMI. We first discuss the comparative performance of our estimator in numerical studies. We then examine the clinical implications of our findings in the empirical application, discuss limitations of the study, and suggest directions for future research.

\subsection{Estimator performance in numerical studies}

Simulation studies demonstrate when estimating pairwise ATEs with multi-valued treatments, our TMLE implementation using a multinomial treatment model yields better coverage than the binomial implementation. This finding is in line with the theoretical properties of doubly-robust estimators, such as TMLE, and is not just a finite sample size finding. These results underscore the importance of using a correct probability distribution for the treatment model. The average coverage probabilities of 95\% confidence intervals for the TMLE-binomial estimator are generally lower compared to the TMLE-multinomial estimator, except in the RCT setting with no treatment effect. However, it is important to note that IPTW and G-estimation also exhibit lower coverage in multiple scenarios. We recognize that coverage is just one aspect of estimator performance and should be considered alongside other metrics like bias and efficiency. The bias yielded by the TMLE-binomial estimator is generally similar to, or smaller than, the bias of TMLE-multinomial. This suggests that the lower coverage observed for TMLE-binomial may be due to bias in the standard error estimates, possibly stemming from errors in the estimated probabilities of treatment used in the influence curve. While it is beyond the scope of this study to correct these standard errors, this is an important avenue for future research, especially since TMLE-binomial was more precise in some cases.

While our simulation results generally show that the TMLE-multinomial estimator achieves closer to nominal coverage, it is important to consider this finding in the context of other performance metrics such as bias and efficiency. Achieving nominal coverage with an estimator that exhibits greater bias and less efficiency may not be universally preferable. For example, the wider confidence intervals observed for TMLE-multinomial could be indicative of overestimation of standard errors. This overestimation could, in turn, result in higher coverage rates that may not necessarily reflect superior estimator performance. The estimator's wider confidence intervals may be advantageous in some scenarios but could also indicate a trade-off between bias and variance. Future work should aim to refine the standard error estimation process for TMLE-multinomial to achieve a more balanced performance across different simulation settings.

In our simulations, we found that the TMLE-binomial estimator often exhibited similar or even smaller bias compared to the TMLE-multinomial estimator, suggesting that its lower coverage rates may stem from biases in the estimated standard errors rather than the point estimates. This was particularly evident in four of the nine scenarios considered, including all three settings with inadequate overlap. Therefore, for analysts primarily concerned with bias, especially in the presence of significant imbalances in the covariate distribution across treatment levels, the binomial approach may be more appealing despite its lower coverage rates.

\subsection{Application: Implications, limitations, and future directions}\label{limitations}

The paper presents, to the best of our knowledge, the first doubly-robust estimates of the relative safety of specific antipsychotic drugs for individuals with SMI. We find a reduction in cardiometabolic risk of a relatively infrequently used FGA (drug A), which has been shown to have a generally low cardiometabolic risk among antipsychotic drugs, relative to a more popular drug, a SGA (Reference drug), thought to have a more favorable safety profile relative to other SGAs. The estimated percentage point reduction of initiating Drug A rather than the Reference drug on 36-month diabetes incidence or death is 1.0 [0.2, 1.8]. This estimate is driven by a reduction in diabetes risk rather than mortality, and is targeted to a clinically meaningful population --- one for which an antipsychotic drug will be prescribed. Below, we outline key limitations of our study and suggest directions for future research.

First, our study, like any observational research, is susceptible to unmeasured confounding and treatment adherence variations. For example, prescribers' subjective drug risk assessments and unobserved patient factors could introduce bias. Prior studies share these limitations and often rely on regression-based methods, making them more vulnerable to confounding and model misspecification. Additionally, in an intent-to-treat study, variations in adherence between oral and injectable forms of the drug are part of the treatment effect and could introduce meaningful biases in the estimated treatment effects.

Second, the target population in our study comprises patients within the same public health insurance system across different states. Our results are therefore conditional on this combined cohort and may not be generalizable to individual state populations. We acknowledge the methodological challenges associated with pooling data from different states into a single analysis. While all patients in our study have the same public insurer and access to the same set of treatments, we recognize that this does not fully account for potential state-specific variations in healthcare practices, patient demographics, or other unobserved confounders. To mitigate this, we included patients' state as a covariate in our models. We understand that this approach may not fully capture the complexity introduced by combining data from different states. Future work could explore more robust methods for accounting for state-level heterogeneity, such as a meta-analysis approach or multi-level modeling.

Third, a limitation of our study is the high rate of censoring, about 27\%. While we have employed Kaplan-Meier survival curves and Cox proportional hazards models to better understand the censoring mechanism and its potential impact on our results, the potential for bias due to censoring remains. The test for the assumption of proportional hazards indicated that this assumption was violated, suggesting that the hazard ratios are not constant over time and that censoring could introduce bias into our estimates. Future studies could explore more advanced methods for handling censoring, such as inverse probability of treatment and censoring weighting (IPTCW),\cite{gruber2022data} to provide more robust estimates.


\section*{Acknowledgments}
Poulos and Normand were supported by QuantumBlack-McKinsey and Company (A42960) to Harvard Medical School. Horvitz-Lennon was partially supported by R01-MH106682 from the National Institute of Mental Health.

\section*{Data availability statement}

The data that support the findings of this study are available from the Centers for Medicare \& Medicaid Services (CMS). Restrictions apply to the availability of these data, which were used under license for this study. Data are available at \url{https://www.cms.gov} with the permission of the CMS. \textsf{R} code to reproduce the results of the numerical studies, as well as code and simulated data to illustrate the empirical application are provided in the public repository: \url{https://github.com/jvpoulos/multi-tmle}.

%

%

%

\section*{Supporting information}

The following supporting information is available as part of the online article:\\
\noindent
	{\bf Web Appendix A}: \emph{Performance metrics used in numerical studies.} We define the three performance metrics used to evaluate the TMLE implementations.\\
	{\bf Web Appendix B}: \emph{Additional descriptive plots and results for numerical studies ($J=6$ treatment levels).} We provide descriptive plots of the DGP for the case of $J=6$ treatment levels, as well as additional simulation results.\\
	{\bf Web Appendix C}: \emph{Numerical studies for $J=3$ treatment levels.} We describe and provide descriptive plots of the DGP for the case of $J=3$ treatment levels, and provide simulation results.\\
        {\bf Web Appendix D}: \emph{Numerical studies with high-dimensional covariate space.} We describe the DGP with 40 and 100 covariates and provide simulation results.\\
        {\bf Web Appendix E}: \emph{Numerical studies with misspecification.} We provide simulation results in scenarios where either the treatment model, outcome model, or both models are misspecified.\\
        {\bf  Web Appendix F}: \emph{Additional descriptive statistics and results from the empirical application.} We provide descriptive statistics of censoring for the initial cohort of $n=64120$ patients, and summary statistics for the cross-validated error and weights for classification algorithms in SL ensembles.

\clearpage
\begin{appendix}

\section{Descriptive summaries for application\label{app1}}

\begin{table}[htbp]
\centering
\begin{adjustbox}{max width=\textwidth}
\begin{threeparttable}
	\caption{Summary statistics of binary baseline covariates, by assigned antipsychotic drug ($n = 38762$).}
	\label{tab:binary}
\begin{tabular}{l|rr|rr|rr|rr|rr|rr|rr}
	\toprule
	\textbf{Variable} & \multicolumn{2}{c}{Reference} & \multicolumn{2}{c}{A} & \multicolumn{2}{c}{B} & \multicolumn{2}{c}{C} & \multicolumn{2}{c}{D} & \multicolumn{2}{c}{E} & \multicolumn{2}{c}{All}\\
	 & $n$ & $\%$ & $n$ & $\%$ & $n$ & $\%$ & $n$ & $\%$ & $n$ & $\%$ & $n$ & $\%$ & $n$ & $\%$ \\ 
	\midrule
	\emph{Sex:} & & & &  &  &  & &  &  & & &  & & \\
	Female & 3889 & 58.2 & 784 & 33.7 & 2160 & 34.3 & 5756 & 55.8 & 3954 & 40.0 & 1917 & 59.1 & 18460 & 47.6 \\ 
        \emph{Payer:} & & & &  &  &  & &  &  & & &  & & \\
	Dual & 5035 & 75.3 & 1629 & 70.0 & 4893 & 77.7 & 7365 & 71.4 & 7399 & 74.8 & 2285 & 70.5 & 28606 & 73.8 \\
	Medicare & 1651 & 24.7 & 699 & 30.0 & 1408 & 22.4 & 2944 & 28.6 & 2498 & 25.2 & 956 & 29.5 & 10156 & 26.2 \\ 
        \emph{Index year:} & & & &  &  &  & &  &  & & &  & & \\
	2008 & 3110 & 46.5 & 1179 & 50.6 & 3634 & 57.7 & 4937 & 47.9 & 5232 & 52.9 & 1628 & 50.2 & 19720 & 50.9 \\ 
        2009 & 2038 & 30.5 & 646 & 27.8 & 1501 & 23.8 & 2939 & 28.5 & 2528 & 25.5 & 895 & 27.6 & 10547 & 27.2 \\ 
        2010$^{\star}$ & 1538 & 23.0 & 503 & 21.6 & 1166 & 18.5 & 2433 & 23.6 & 2137 & 21.6 & 718 & 22.1 & 8495 & 21.9 \\
        \emph{State:} & & & &  &  &  & &  &  & & &  & & \\
	California & 3765 & 56.3 & 1180 & 50.7 & 3904 & 62.0 & 5470 & 53.1 & 5314 & 53.7 & 1563 & 48.2 & 21196 & 54.7 \\ 
	Georgia &  635 & 9.5 & 423 & 18.2 & 800 & 12.7 & 1544 & 15.0 & 1695 & 17.1 & 521 & 16.1 & 5618 & 14.5 \\  
	Iowa & 723 & 10.8 & 211 & 9.1 & 377 & 6.0 & 762 & 7.4 & 764 & 7.7 & 275 & 8.5 & 3112 & 8.0 \\
	Mississippi & 435 & 6.5 & 283 & 12.2 & 406 & 6.4 & 674 & 6.5 & 743 & 7.5 & 283 & 8.7 & 2824 & 7.3 \\ 
	Oklahoma & 704 & 10.5 & 143 & 6.1 & 437 & 6.9 & 1026 & 9.9 & 820 & 8.3 & 305 & 9.4 & 3435 & 8.9 \\ 
	South Dakota & 126 & 1.9 & 16 & 0.7 & 85 & 1.4 & 161 & 1.6 & 144 & 1.4 & 44 & 1.4 & 576 & 1.5 \\
	West Virginia & 298 & 4.5 & 72 & 3.1 & 292 & 4.6 & 672 & 6.5 & 417 & 4.2 & 250 & 7.7 & 2001 & 5.2 \\
        \emph{Race/ethnicity:} & & & &  &  &  & &  &  & & &  & & \\
        Black & 803 & 12.0 & 768 & 33.0 & 1062 & 16.9 & 1370 & 13.3 & 2132 & 21.5 & 514 & 15.9 & 6649 & 17.1 \\ 
	Latino & 749 & 11.2 & 268 & 11.5 & 695 & 11.0 & 1048 & 10.2 & 1102 & 11.1 & 325 & 10.0 & 4187 & 10.8 \\ 
	Other/missing &  448 & 6.7 & 146 & 6.3 & 514 & 8.2 & 563 & 5.5 & 728 & 7.4 & 178 & 5.5 & 2577 & 6.7 \\ 
	White & 4686 & 70.1 & 1146 & 49.2 & 4030 & 64.0 & 7328 & 71.1 & 5935 & 60.0 & 2224 & 68.6 & 25349 & 65.4 \\ 
        \emph{Primary diagnosis:} & & & &  &  &  & &  &  & & &  & & \\
        Bipolar I & 2042 & 30.5 & 178 & 7.7 & 1167 & 18.5 & 3642 & 35.3 & 1639 & 16.6 & 1053 & 32.5 & 9721 & 25.1 \\
	MDD & 1782 & 26.6 & 124 & 5.3 & 738 & 11.7 & 3064 & 29.7 & 1459 & 14.7 & 573 & 17.7 & 7740 & 20.0 \\ 
	Schiz. & 2862 & 42.8 & 2026 & 87.0 & 4396 & 69.8 & 3603 & 35.0 & 6799 & 68.7 & 1615 & 49.8 & 21301 & 55.0 \\ 
	\emph{Health status:} & & & &  &  &  & &  &  & & &  & & \\
	Psychiatric comorbidity  & 1250 & 18.7 & 271 & 11.6 & 886 & 14.1 & 2255 & 21.9 & 1525 & 15.4 & 587 & 18.1 & 6774 & 17.5  \\ 
	Metabolic risk & 178 & 2.7 & 27 & 1.2 & 77 & 1.2 & 195 & 1.9 & 166 & 1.7 & 75 & 2.3 & 718 & 1.9 \\ 
	Other chronic conditions  & 1548 & 23.1 & 324 & 13.9 & 1168 & 18.5 & 2712 & 26.3 & 1934 & 19.5 & 775 & 23.9 & 8461 & 21.8 \\ 
	\emph{Metabolic testing:} & & & &  &  &  & &  &  & & &  & & \\
	Lipid or glucose lab tests  & 1246 & 18.6 & 300 & 12.9 & 989 & 15.7 & 2247 & 21.8 & 1681 & 17.0 & 634 & 19.6 & 7097 & 18.3 \\
	\emph{Drug use:} & & & &  &  &  & &  &  & & &  & & \\
	Antidiabetic & 364 & 5.4 & 134 & 5.8 & 185 & 2.9 & 527 & 5.1 & 543 & 5.5 & 200 & 6.2 & 1953 & 5.0 \\ 
	Cardiometabolic disorders & 1798 & 26.9 & 516 & 22.2 & 1435 & 22.8 & 2866 & 27.8 & 2295 & 23.2 & 904 & 27.9 & 9814 & 25.3 \\ 
	Cardiometabolic effects & 4775 & 71.4 & 1073 & 46.1 & 3611 & 57.3 & 7508 & 72.8 & 5812 & 58.7 & 2338 & 72.1 & 25117 & 64.8\\  
\bottomrule
\end{tabular}
  \begin{tablenotes}[flushleft]
\item \emph{Notes:} 2010$^{\star}$ indicates summary statistics for the index years of 2010 and 2011. 
\end{tablenotes}
\end{threeparttable}
\end{adjustbox}
\end{table}

\begin{table}[tbhp]
\centering
\begin{adjustbox}{max width=\textwidth}
\begin{threeparttable}
	\caption{Summary statistics of selected continuous baseline covariates ($n = 38762$).} \label{tab:continous} 
 \begin{tabular}{lcccccc}
	\toprule
 \textbf{Variable} & \textbf{Drug} & $\mathbf{n_j}$ & \textbf{Min.} & \textbf{Mean} & \textbf{Max.} & $\mathbf{S.d.}$ \\   \midrule
Age & Reference &  6686 & 19.9 &   43.7 &    64.0 &   10.2 \\  
   & A &  2328 & 20.8 &   45.4 &    63.7 &   10.0 \\ 
   & B &  6301 & 20.2 &   44.9 &    64.2 &   10.1 \\ 
   & C & 10309 & 20.1 &   45.0 &    64.1 &    9.9 \\  
   & D & 9897 & 20.0 &   44.2 &    64.5 &   10.4 \\
   & E &  3241 & 20.0 &   43.6 &    64.1 &   10.1 \\ 
   \hline
 & All & 38762 & 19.9 &   44.5 &    64.5 &   10.2 \\ 
   \hline
Antipsychotic drug use  & Reference &  6686 &  0.0 &   96.0 &   183.0 &   71.2 \\  
   (days)& A &  2328 &  0.0 &  105.4 &   183.0 &   68.0 \\
   & B &  6301 &  0.0 &  124.6 &   183.0 &   65.2 \\  
   & C & 10309 &  0.0 &  102.2 &   183.0 &   70.8 \\ 
   & D & 9897 &  0.0 &  114.7 &   183.0 &   68.7 \\
   & E &  3241 &  0.0 &  115.7 &   183.0 &   68.2 \\
   \hline
 & All & 38762 &  0.0 &  109.3 &   183.0 &   69.7 \\ 
   \hline
Psychiatric ER visits & Reference & 6686 &  0.0 &    0.1 &     9.0 &    0.5 \\ 
   & A &  2328 &  0.0 &    0.2 &    10.0 &    0.6 \\ 
   & B &  6301 &  0.0 &    0.1 &    16.0 &    0.6 \\ 
   & C & 10309 &  0.0 &    0.2 &    13.0 &    0.6 \\ 
   & D &  9897 &  0.0 &    0.1 &    33.0 &    0.7 \\ 
   & E &  3241 &  0.0 &    0.1 &     8.0 &    0.5 \\ 
   \hline
 & All & 38762 &  0.0 &    0.1 &    33.0 &    0.6 \\ 
   \hline
Psychiatric outpatient & Reference &  6686 &  0.0 &    6.4 &   172.0 &   12.2 \\ 
 visits    & A &  2328 &  0.0 &    6.7 &   183.0 &   15.5 \\ 
   & B &  6301 &  0.0 &    6.1 &   183.0 &   13.8 \\ 
   & C & 10309 &  0.0 &    5.3 &   183.0 &   10.0 \\ 
   & D &  9897 &  0.0 &    7.1 &   183.0 &   16.5 \\ 
   & E &  3241 &  0.0 &    6.2 &   180.0 &   12.5 \\ 
   \hline
 & All & 38762 &  0.0 &    6.3 &   183.0 &   13.4 \\ 
   \hline
Psychiatric inpatient  & Reference &  6686 &  0.0 &    1.4 &   183.0 &    7.8 \\ 
 days  & A &  2328 &  0.0 &    2.7 &   106.0 &    9.6 \\ 
   & B &  6301 &  0.0 &    2.0 &   183.0 &    8.9 \\ 
   & C & 10309 &  0.0 &    1.9 &   183.0 &    7.8 \\ 
   & D &  9897 &  0.0 &    2.3 &   183.0 &    9.4 \\ 
   & E &  3241 &  0.0 &    1.7 &   165.0 &    8.1 \\ 
   \hline
 & All & 38762 &  0.0 &    2.0 &   183.0 &    8.6 \\ 
\bottomrule
\end{tabular}
  \begin{tablenotes}[flushleft]
\item \emph{Notes:} non-psychiatric or injury-related ER visits, outpatient visits, and inpatient days not shown due to space constraints.
\end{tablenotes}
\end{threeparttable}
\end{adjustbox}
\end{table}

\end{appendix}

\clearpage
\bibliography{references}%

\clearpage

\setcounter{table}{0} 

\begin{table}[tbhp]
  \centering
	\caption{Three-year safety outcomes. Number (percent) having outcome.} \label{tb:outcomes}
\begin{tabular}{lrrrr}
	\toprule
	\textbf{Antipsychotic} & Number of Patients    & Diabetes or Death  & Diabetes  & All-Cause Death  \\ 
	\midrule
	Reference & 6686 (17.2)  & 891 (13.3)   & 679 (10.2)  & 225 (3.4)  \\
	A & 2328 (6.0)    & 309 (13.3)   & 217 (9.3)  & 103 (4.4) \\ 
	B  &  6301 (16.2)  & 714 (11.3)   & 421 (6.7)  & 313 (5.0) \\
	C  & 10309 (26.5)  & 1602 (15.5)   & 989 (9.6) & 662 (6.4) \\
	D & 9897 (25.5)   & 1360 (13.7)   & 941 (9.5) & 470 (4.8) \\
	E & 3241 (8.3)    & 508 (15.7)   & 357 (11.0)  & 166 (5.1) \\ \hline
	All         & 38762 (100)   & 5384 (13.9)  & 3604 (9.3) & 1939 (5.0) \\
	 \bottomrule
\end{tabular}
\end{table}

\begin{table}[tbhp]
  \centering
	\caption{Summary statistics and ESS of estimated multinomial or binomial treatment probabilities using SL.} \label{tb:ESS}
	\resizebox{\columnwidth}{!}{
\begin{tabular}{lcccccccccccc}
	\toprule
	 & \multicolumn{6}{c}{TMLE-Multinomial} & \multicolumn{6}{c}{TMLE-Binomial} \\
	 \textbf{Antipsychotic} & Min. & Mean & Max. & S.d. & $ESS_j$ & $\frac{ESS_j}{n_j}$ & Min. & Mean & Max. & S.d. & $ESS_j$ & $\frac{ESS_j}{n_j}$ \\
	\midrule
	Reference & 0.023 & 0.172 & 0.545 & 0.076 & 5931 & 0.887 & 0.013 & 0.169 & 0.607 & 0.078 & 5863 & 0.877 \\
	A & 0.004 & 0.061 & 0.480 & 0.053 & 1906 & 0.818 & 0.003 & 0.058 & 0.496 & 0.054 & 1359 & 0.584  \\ 
	B & 0.019 & 0.163 & 0.466 & 0.076 & 5698 & 0.904 & 0.003 & 0.160 & 0.417 & 0.075 & 5331 & 0.846 \\
	C & 0.044 & 0.265 & 0.845 & 0.122 & 8734 & 0.847 & 0.033 & 0.263 & 0.845 & 0.133 & 8813 & 0.855 \\
	D & 0.026 & 0.254 & 0.615 & 0.087 & 9208 & 0.930 & 0.029 & 0.252 & 0.631 & 0.091 & 9186 & 0.928 \\
	E & 0.022 & 0.084 & 0.419 & 0.034 & 2959 & 0.913 & 0.021 & 0.080 & 0.340 & 0.033 & 2943 & 0.908 \\ 
	 \bottomrule
\end{tabular}
}
\end{table}

\end{document}


\begin{singlespacing}
	\maketitle  
\end{singlespacing}

\thispagestyle{empty}

\renewcommand{\abstractname}{Summary}
\begin{abstract}
	\noindent 
	{\bf Web Appendix A}: \emph{Performance metrics used in numerical studies.} We define the three performance metrics used to evaluate the TMLE implementations.\\
	{\bf Web Appendix B}: \emph{Additional descriptive plots and results for numerical studies ($J=6$ treatment levels).} We provide descriptive plots of the DGP for the case of $J=6$ treatment levels, as well as additional simulation results.\\
	{\bf Web Appendix C}: \emph{Numerical studies for $J=3$ treatment levels.} We describe and provide descriptive plots of the DGP for the case of $J=3$ treatment levels, and provide simulation results.\\
 	{\bf Web Appendix D}: \emph{Numerical studies with high-dimensional covariate space.} We describe the DGP with 40 and 100 covariates and provide simulation results.\\
   	{\bf Web Appendix E}: \emph{Numerical studies with misspecification.} We provide simulation results in scenarios where either the treatment model, outcome model, or both models are misspecified.\\
	{\bf  Web Appendix F}: \emph{Additional descriptive statistics and results from the empirical application.} We provide descriptive statistics of censoring for the initial cohort of $n=64120$ patients, and summary statistics for the cross-validated error and weights for classification algorithms in super learner ensembles.
\end{abstract}



\renewcommand{\figurename}{Web Figure}
\renewcommand{\tablename}{Web Table}

\pagenumbering{roman}
\pagebreak
\pagenumbering{arabic}

\section*{Web Appendix A: Performance metrics used in numerical studies}

We consider three metrics to evaluate the ability of each TMLE implementation to recover the true ATE. The first metric focuses on bias. 
\begin{definition}
{\em \textbf{Mean absolute bias}. The mean absolute difference between the true ATE and the estimated ATE when comparing reference $j_{\text{Ref}}$ to any other treatment $j_{\text{Alt}}$, averaged over $H$ simulations.} 
\begin{align*}
\text{Absolute bias} &= \frac{1}{H} \sum_{h=1}^{H} \left\{ \left| \widehat{\text{ATE}}^{(h)}_{j_{\text{Ref}}, \, j_{\text{Alt}}} - \text{ATE}^{(h)}_{j_{\text{Ref}}, \, j_{\text{Alt}}} \right| \right\} \\
&= \frac{1}{H} \sum_{h=1}^{H} \left\{ \left| \left(\hat{\mu}^{(h)}_{j_{\text{Alt}}} - \hat{\mu}^{(h)}_{j_{\text{Ref}}}\right) - \left(\mu^{(h)}_{j_{\text{Alt}}} - \mu^{(h)}_{j_{\text{Ref}}}\right)\right| \right\}; ~~ j_{\text{Alt}} \neq j_{\text{Ref}}, 
\end{align*}
where $\mu^{(h)}_{j_{\text{Alt}}}$ and $\mu^{(h)}_{j_{\text{Ref}}}$ are the averages of the true potential outcomes under treatment and reference, respectively, generated according to the Bernoulli model.
\end{definition}

The second metric assesses the performance of the influence function in terms of coverage of 95\% confidence intervals for the estimated ATE, $\widehat{\text{CI}}^{(h)} = \widehat{\text{ATE}}^{(h)}_{j_{\text{Ref}}, \, j_{\text{Alt}}} \pm  1.96 \hat{\sigma}^{(h)}$, where $\hat{\sigma}^{(h)}$ is the standard error for the ATE in simulation $h$.

\begin{definition}
{\em \textbf{Coverage probability}. The proportion of the estimated 95\% confidence intervals in the $H$ simulations that contain the true ATE.} 
\begin{equation*}
    \text{Coverage probability} = \frac{1}{H} \sum_{h=1}^{H}\mathbbm{1}\left\{\text{ATE}_{j_{\text{Ref}}, \, j_{\text{Alt}}}^{(h)}\in\widehat{\text{CI}}^{(h)}\right\} ; ~~ j_{\text{Alt}} \neq j_{\text{Ref}}. \label{eq:coverage}
\end{equation*}
\end{definition}
We also average the coverage probability and absolute bias metrics over all pairwise comparisons to provide a more concise summary of the implementations' performance.

The third performance metric is confidence interval width, which provides a measure of the variability of estimated ATE. 
\begin{definition}
{\em \textbf{Confidence interval width}. The difference between the upper and lower bounds of the estimated 95\% confidence intervals over the $H$ simulations.} 
\begin{equation*}
    \text{Confidence interval width} = \frac{1}{H}
    \sum_{h=1}^{H}\left\{ 2 \times 1.96 \hat{\sigma}_{n}^{(h)} \right\} ; ~~ j_{\text{Alt}} \neq j_{\text{Ref}}.
\end{equation*}
\end{definition}

\clearpage
\section*{Web Appendix B: Numerical studies descriptive plots and results for $J=6$ treatment levels}

\begin{figure}
     \centering
     \begin{subfigure}[b]{0.49\textwidth}
         \centering
         \includegraphics[width=\textwidth]{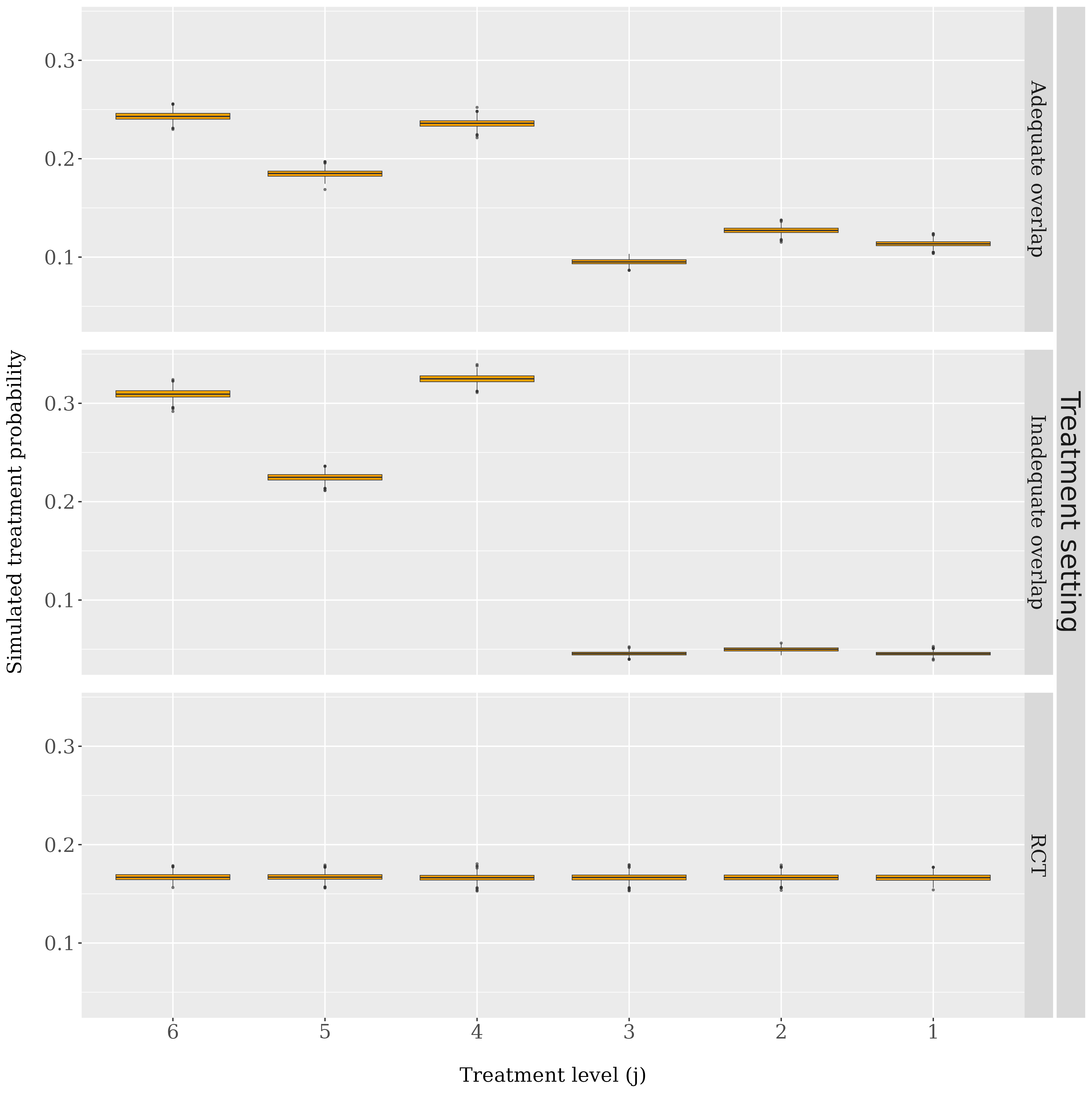}
         \caption{Simulated treatment probabilities}
         \label{simulation_A}
     \end{subfigure}
     \hfill
     \begin{subfigure}[b]{0.49\textwidth}
         \centering
         \includegraphics[width=\textwidth]{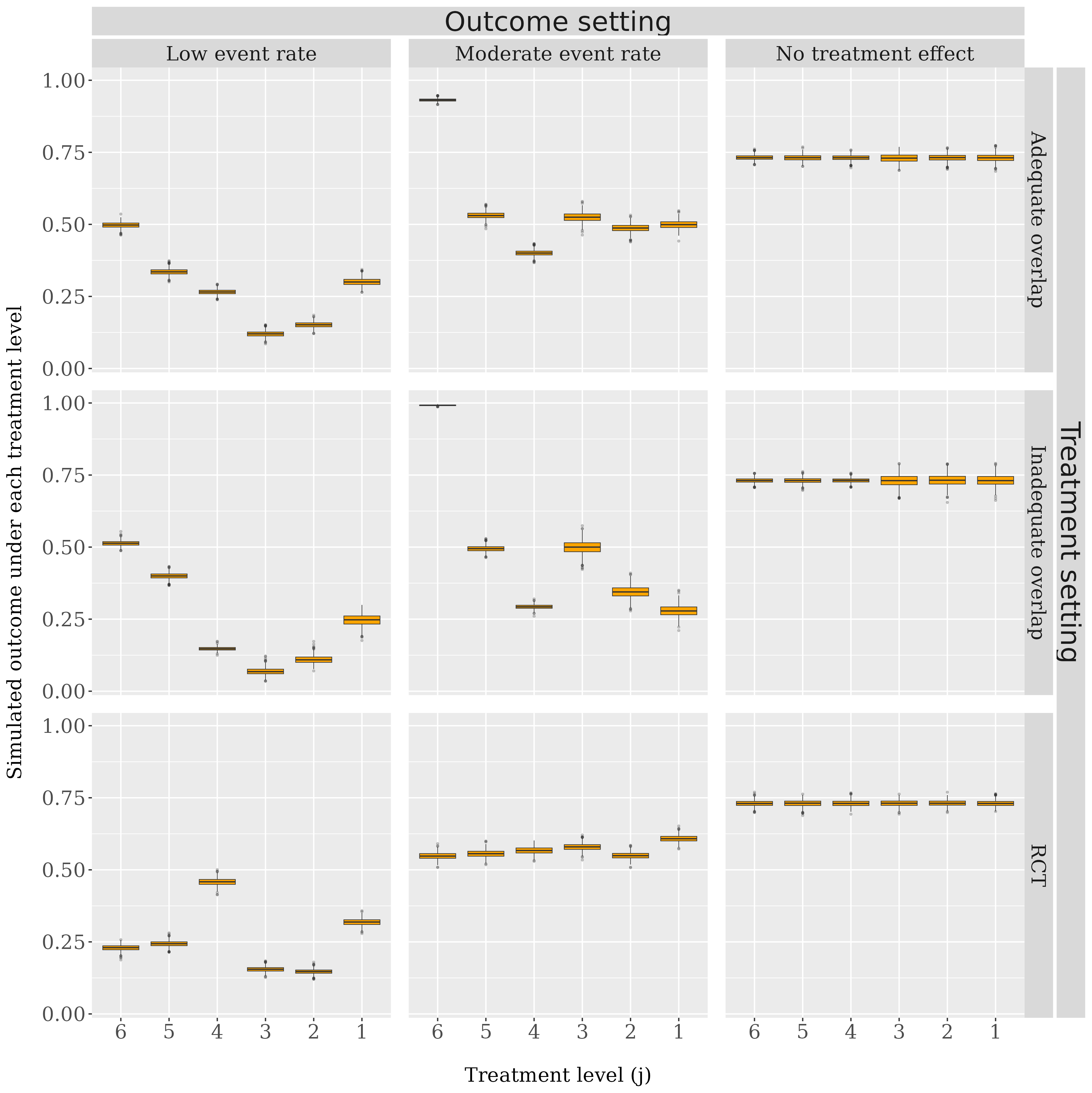}
         \caption{Simulated event rate}
         \label{simulation_Y}
     \end{subfigure}
        \caption{Box and whisker plots of simulated treatment probabilities (A) and event rates (B) under each treatment level, summarizing the median, the first and third quartiles, and outlying points of the distribution across 1000 simulation runs.}
        \label{simulation_A_Y}
\end{figure}

\begin{figure}
     \centering
     \begin{subfigure}[b]{0.49\textwidth}
         \centering
         \includegraphics[width=\textwidth]{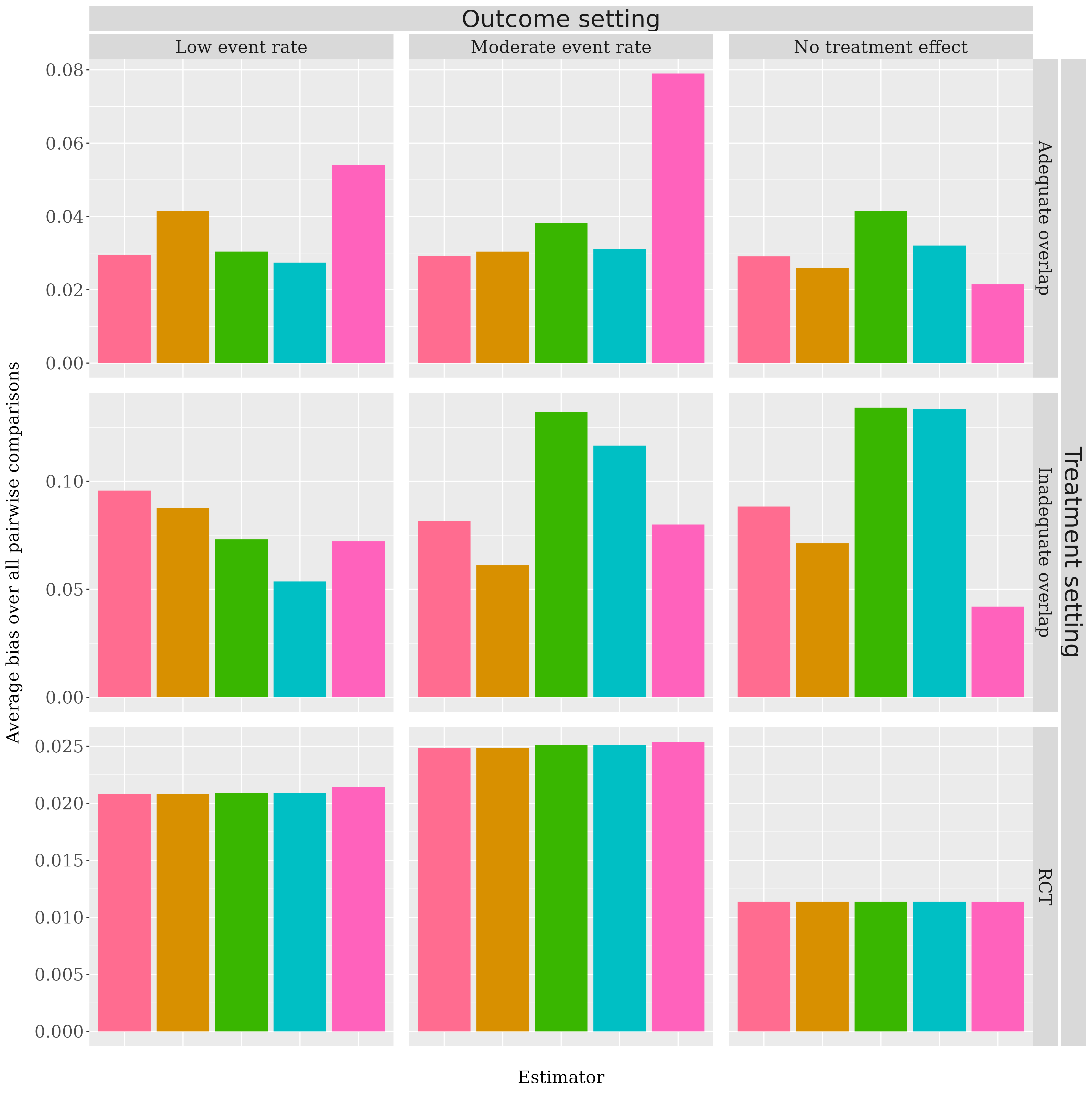}
         \caption{Average bias}
         \label{bias_average_6_GLM}
     \end{subfigure}
      \hfill 
           \begin{subfigure}[b]{0.49\textwidth}
         \centering
         \includegraphics[width=\textwidth]{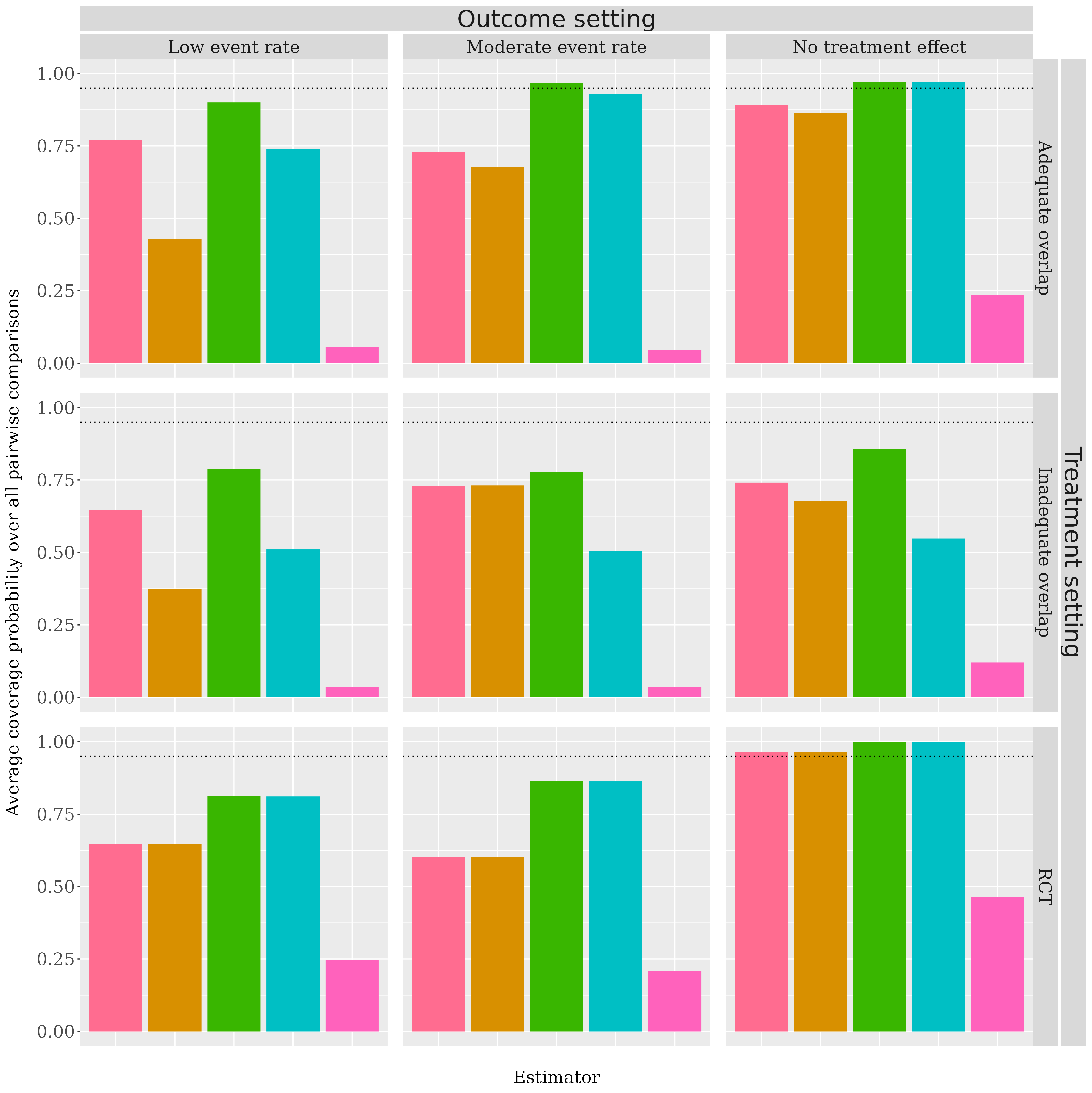}
         \caption{Average coverage probability}
         \label{cp_average_6_GLM}
     \end{subfigure}
     \hfill \vspace{5mm}
          \begin{subfigure}[b]{0.49\textwidth}
         \centering
         \includegraphics[width=\textwidth]{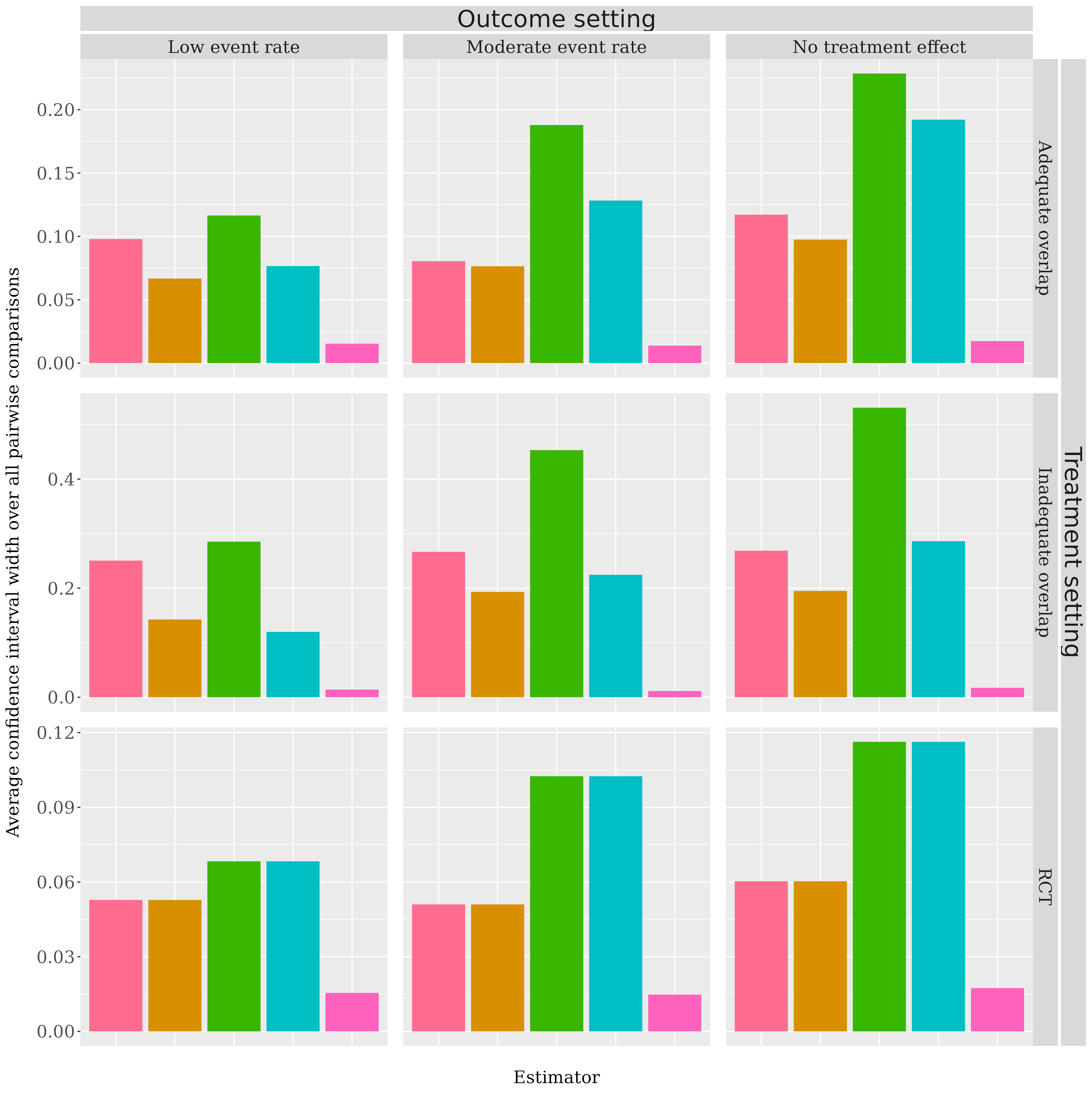}
         \caption{Average confidence interval widths}
         \label{ciw_average_6_GLM}
     \end{subfigure}
        \caption{Average bias (A), coverage probability (B), and confidence interval widths (C) for the ATE over all 15 pairwise comparisons and 1000 simulated datasets, using GLM to estimate the treatment and outcome models rather than super learner. Estimator: \hspace{1mm}
		{\protect\tikz \protect\draw[color={Rsalmon}] (0,0) -- plot[mark=square*, mark options={scale=2}] (0,0) -- (0,0);}\, TMLE-multi. (GLM); \hspace{1mm}
	{\protect\tikz \protect\draw[color={Rgold2}] (0,0) -- plot[mark=square*, mark options={scale=2}] (0,0) -- (0,0);}\, TMLE-bin. (GLM); \hspace{1mm}
	{\protect\tikz \protect\draw[color={Rgreen2}] (0,0) -- plot[mark=square*, mark options={scale=2}] (0,0) -- (0,0);}\, IPTW-multi. (GLM); \hspace{1mm}
 		{\protect\tikz \protect\draw[color={Rteal}] (0,0) -- plot[mark=square*, mark options={scale=2}] (0,0) -- (0,0);}\, IPTW-bin. (GLM); \hspace{1mm}
   		{\protect\tikz \protect\draw[color={Rmagenta}] (0,0) -- plot[mark=square*, mark options={scale=2}] (0,0) -- (0,0);}\, G-comp. (GLM).}
        \label{bias_cp_ciw_average_6_GLM}
\end{figure}

\begin{figure}[htbp]
	\centering
	\includegraphics[width=\columnwidth]{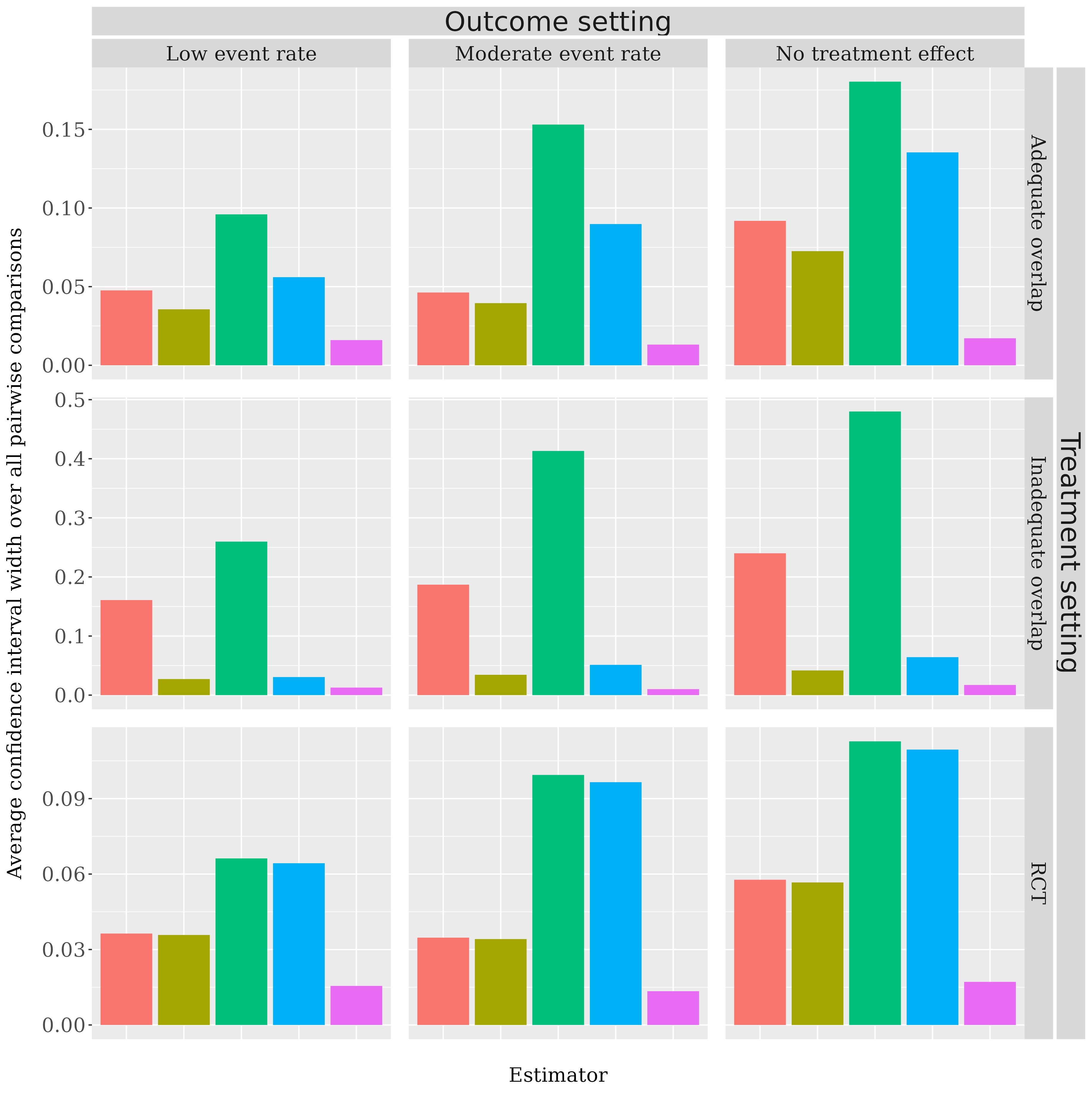}
	\caption{Average confidence interval widths for the ATE over all 15 pairwise comparisons and 1000 simulated datasets. Estimator: \hspace{1mm}
		{\protect\tikz \protect\draw[color={Rred}] (0,0) -- plot[mark=square*, mark options={scale=2}] (0,0) -- (0,0);}\, TMLE-multi. (SL); \hspace{1mm}
	{\protect\tikz \protect\draw[color={Rgold}] (0,0) -- plot[mark=square*, mark options={scale=2}] (0,0) -- (0,0);}\, TMLE-bin. (SL); \hspace{1mm}
	{\protect\tikz \protect\draw[color={Rgreen}] (0,0) -- plot[mark=square*, mark options={scale=2}] (0,0) -- (0,0);}\, IPTW-multi. (SL); \hspace{1mm}
 		{\protect\tikz \protect\draw[color={Rblue}] (0,0) -- plot[mark=square*, mark options={scale=2}] (0,0) -- (0,0);}\, IPTW-bin. (SL); \hspace{1mm}
   		{\protect\tikz \protect\draw[color={Rpink}] (0,0) -- plot[mark=square*, mark options={scale=2}] (0,0) -- (0,0);}\, G-comp. (SL).} \label{ciw_average_6}
\end{figure}

\begin{figure}[htbp]
	\centering
	\includegraphics[width=\columnwidth]{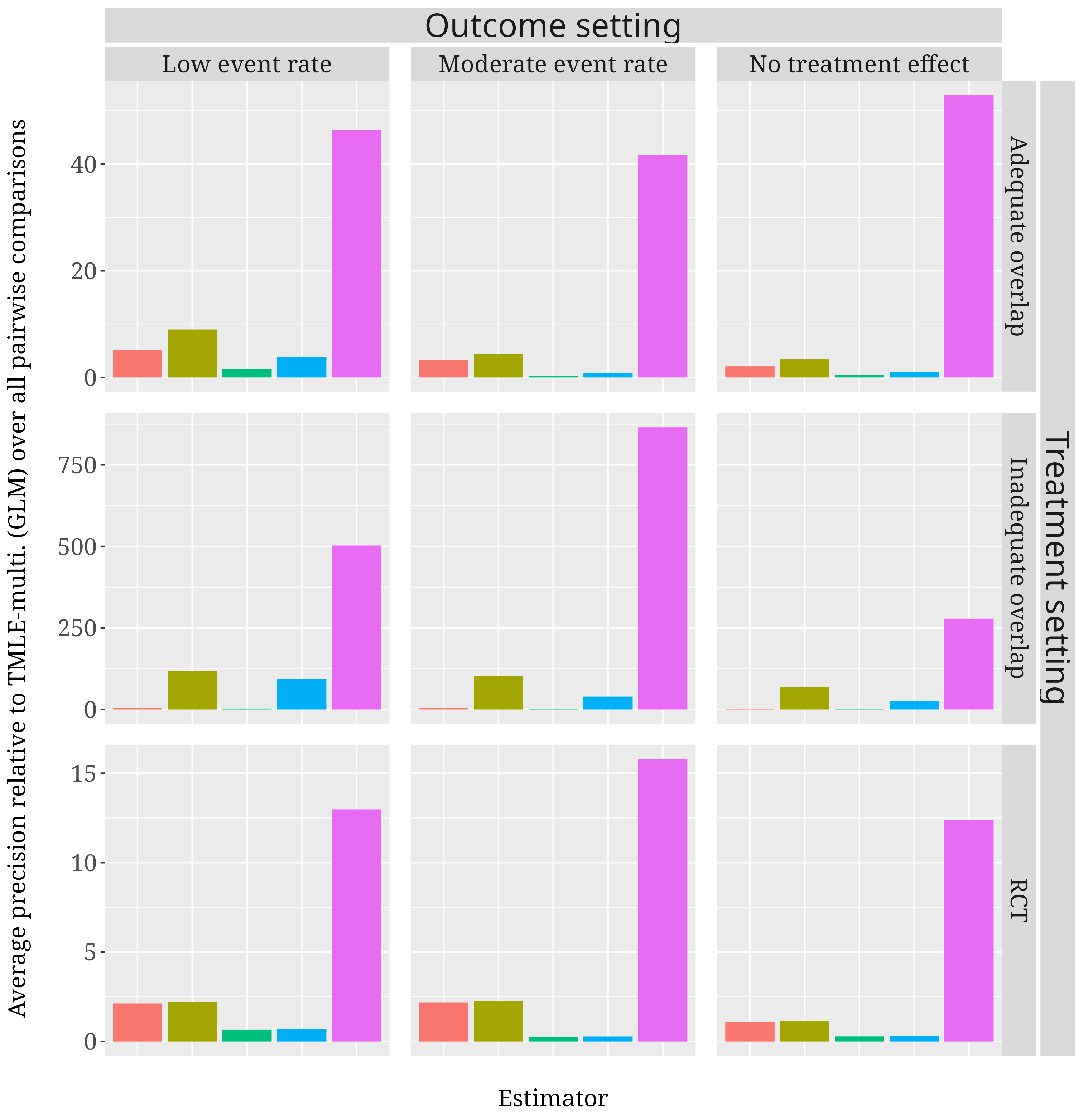}
	\caption{Average precision relative to TMLE-multi. (GLM) over all 15 pairwise comparisons and 1000 simulated datasets. Relative precision is  calculated as the variance of TMLE-multi. (GLM) divided by the variance of the comparison estimator. Estimator: \hspace{1mm}
		{\protect\tikz \protect\draw[color={Rred}] (0,0) -- plot[mark=square*, mark options={scale=2}] (0,0) -- (0,0);}\, TMLE-multi. (SL); \hspace{1mm}
	{\protect\tikz \protect\draw[color={Rgold}] (0,0) -- plot[mark=square*, mark options={scale=2}] (0,0) -- (0,0);}\, TMLE-bin. (SL); \hspace{1mm}
	{\protect\tikz \protect\draw[color={Rgreen}] (0,0) -- plot[mark=square*, mark options={scale=2}] (0,0) -- (0,0);}\, IPTW-multi. (SL); \hspace{1mm}
 		{\protect\tikz \protect\draw[color={Rblue}] (0,0) -- plot[mark=square*, mark options={scale=2}] (0,0) -- (0,0);}\, IPTW-bin. (SL); \hspace{1mm}
   		{\protect\tikz \protect\draw[color={Rpink}] (0,0) -- plot[mark=square*, mark options={scale=2}] (0,0) -- (0,0);}\, G-comp. (SL).} \label{rel_precision_6}
\end{figure}

\begin{figure}
     \centering
     \begin{subfigure}[b]{0.49\textwidth}
         \centering
         \includegraphics[width=\textwidth]{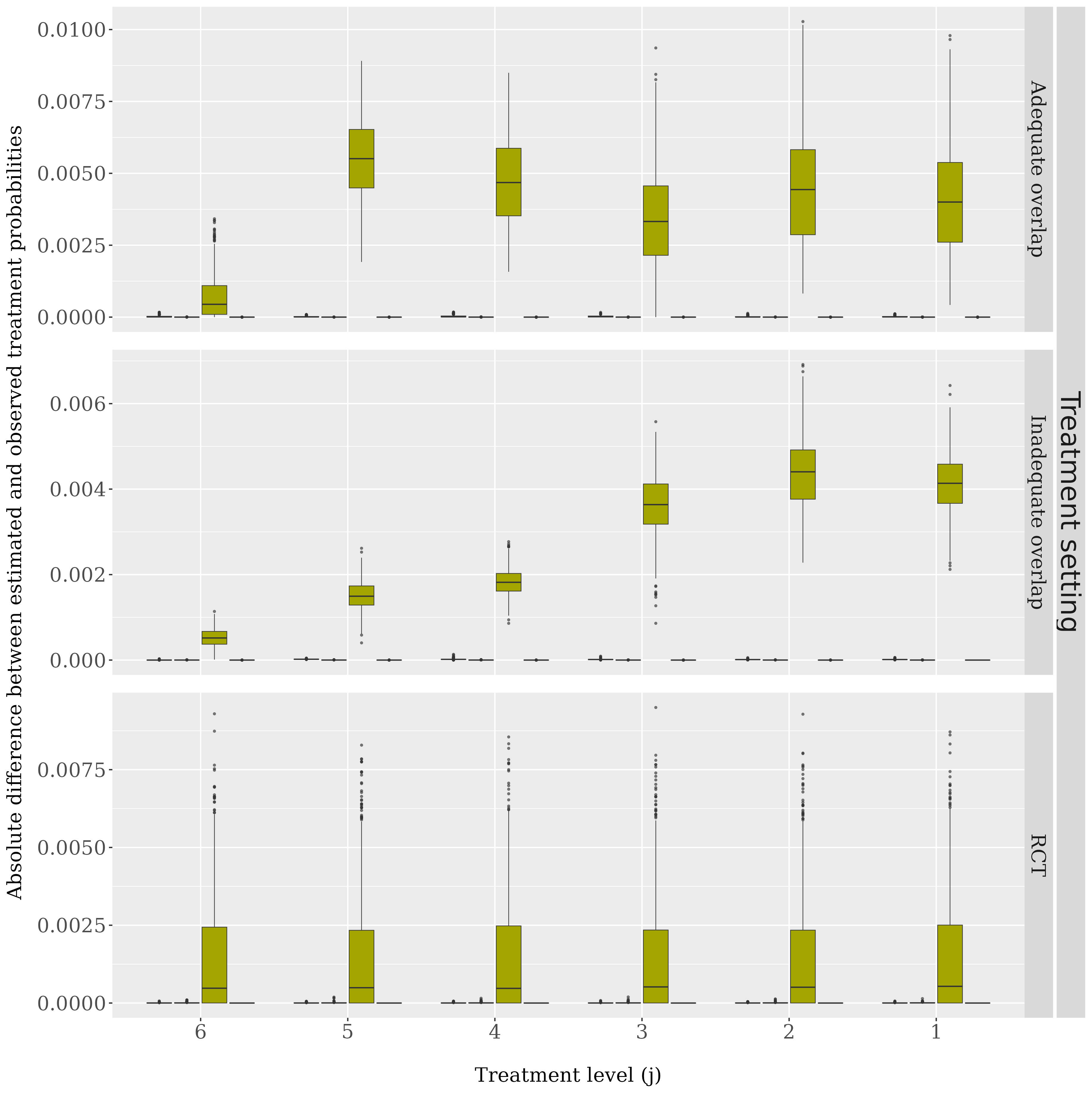}
         \caption{Accuracy of treatment estimation}
         \label{simulation_est_A_diff}
     \end{subfigure}
     \hfill
     \begin{subfigure}[b]{0.49\textwidth}
         \centering
         \includegraphics[width=\textwidth]{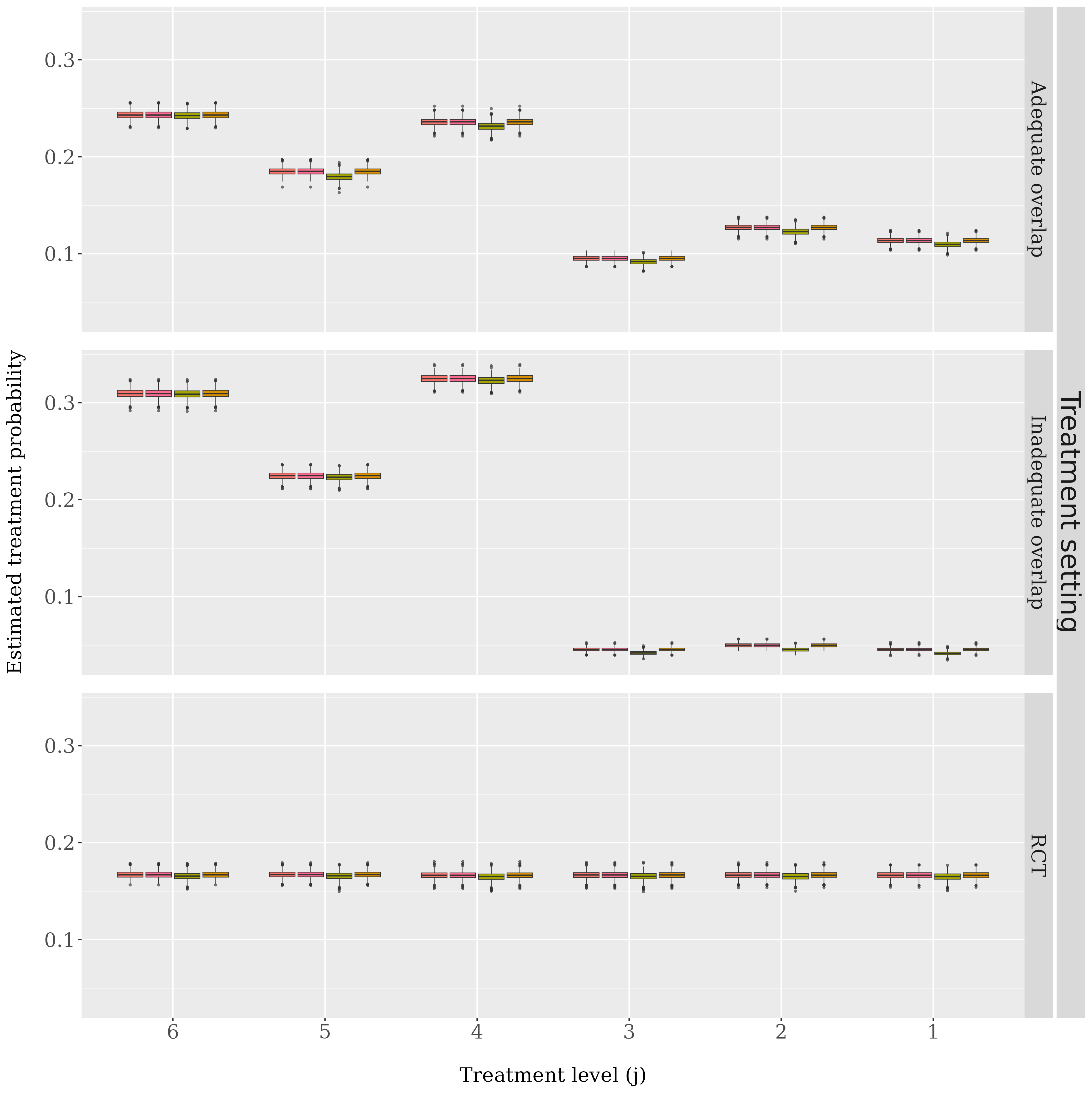}
         \caption{Estimated treatment probabilities}
         \label{simulation_est_A}
     \end{subfigure}
          \hfill \vspace{5mm}
     \begin{subfigure}[b]{0.49\textwidth}
         \centering
         \includegraphics[width=\textwidth]{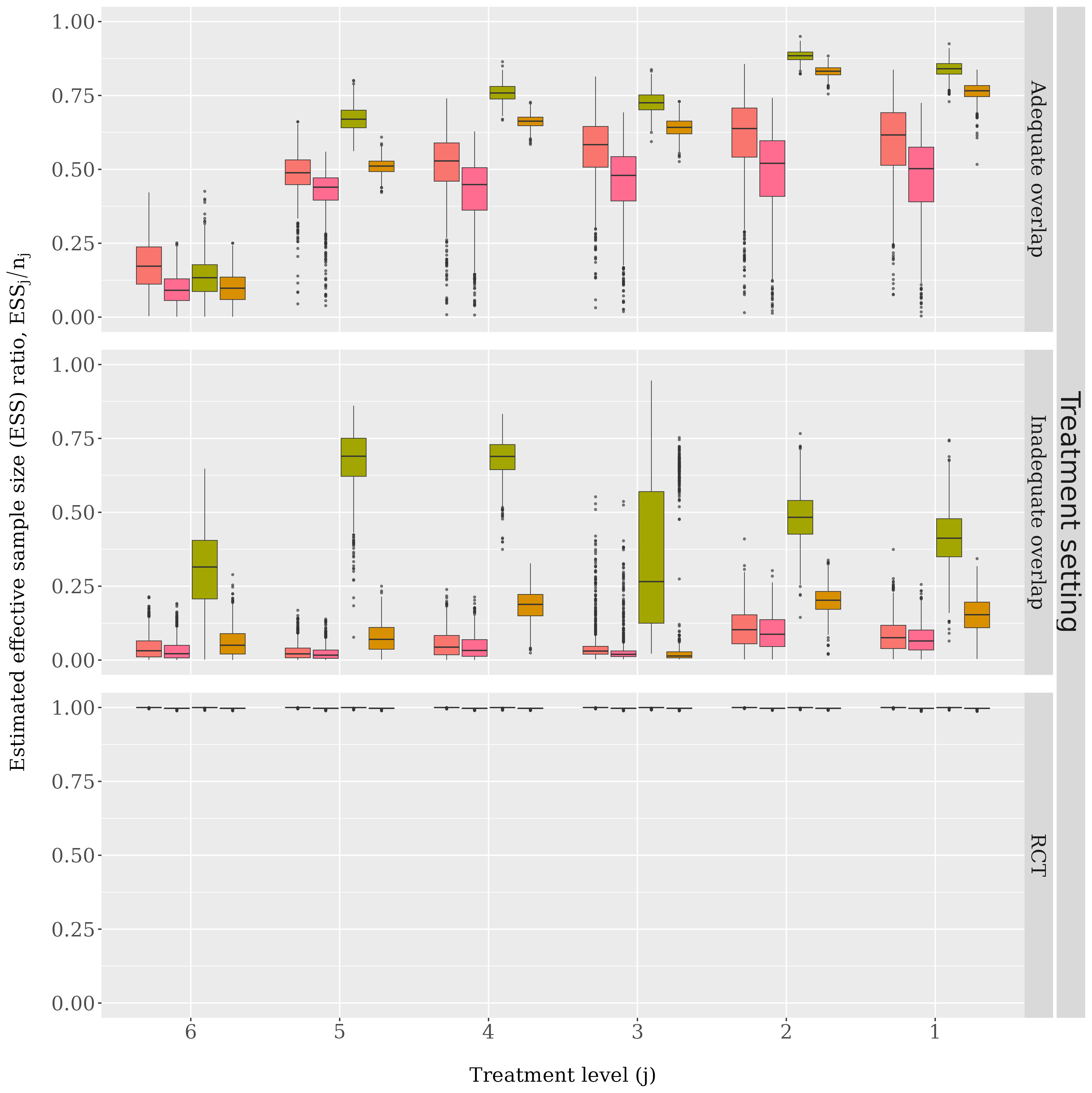}
         \caption{Estimated ESS ratio}
         \label{simulation_est_ESS_ratio}
     \end{subfigure}
        \caption{Accuracy of treatment estimation in terms of the absolute difference between the observed and estimated treatment probabilities (A); estimated treatment probabilities (B); and estimated effective sample size (ESS) ratio, $ESS_j/n_j$, for the ATE (C). Treatment model: \hspace{1mm}
		{\protect\tikz \protect\draw[color={Rred}] (0,0) -- plot[mark=square*, mark options={scale=2}] (0,0) -- (0,0);}\, Multinomial (SL); \hspace{1mm}
  		{\protect\tikz \protect\draw[color={Rsalmon}] (0,0) -- plot[mark=square*, mark options={scale=2}] (0,0) -- (0,0);}\, Multinomial (GLM); \hspace{1mm}
	{\protect\tikz \protect\draw[color={Rgold}] (0,0) -- plot[mark=square*, mark options={scale=2}] (0,0) -- (0,0);}\, Binomial (SL); \hspace{1mm}
 	{\protect\tikz \protect\draw[color={Rgold2}] (0,0) -- plot[mark=square*, mark options={scale=2}] (0,0) -- (0,0);}\, Binomial (GLM).}
        \label{simulation_est_A_diff_ESS}
\end{figure}


\section*{Web Appendix C: Numerical studies for $J=3$ treatment levels} 

In the simulation design with $J=3$, the total sample size is $n=5000$. The treatment model coefficients are given by
\begin{eqnarray*}
\beta_{1}^{\top} &= & (0,0,0,0,0,0,0)\\
\beta_{2}^{\top} & = & \kappa_{2} \times(0, 1, 1, 1, -1, 1, 1)\\ \beta_{3}^{\top} & = & \kappa_{3} \times(0, 1, 1, 1, 1, 1, 1)
\end{eqnarray*}
with $\left(\kappa_{2}, \kappa_{3}\right)=(0.2, 0.1)$ to simulate the ``adequate overlap'' scenario, $\left(\kappa_{2}, \kappa_{3}\right)=(0.7, 0.4)$ to simulate the `inadequate overlap'' scenario, and $\left(\kappa_{2}, \kappa_{3}\right)=(0, 0)$ for the RCT setting. The outcome model settings for moderate event rates are
\begin{eqnarray*}
\gamma_{1}^{\top} & = & (-1.5, 1, 1, 1, 1, 1, 1),\\
\gamma_{2}^{\top} & = & (-3, 2, 3, 1, 2, 2, 2), \text{and}\\
\gamma_{3}^{\top}&  = & (1.5, 3, 1, 2, -1, -1, -1).
\end{eqnarray*}
For low event rates, 
\begin{eqnarray*}
\gamma_{1}^{\top} & = & (-4, 1, -2, -1, 1, 1, 1),\\
\gamma_{2}^{\top} & = & (-2, 1, -1, -1, -1, -1, -4), \text{and}\\
\gamma_{3}^{\top} & = & (3, 3, -1, 1, -2, -1, -2).
\end{eqnarray*}
Finally, in the  ``no treatment effect'' setting, we specify $\gamma_{1}^{\top}, \gamma_{2}^{\top}, \gamma_{3}^{\top} = (0, 0, 0, 0, 0, 0, 0)$.

\begin{figure}
     \centering
     \begin{subfigure}[b]{0.49\textwidth}
         \centering
         \includegraphics[width=\textwidth]{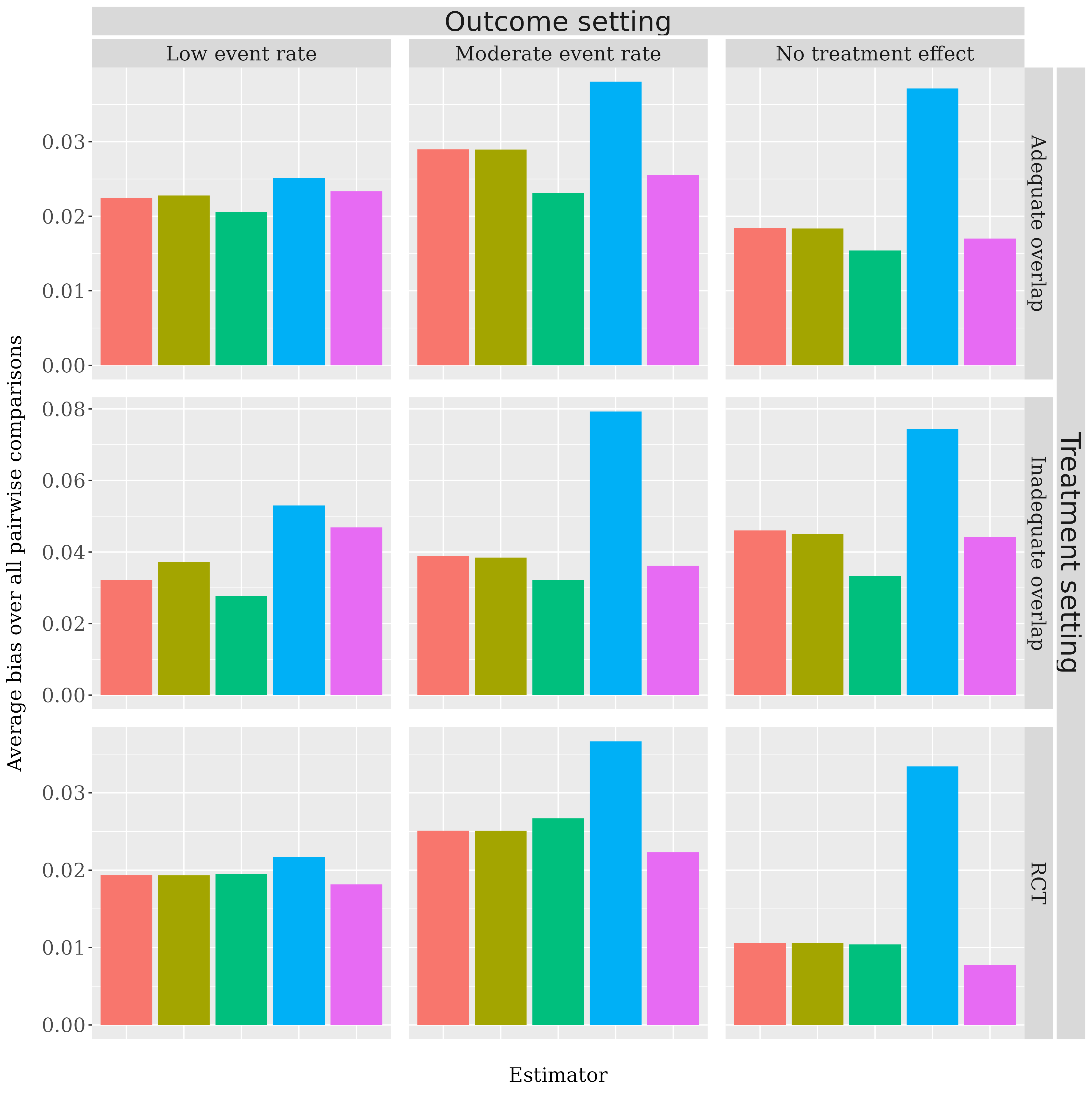}
         \caption{Average bias}
         \label{bias_average_3}
     \end{subfigure}
     \hfill 
          \begin{subfigure}[b]{0.49\textwidth}
         \centering
         \includegraphics[width=\textwidth]{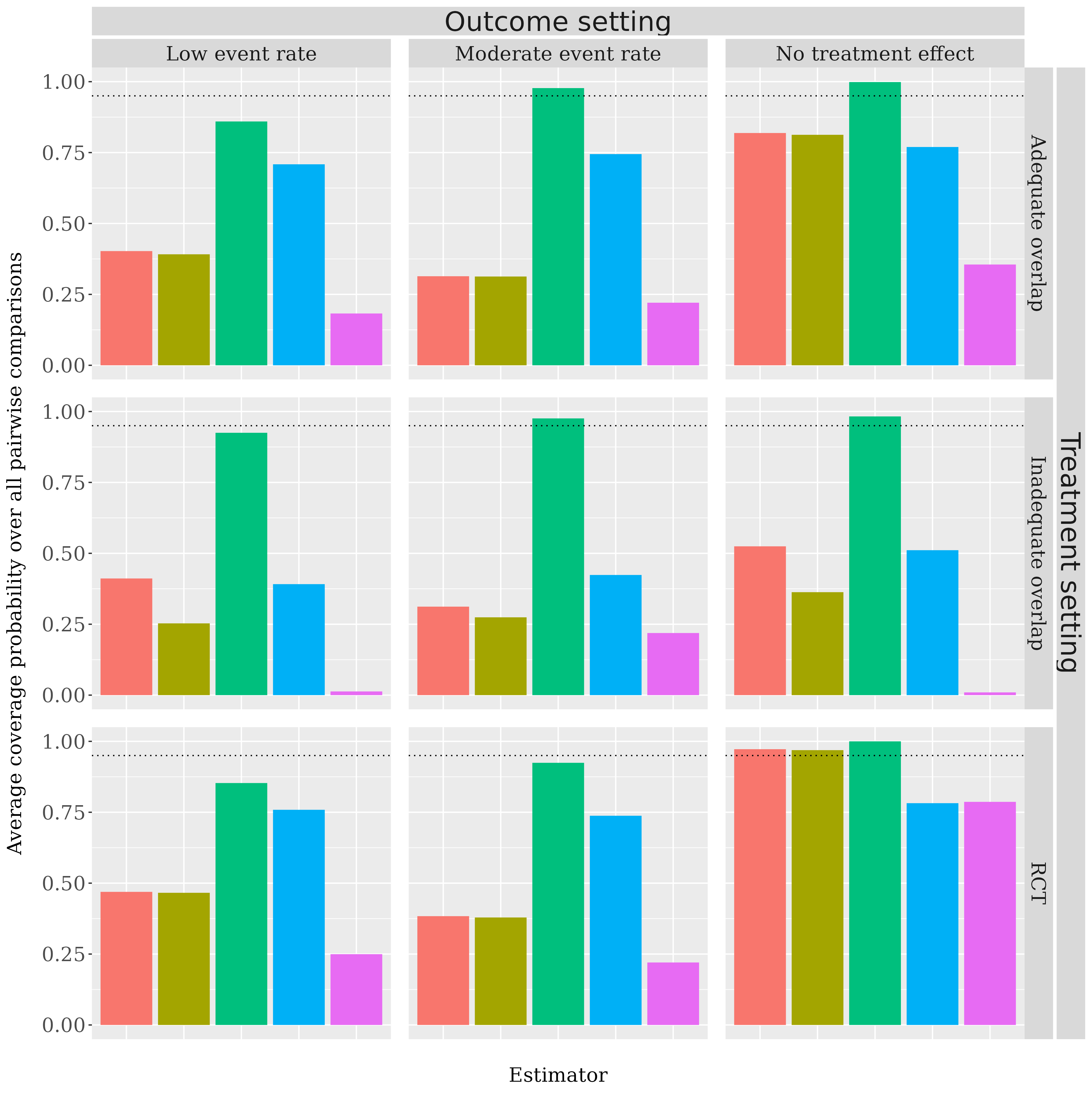}
         \caption{Average coverage probability}
         \label{cp_average_3}
     \end{subfigure}
       \hfill \vspace{5mm}
     \begin{subfigure}[b]{0.49\textwidth}
         \centering
         \includegraphics[width=\textwidth]{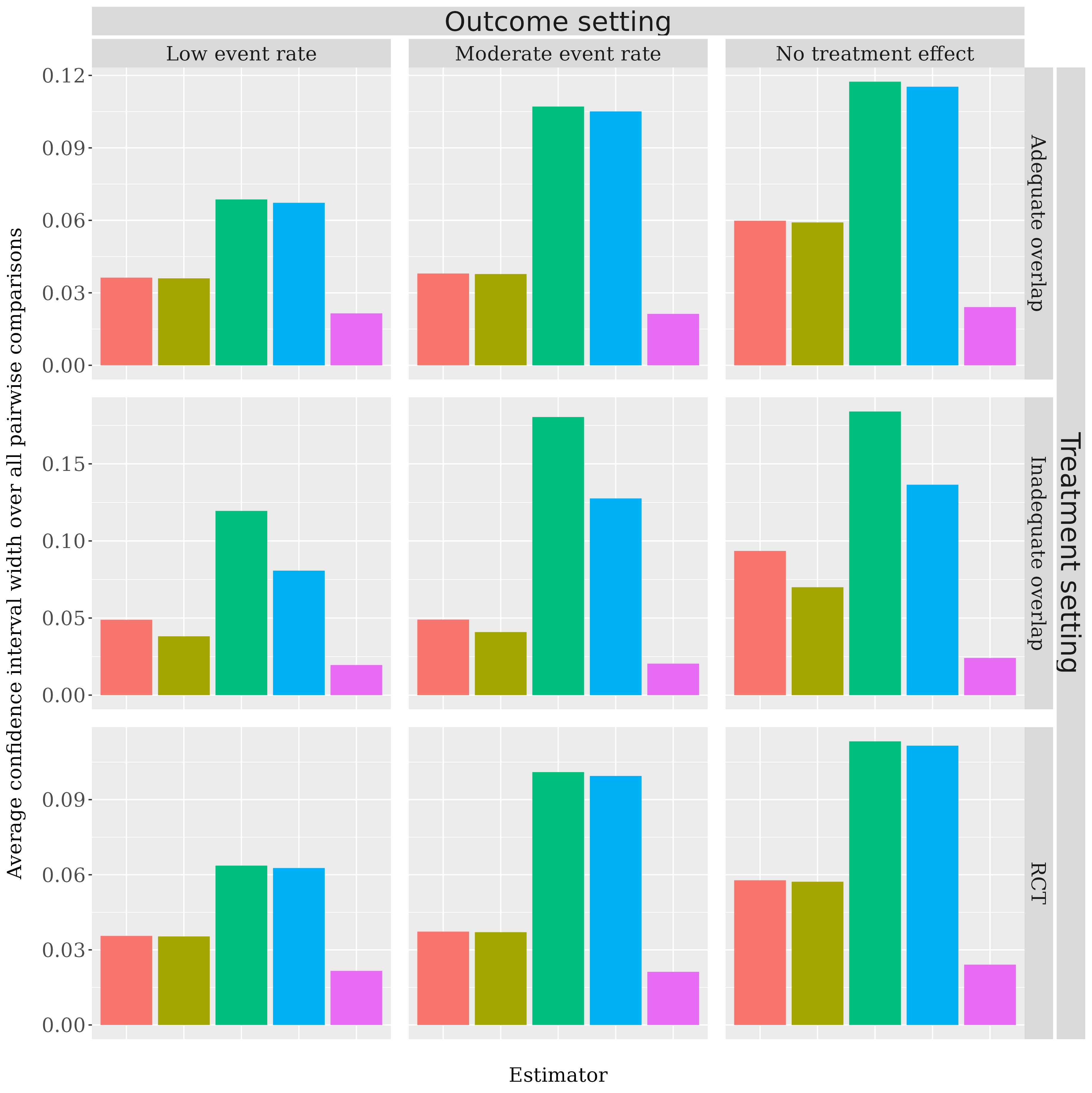}
         \caption{Average confidence interval widths}
         \label{ciw_average_3}
     \end{subfigure}
        \caption{Average bias (A), coverage probability (B), and confidence interval widths (C) for the ATE over all 3 pairwise comparisons and 1000 simulated datasets. Estimator: \hspace{1mm}
		{\protect\tikz \protect\draw[color={Rred}] (0,0) -- plot[mark=square*, mark options={scale=2}] (0,0) -- (0,0);}\, TMLE-multi. (SL); \hspace{1mm}
	{\protect\tikz \protect\draw[color={Rgold}] (0,0) -- plot[mark=square*, mark options={scale=2}] (0,0) -- (0,0);}\, TMLE-bin. (SL); \hspace{1mm}
	{\protect\tikz \protect\draw[color={Rgreen}] (0,0) -- plot[mark=square*, mark options={scale=2}] (0,0) -- (0,0);}\, IPTW-multi. (SL); \hspace{1mm}
 		{\protect\tikz \protect\draw[color={Rblue}] (0,0) -- plot[mark=square*, mark options={scale=2}] (0,0) -- (0,0);}\, IPTW-bin. (SL); \hspace{1mm}
   		{\protect\tikz \protect\draw[color={Rpink}] (0,0) -- plot[mark=square*, mark options={scale=2}] (0,0) -- (0,0);}\, G-comp. (SL).}
        \label{bias_cp_ciw_average_3}
\end{figure}

\section*{Web Appendix D: Numerical studies with high-dimensional covariate space}

This section describes the DGP for the additional simulations with 40 and 100 covariates. The DGP with standard six covariates is described in Sections \ref{treatment-gen} and \ref{outcome-gen} in the paper.

\subsection*{Multinomial Treatment Assignment}\label{treatment-gen-extended}

The first three covariates are generated in the same way as in the six-covariate case. For the extended set of covariates \(x_{7i}, x_{8i}, \ldots, x_{40i}\) or \(x_{7i}, x_{8i}, \ldots, x_{100i}\), they are generated using a diverse set of statistical distributions, including truncated forms of exponential, geometric, hypergeometric, logistic, and Poisson distributions. This wide variety allows for a rich, multidimensional dataset that can be used to assess treatment effect estimations under different conditions.

The treatment model follows the multinomial logistic model just as in the six-covariate case. For 40 covariates, the \( \beta \) values for each treatment group are specified as follows: 
\begin{eqnarray*}
\beta_{1}^{\top} & = & (0, 0, \ldots, 0) \\
\beta_{2}^{\top} & = &  \kappa_{2} \times(0, 0.5, 0.5, 1, 0.5, 0.5, 0.5, \ldots, 0.5) \\
\beta_{3}^{\top} & = &  \kappa_{3} \times(0, 0.15, 0.15, \ldots, 0.15, 0, -1, 0.5) \\
\beta_{4}^{\top} & = &  \kappa_{4} \times(0, 0.15, 0.15, \ldots, 0.15, 0, 1, 0.5) \\
\beta_{5}^{\top} & = &  \kappa_{5} \times(0, 0.15, 0.15, \ldots, 0.15, -2, 1, 1) \\
\beta_{6}^{\top} & = &  \kappa_{6} \times(0.25, 0.15, 0.15, \ldots, 0.15, -1, -1, -1).
\end{eqnarray*}
\noindent
These \( \beta \) values define the linear relationship between the covariates and the log-odds of each treatment group in the multinomial logistic model. The $\beta$ values for the 100-covariate case follows a similar pattern as the 40-covariate case, except $$\beta_{2}^{\top} = \kappa_{2} \times(0, 0.15, 0.15, 1, 0.15, 0.15, 0.15, \ldots, 0.15)$$ when there are 100 covariates. The \(\kappa\) values, which control the level of overlap or similarity in the distributions of propensity scores across treatment levels, are identical to the six-covariate case.

\subsection*{Outcome Generation}\label{outcome-gen-extended}

The outcome generation process is similar to the six-covariate scenario but extended to accommodate 40 or 100 covariates by repeating the pattern for each treatment group. For 40 covariates, the \(\gamma_j\) vector is extended to \(\gamma_j^\top = (c_1, c_2, \ldots, c_{40})\), where \(c_i\) are constants specific to each simulation setting and each treatment level \(j\). For 100 covariates, \(\gamma_j\) becomes \(\gamma_j^\top = (c_1, c_2, \ldots, c_{100})\).

Three different settings are used to vary the event rate, similar to the six-covariate scenario. These settings are ``low event rate," ``moderate event rate," and ``no treatment effect," each defined by specific values of \(\gamma_j\), as per Section \ref{outcome-gen} in the paper. An exception is made in the 100 covariate case, where we reduce some of the $\gamma_j$ values in the `low event rate'' setting:
\begin{eqnarray*}
\gamma_{6}^{\top} = ( -3, -2, -1, 1, -2, -1, -2).
\end{eqnarray*}
\noindent
Results under the “moderate event rate” setting are not provided in the case of 100 covariates due to computational issues.


\begin{figure}
     \centering
     \begin{subfigure}[b]{0.49\textwidth}
         \centering
         \includegraphics[width=\textwidth]{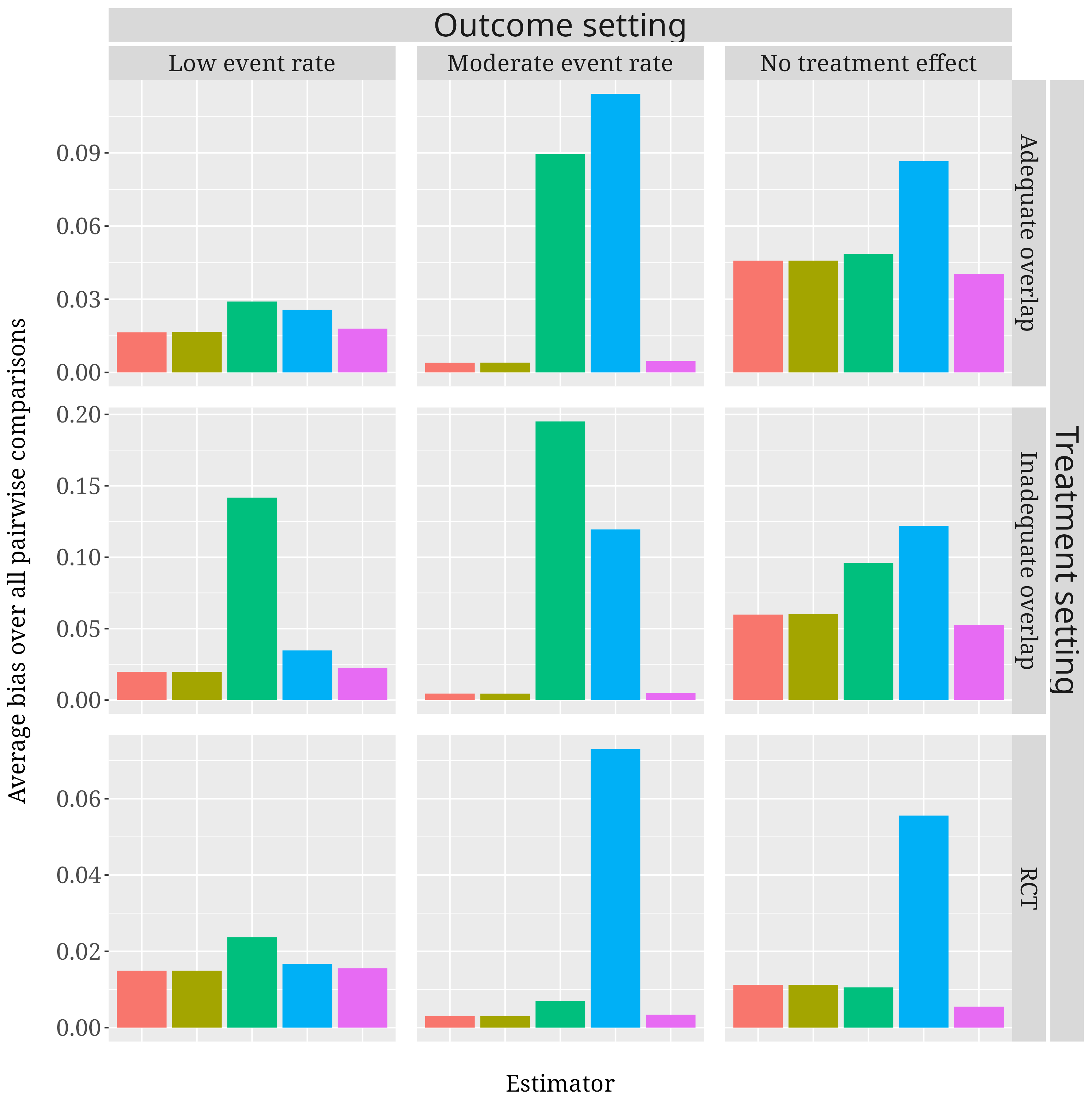}
         \caption{Average bias}
         \label{bias_average_6_40}
     \end{subfigure}
     \hfill
     \begin{subfigure}[b]{0.49\textwidth}
         \centering
         \includegraphics[width=\textwidth]{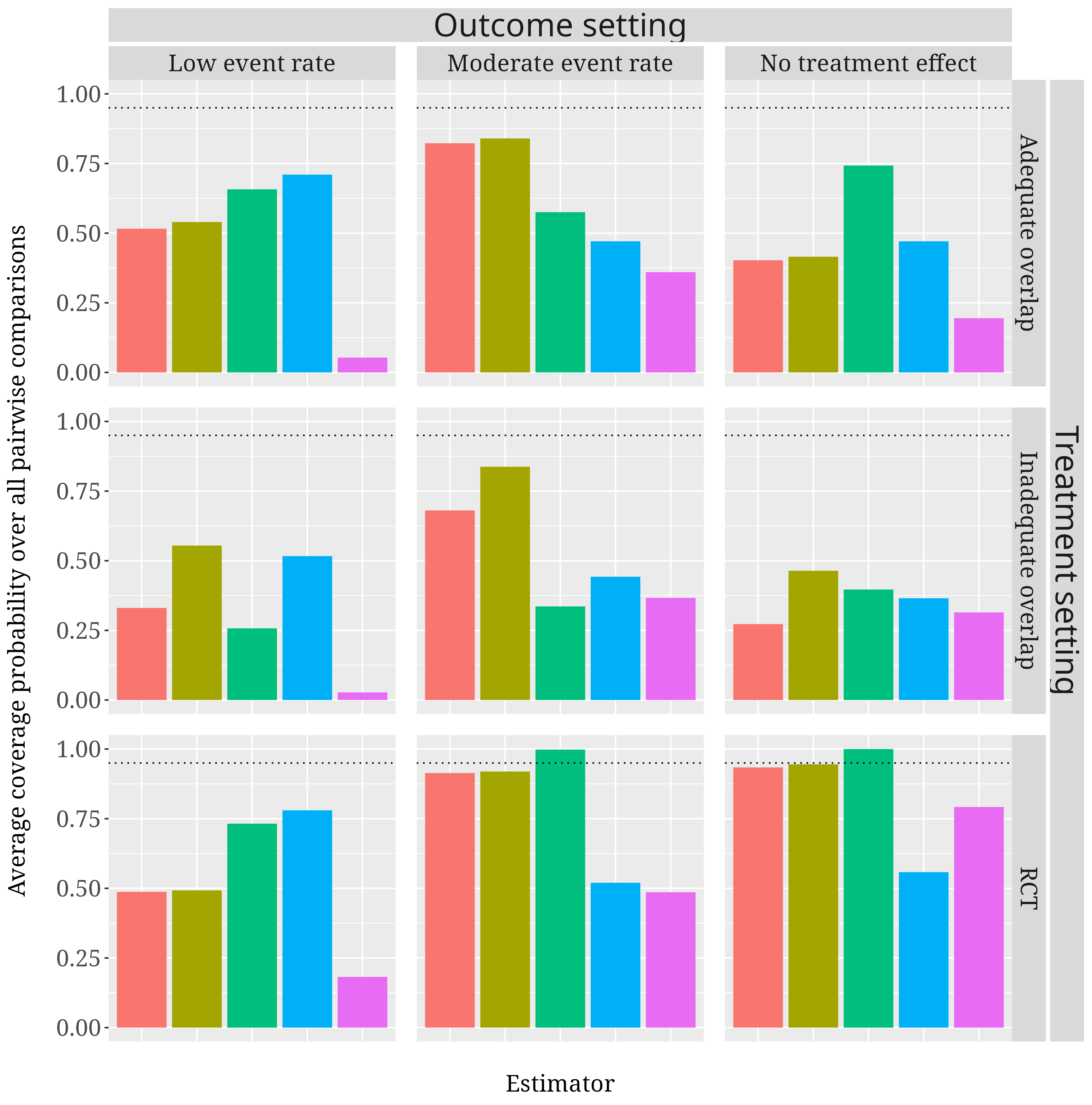}
         \caption{Average coverage probability}
         \label{cp_average_6_40}
     \end{subfigure}
    \hfill \vspace{5mm}
     \begin{subfigure}[b]{0.49\textwidth}
         \centering
         \includegraphics[width=\textwidth]{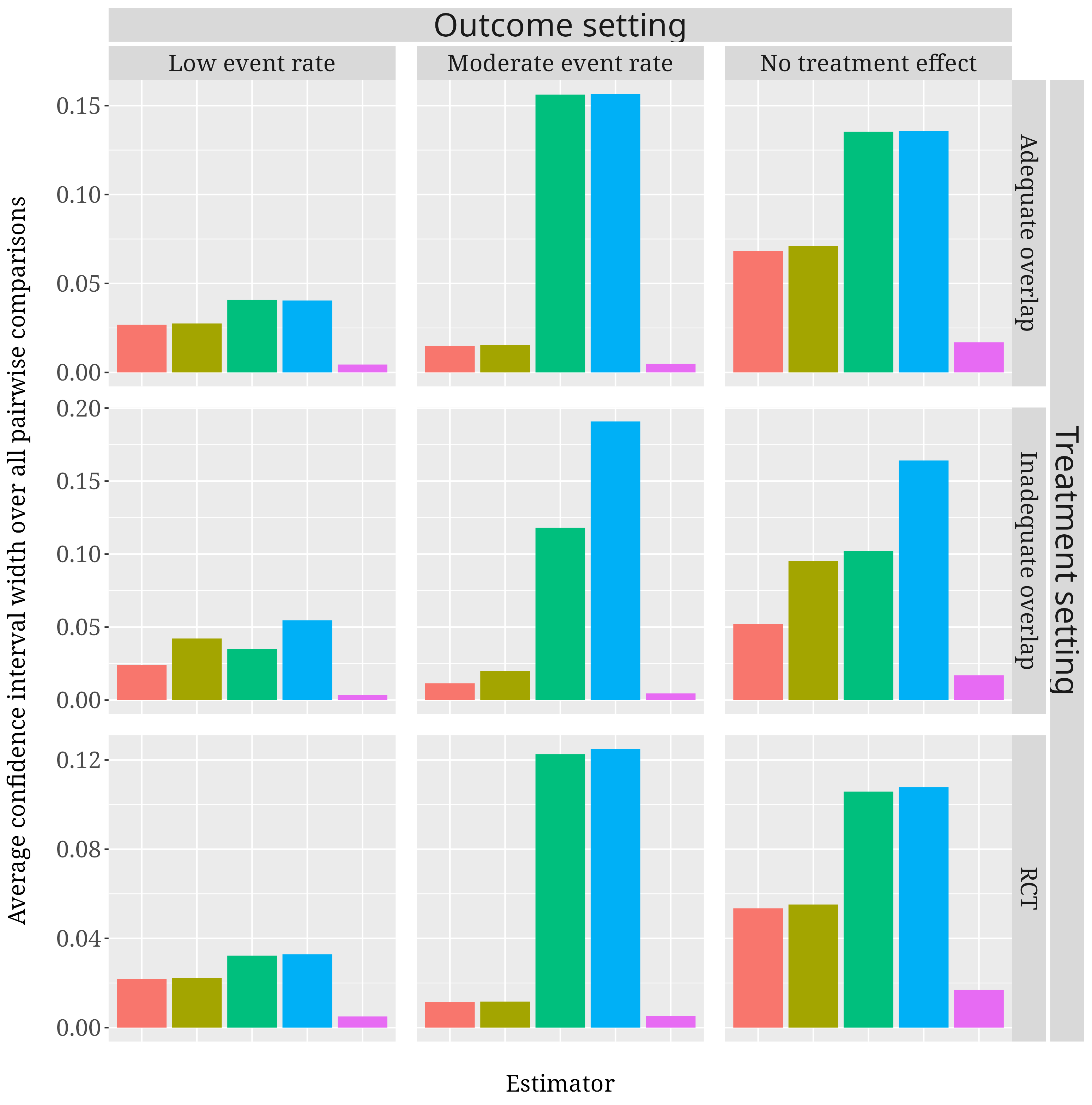}
         \caption{Average confidence interval widths}
         \label{ciw_average_6_40}
     \end{subfigure}
        \caption{Average bias (A), coverage probability (B), and confidence interval widths (C) for the ATE over all 15 pairwise comparisons and 100 simulated datasets, in a high-dimensional scenario with 40 covariates. Estimator: \hspace{1mm}
		{\protect\tikz \protect\draw[color={Rred}] (0,0) -- plot[mark=square*, mark options={scale=2}] (0,0) -- (0,0);}\, TMLE-multi. (SL); \hspace{1mm}
	{\protect\tikz \protect\draw[color={Rgold}] (0,0) -- plot[mark=square*, mark options={scale=2}] (0,0) -- (0,0);}\, TMLE-bin. (SL); \hspace{1mm}
	{\protect\tikz \protect\draw[color={Rgreen}] (0,0) -- plot[mark=square*, mark options={scale=2}] (0,0) -- (0,0);}\, IPTW-multi. (SL); \hspace{1mm}
 		{\protect\tikz \protect\draw[color={Rblue}] (0,0) -- plot[mark=square*, mark options={scale=2}] (0,0) -- (0,0);}\, IPTW-bin. (SL); \hspace{1mm}
   		{\protect\tikz \protect\draw[color={Rpink}] (0,0) -- plot[mark=square*, mark options={scale=2}] (0,0) -- (0,0);}\, G-comp. (SL).}
        \label{bias_cp_ciw_average_6_40}
\end{figure}


\begin{figure}
     \centering
     \begin{subfigure}[b]{0.49\textwidth}
         \centering
         \includegraphics[width=\textwidth]{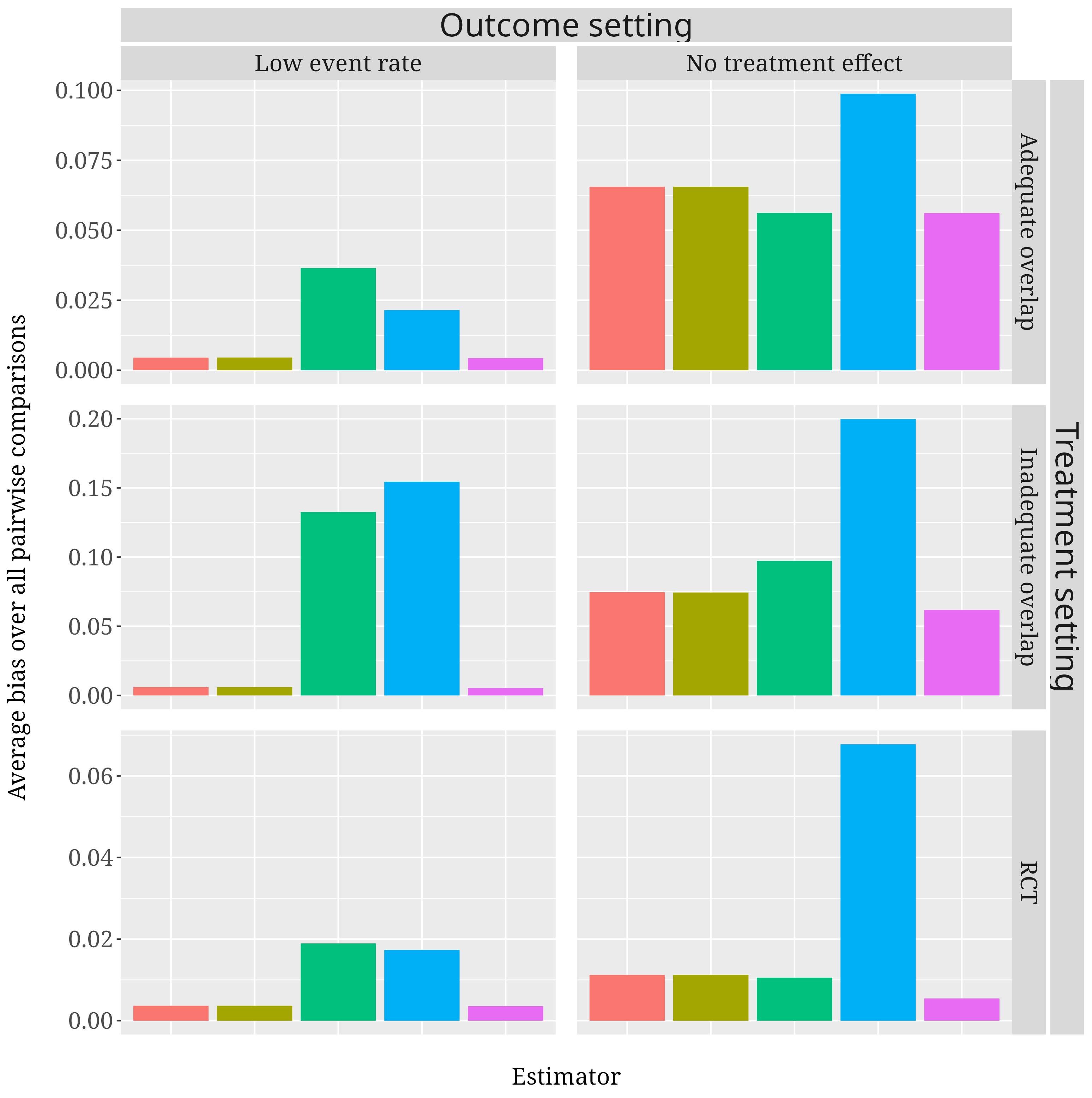}
         \caption{Average bias}
         \label{bias_average_6_100}
     \end{subfigure}
     \hfill
          \begin{subfigure}[b]{0.49\textwidth}
         \centering
         \includegraphics[width=\textwidth]{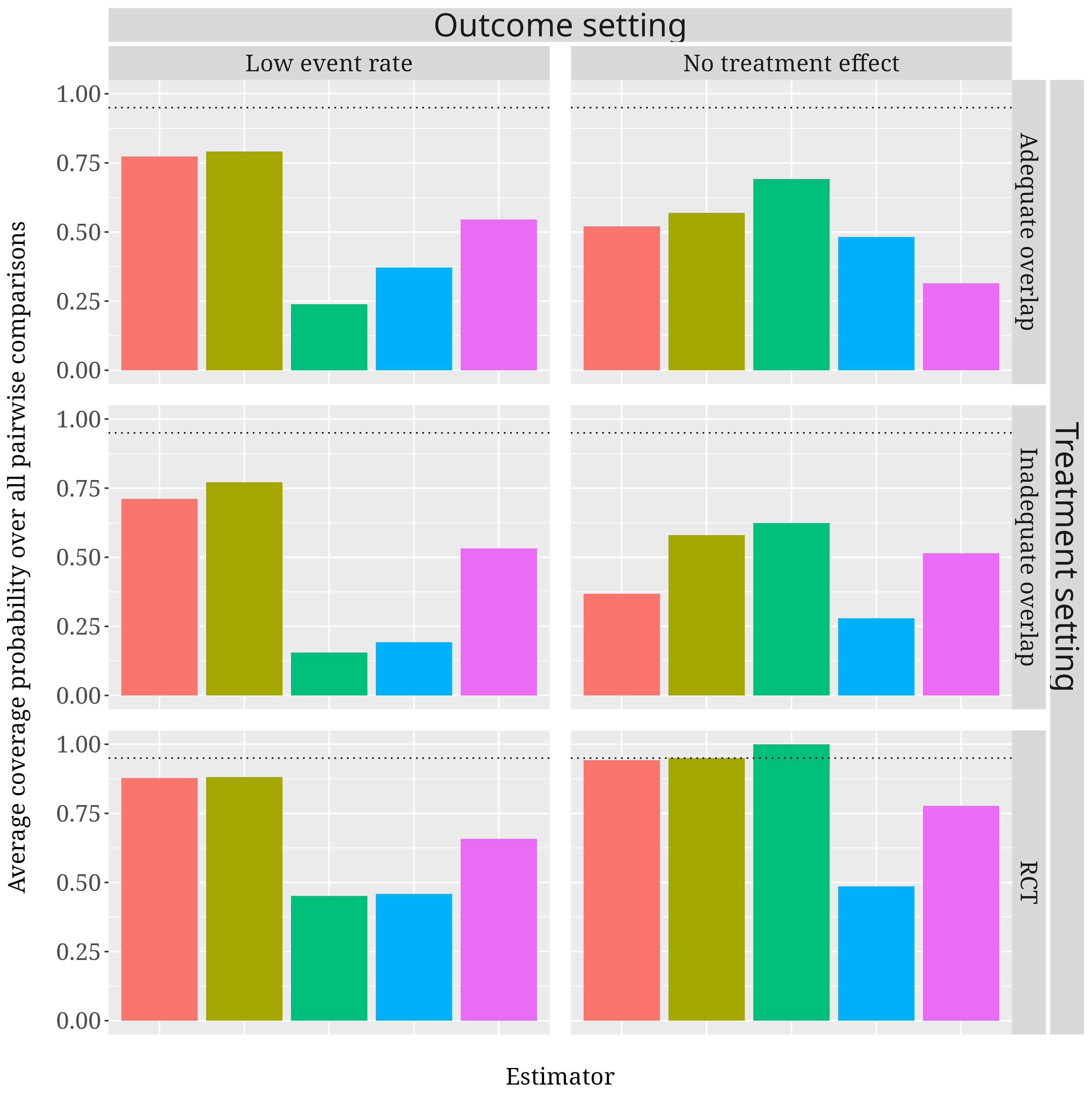}
         \caption{Average coverage probability}
         \label{cp_average_6_100}
     \end{subfigure}
        \hfill \vspace{5mm}
     \begin{subfigure}[b]{0.49\textwidth}
         \centering
         \includegraphics[width=\textwidth]{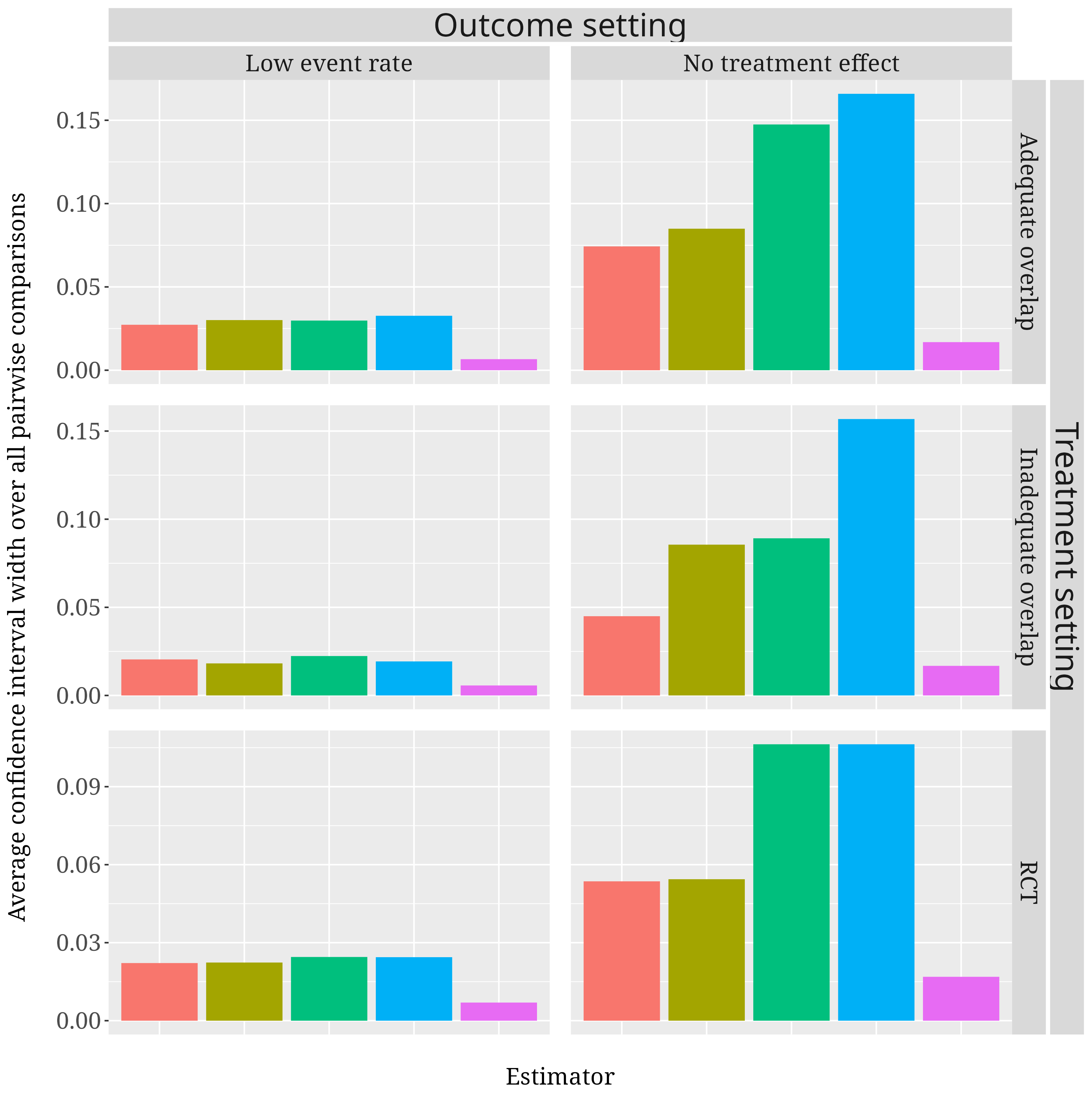}
         \caption{Average confidence interval widths}
         \label{ciw_average_6_100}
     \end{subfigure}
        \caption{Average bias (A), coverage probability (B), and confidence interval widths (C) for the ATE over all 15 pairwise comparisons and 100 simulated datasets, in a high-dimensional scenario with 100 covariates. Results under the ``moderate event rate" setting are not provided due to computational issues. Estimator: \hspace{1mm}
		{\protect\tikz \protect\draw[color={Rred}] (0,0) -- plot[mark=square*, mark options={scale=2}] (0,0) -- (0,0);}\, TMLE-multi. (SL); \hspace{1mm}
	{\protect\tikz \protect\draw[color={Rgold}] (0,0) -- plot[mark=square*, mark options={scale=2}] (0,0) -- (0,0);}\, TMLE-bin. (SL); \hspace{1mm}
	{\protect\tikz \protect\draw[color={Rgreen}] (0,0) -- plot[mark=square*, mark options={scale=2}] (0,0) -- (0,0);}\, IPTW-multi. (SL); \hspace{1mm}
 		{\protect\tikz \protect\draw[color={Rblue}] (0,0) -- plot[mark=square*, mark options={scale=2}] (0,0) -- (0,0);}\, IPTW-bin. (SL); \hspace{1mm}
   		{\protect\tikz \protect\draw[color={Rpink}] (0,0) -- plot[mark=square*, mark options={scale=2}] (0,0) -- (0,0);}\, G-comp. (SL).}
        \label{bias_cp_ciw_average_6_100}
\end{figure}

\section*{{\large Web Appendix E: Numerical studies with misspecification}}

\begin{figure}
     \centering
     \begin{subfigure}[b]{0.49\textwidth}
         \centering
         \includegraphics[width=\textwidth]{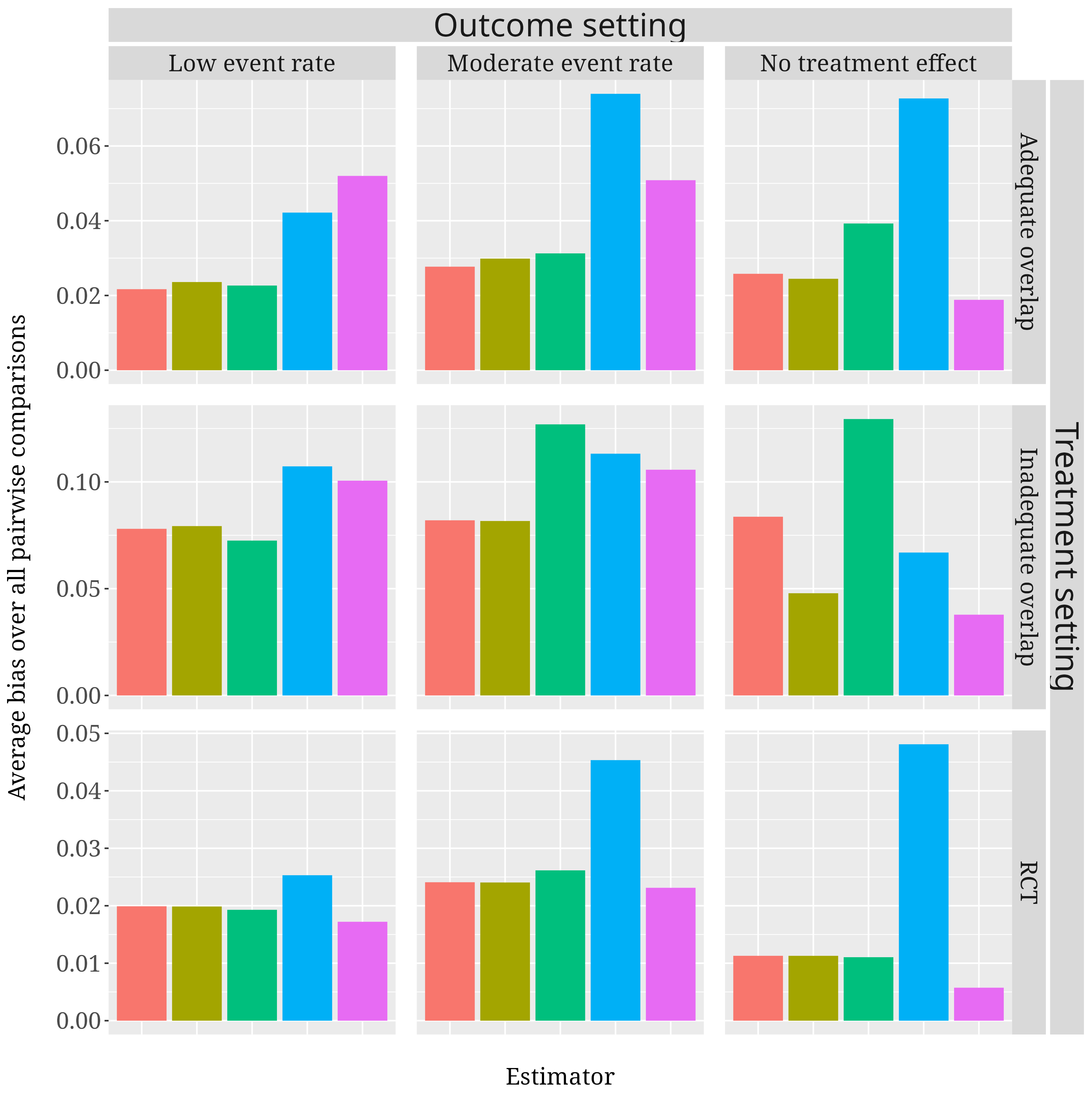}
         \caption{Average bias}
         \label{bias_average_6_misOut}
     \end{subfigure}
     \hfill
     \begin{subfigure}[b]{0.49\textwidth}
         \centering
         \includegraphics[width=\textwidth]{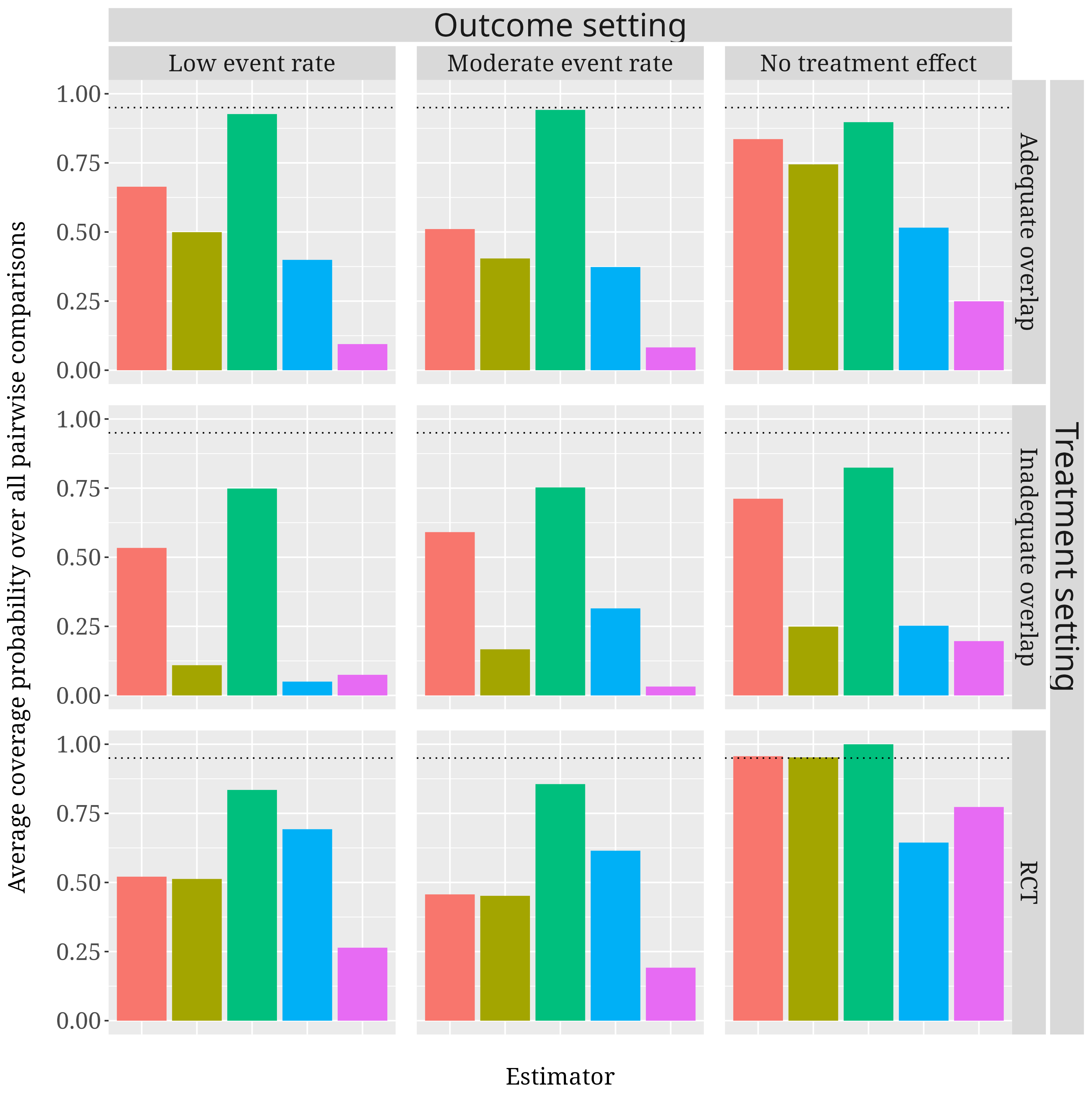}
         \caption{Average coverage probability}
         \label{cp_average_6_misOut}
     \end{subfigure}
         \hfill \vspace{5mm}
     \begin{subfigure}[b]{0.49\textwidth}
         \centering
         \includegraphics[width=\textwidth]{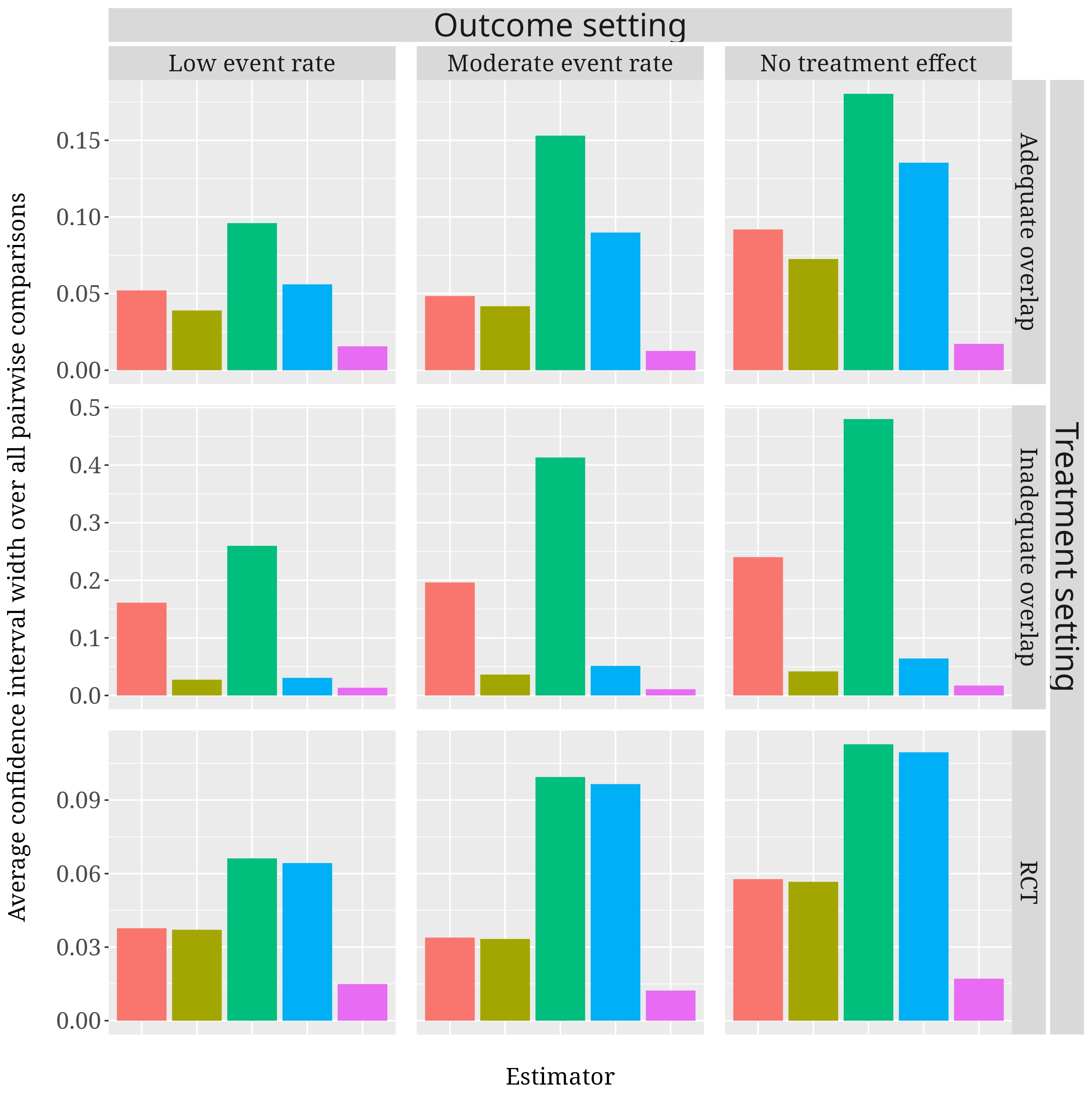}
         \caption{Average confidence interval widths}
         \label{ciw_average_6_misOut}
     \end{subfigure}
        \caption{Average bias (A), coverage probability (B), and confidence interval widths (C) for the ATE over all 15 pairwise comparisons and 1000 simulated datasets, in a scenario with a misspecified outcome model. Estimator: \hspace{1mm}
		{\protect\tikz \protect\draw[color={Rred}] (0,0) -- plot[mark=square*, mark options={scale=2}] (0,0) -- (0,0);}\, TMLE-multi. (SL); \hspace{1mm}
	{\protect\tikz \protect\draw[color={Rgold}] (0,0) -- plot[mark=square*, mark options={scale=2}] (0,0) -- (0,0);}\, TMLE-bin. (SL); \hspace{1mm}
	{\protect\tikz \protect\draw[color={Rgreen}] (0,0) -- plot[mark=square*, mark options={scale=2}] (0,0) -- (0,0);}\, IPTW-multi. (SL); \hspace{1mm}
 		{\protect\tikz \protect\draw[color={Rblue}] (0,0) -- plot[mark=square*, mark options={scale=2}] (0,0) -- (0,0);}\, IPTW-bin. (SL); \hspace{1mm}
   		{\protect\tikz \protect\draw[color={Rpink}] (0,0) -- plot[mark=square*, mark options={scale=2}] (0,0) -- (0,0);}\, G-comp. (SL).}
        \label{bias_cp_ciw_average_6_misOut}
\end{figure}

\begin{figure}
     \centering
     \begin{subfigure}[b]{0.49\textwidth}
         \centering
         \includegraphics[width=\textwidth]{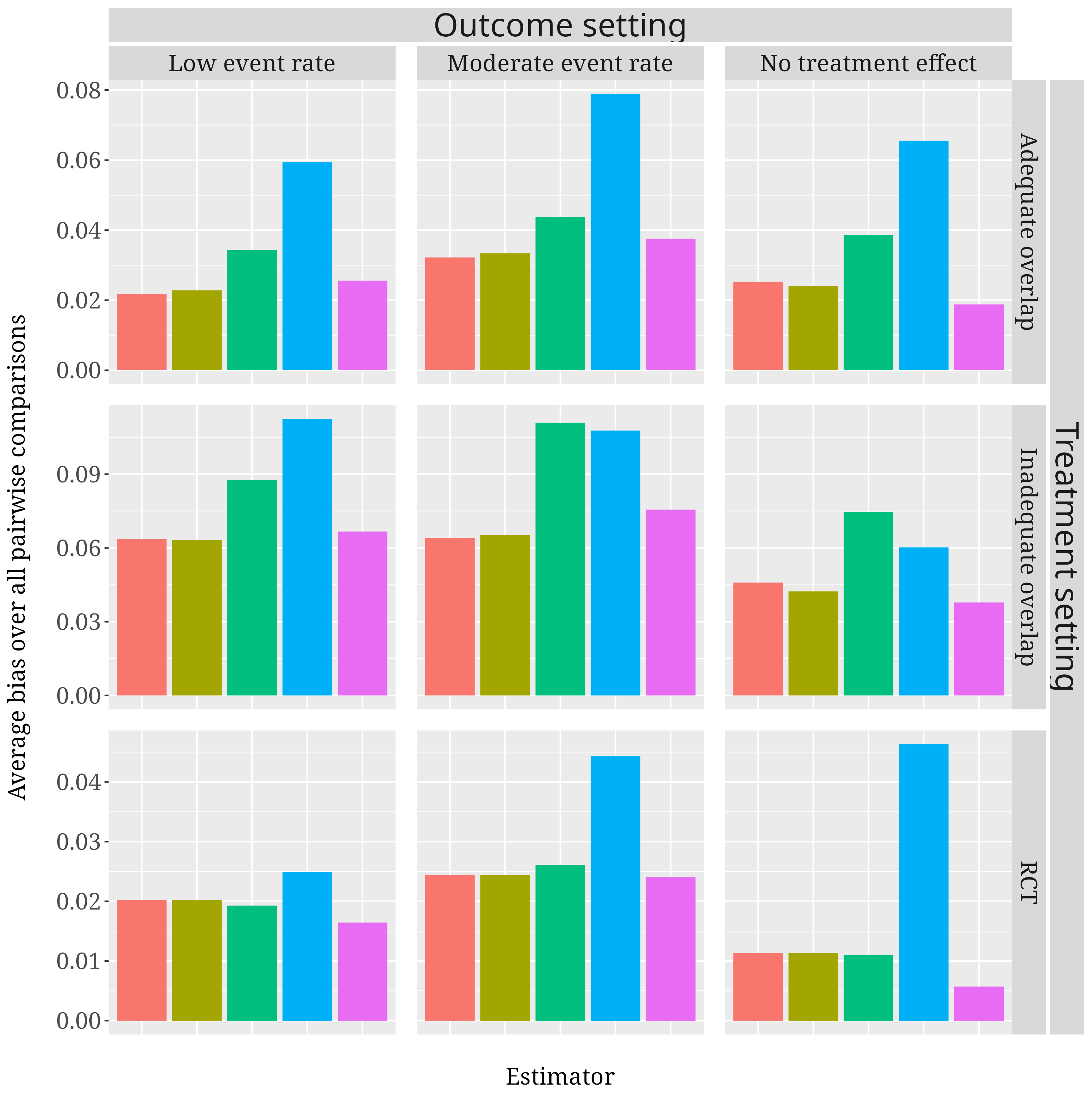}
         \caption{Average bias}
         \label{bias_average_6_misTreat}
     \end{subfigure}
     \hfill
     \begin{subfigure}[b]{0.49\textwidth}
         \centering
         \includegraphics[width=\textwidth]{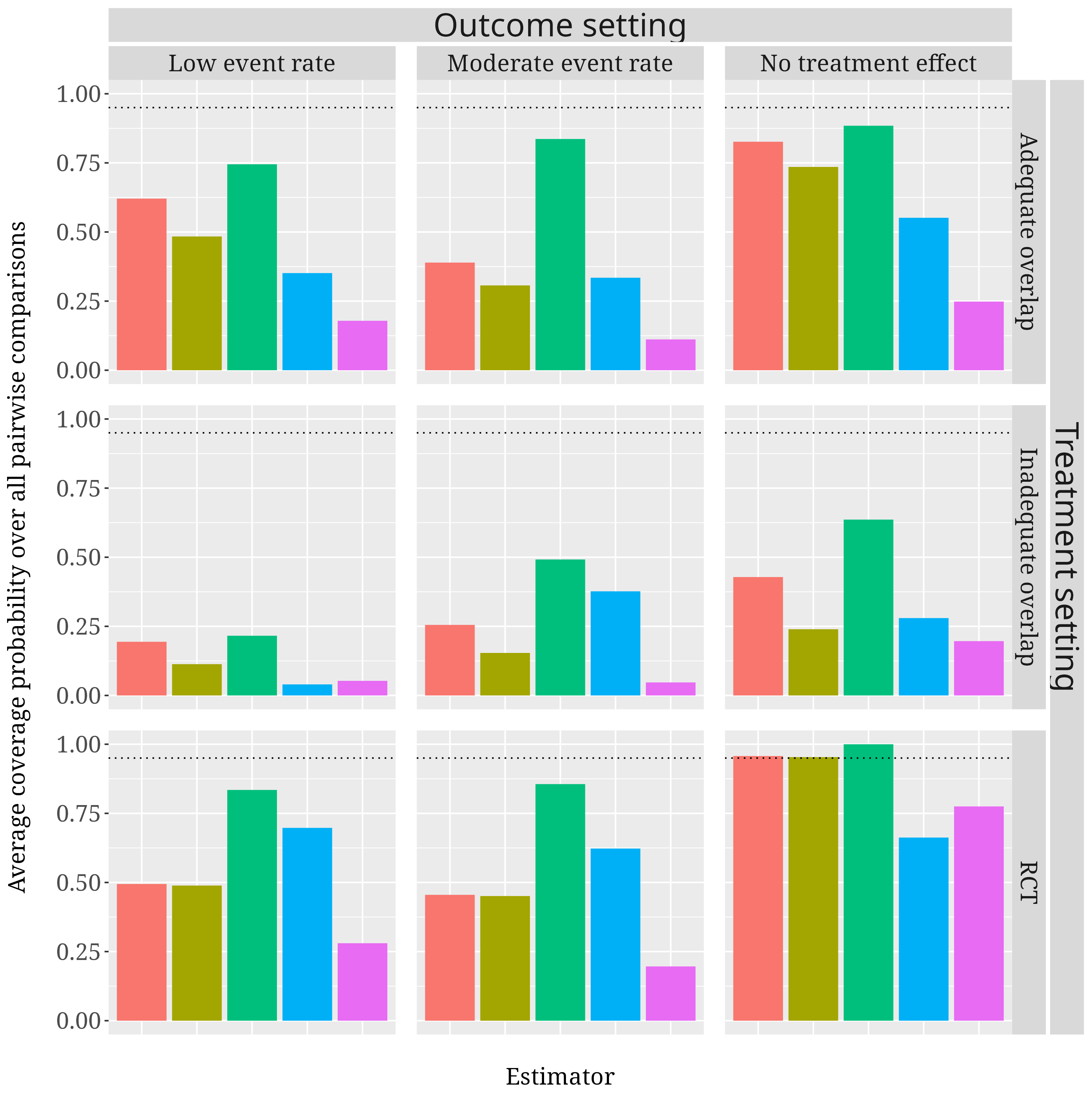}
         \caption{Average coverage probability}
         \label{cp_average_6_misTreat}
     \end{subfigure}
    \hfill \vspace{5mm}
     \begin{subfigure}[b]{0.49\textwidth}
         \centering
         \includegraphics[width=\textwidth]{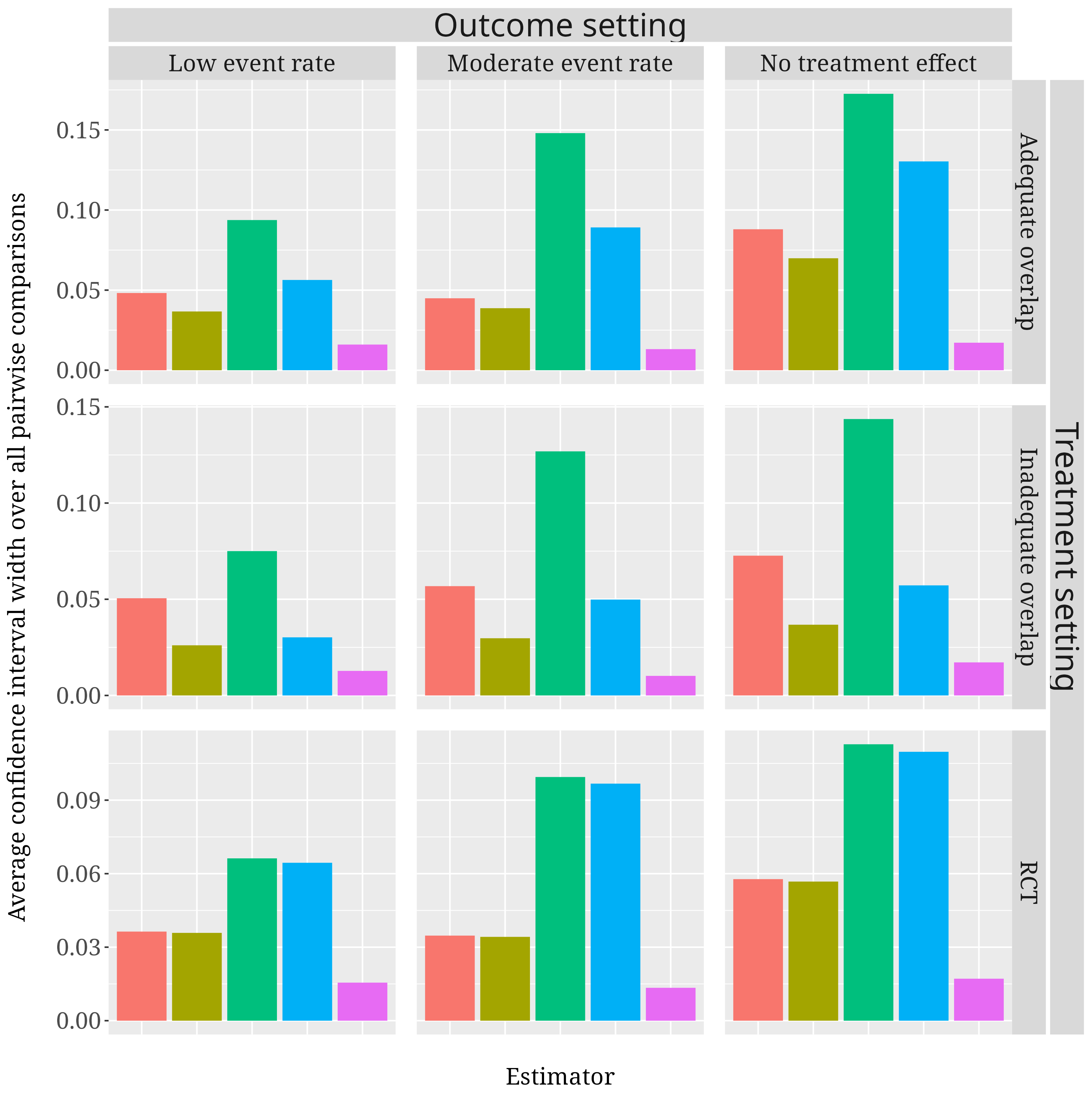}
         \caption{Average confidence interval widths}
         \label{ciw_average_6_misTreat}
     \end{subfigure}
        \caption{Average bias (A), coverage probability (B), confidence interval widths (C) for the ATE over all 15 pairwise comparisons and 1000 simulated datasets, in a scenario with a misspecified treatment model. Estimator: \hspace{1mm}
		{\protect\tikz \protect\draw[color={Rred}] (0,0) -- plot[mark=square*, mark options={scale=2}] (0,0) -- (0,0);}\, TMLE-multi. (SL); \hspace{1mm}
	{\protect\tikz \protect\draw[color={Rgold}] (0,0) -- plot[mark=square*, mark options={scale=2}] (0,0) -- (0,0);}\, TMLE-bin. (SL); \hspace{1mm}
	{\protect\tikz \protect\draw[color={Rgreen}] (0,0) -- plot[mark=square*, mark options={scale=2}] (0,0) -- (0,0);}\, IPTW-multi. (SL); \hspace{1mm}
 		{\protect\tikz \protect\draw[color={Rblue}] (0,0) -- plot[mark=square*, mark options={scale=2}] (0,0) -- (0,0);}\, IPTW-bin. (SL); \hspace{1mm}
   		{\protect\tikz \protect\draw[color={Rpink}] (0,0) -- plot[mark=square*, mark options={scale=2}] (0,0) -- (0,0);}\, G-comp. (SL).}
        \label{bias_cp_ciw_average_6_misTreat}
\end{figure}

\begin{figure}
     \centering
     \begin{subfigure}[b]{0.49\textwidth}
         \centering
         \includegraphics[width=\textwidth]{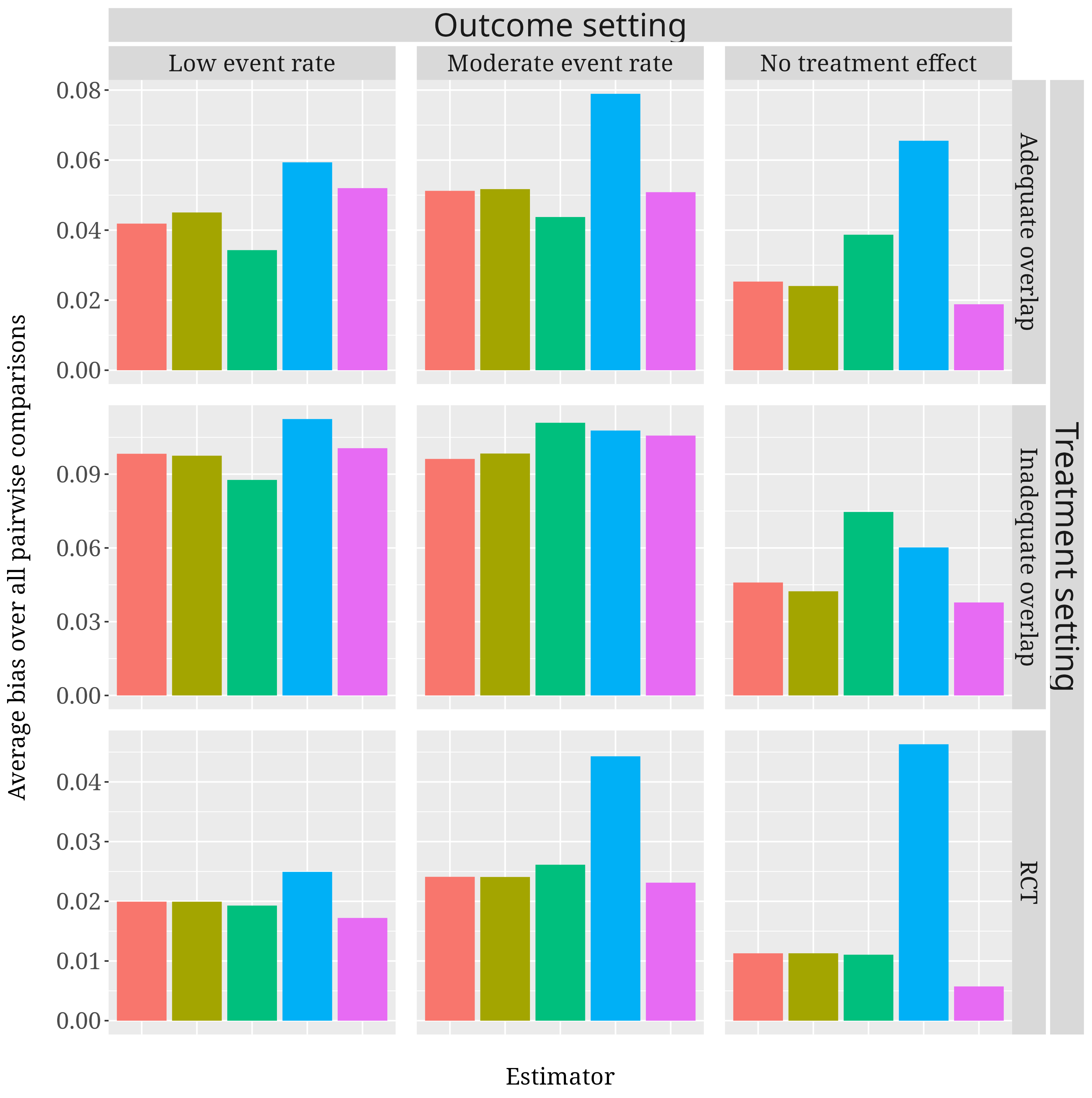}
         \caption{Average bias}
         \label{bias_average_6_misBoth}
     \end{subfigure}
     \hfill
     \begin{subfigure}[b]{0.49\textwidth}
         \centering
         \includegraphics[width=\textwidth]{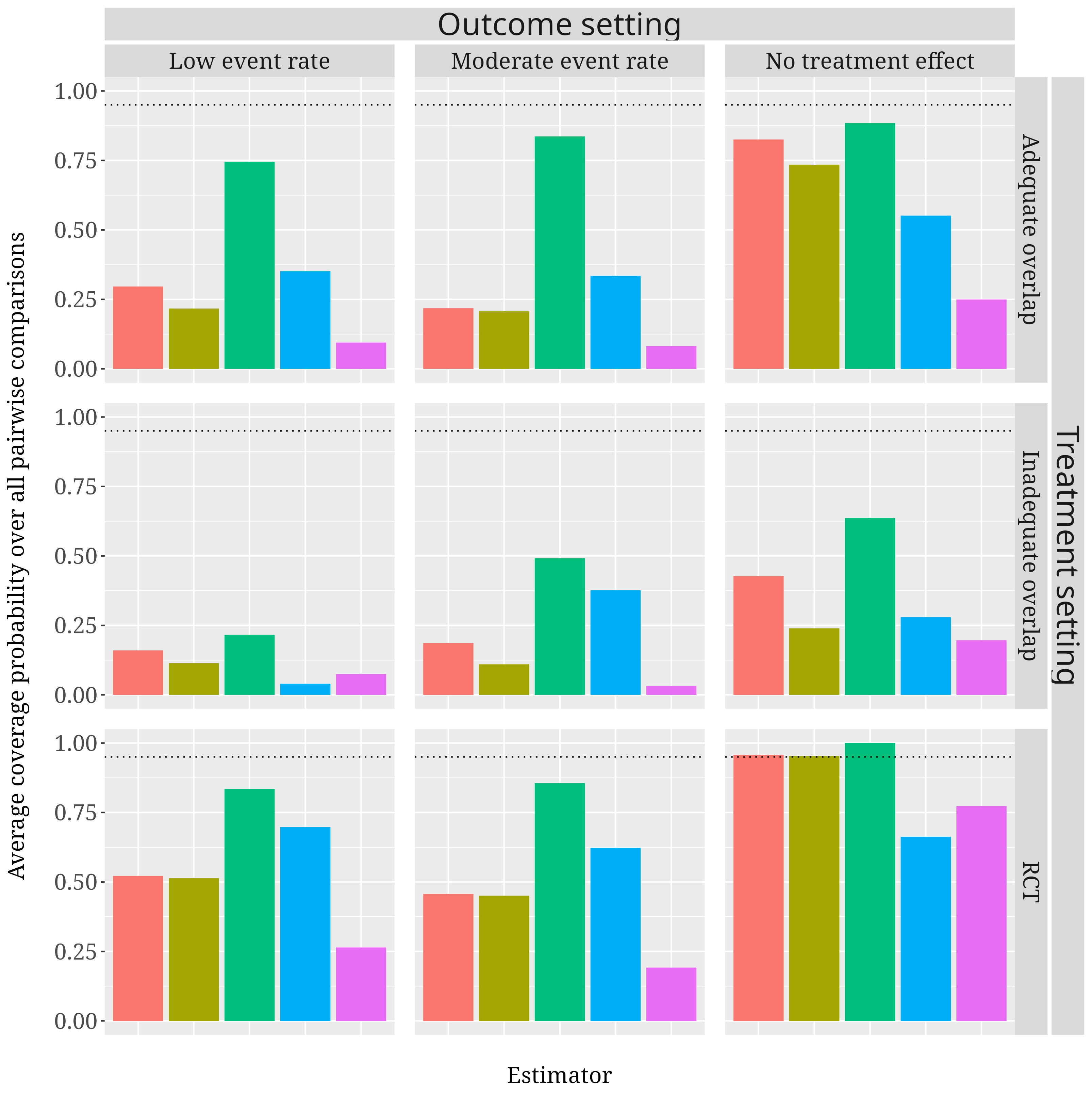}
         \caption{Average coverage probability}
         \label{cp_average_6_misBoth}
     \end{subfigure}
        \hfill \vspace{5mm}
     \begin{subfigure}[b]{0.49\textwidth}
         \centering
         \includegraphics[width=\textwidth]{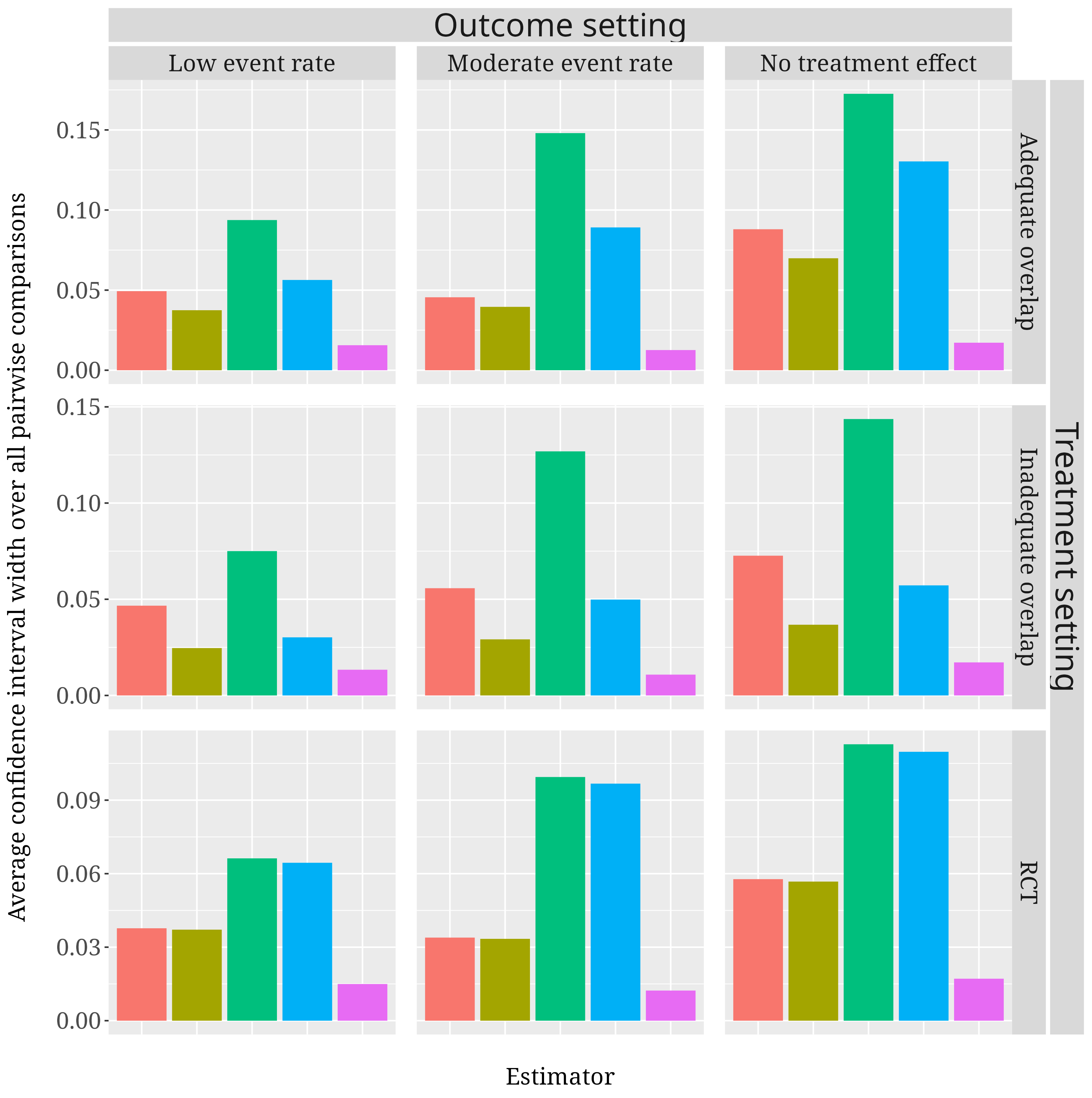}
         \caption{Average confidence interval widths}
         \label{ciw_average_6_misBoth}
     \end{subfigure}
        \caption{Average bias (A), coverage probability (B), and confidence interval widths (C) for the ATE over all 15 pairwise comparisons and 1000 simulated datasets, in a scenario with both outcome model and treatment model misspecified. Estimator: \hspace{1mm}
		{\protect\tikz \protect\draw[color={Rred}] (0,0) -- plot[mark=square*, mark options={scale=2}] (0,0) -- (0,0);}\, TMLE-multi. (SL); \hspace{1mm}
	{\protect\tikz \protect\draw[color={Rgold}] (0,0) -- plot[mark=square*, mark options={scale=2}] (0,0) -- (0,0);}\, TMLE-bin. (SL); \hspace{1mm}
	{\protect\tikz \protect\draw[color={Rgreen}] (0,0) -- plot[mark=square*, mark options={scale=2}] (0,0) -- (0,0);}\, IPTW-multi. (SL); \hspace{1mm}
 		{\protect\tikz \protect\draw[color={Rblue}] (0,0) -- plot[mark=square*, mark options={scale=2}] (0,0) -- (0,0);}\, IPTW-bin. (SL); \hspace{1mm}
   		{\protect\tikz \protect\draw[color={Rpink}] (0,0) -- plot[mark=square*, mark options={scale=2}] (0,0) -- (0,0);}\, G-comp. (SL).}
        \label{bias_cp_ciw_average_6_misBoth}
\end{figure}

\section*{Web Appendix F: Additional descriptive statistics and results from the empirical application}

\begin{table}[htbp] 
\centering
\begin{threeparttable}
\caption{Cross-validated error and weights for classification algorithms in super learner ensembles.\label{tab:ensemble}}
  \begin{tabularx}{\linewidth}{l*{3}{Y}}
      \toprule
     Algorithm  & Weight & NLL \\ 
	\rowcolor{Gray}\multicolumn{3}{l}{\textbf{Binomial outcome model (diabetes or death)}} \\
	Gradient boosting (\texttt{xgboost}) & 0.435 & 0.344 \\ 
	Elastic net regression, $\alpha=0.25$ (\texttt{glmnet}) & 0.106 & 0.345  \\ 
	Elastic net regression, $\alpha=0.50$ (\texttt{glmnet}) & 0.106 & 0.345  \\ 
	Elastic net regression, $\alpha=0.75$ (\texttt{glmnet}) & 0.108 & 0.345  \\ 
    Lasso regression, $\alpha=1$ (\texttt{glmnet}) & 0.109 & 0.345  \\ 
    Random forests, $\text{num.trees} = 100$ (\texttt{ranger}) & 0.000 & 0.351  \\ 
    Random forests, $\text{num.trees} = 500$ (\texttt{ranger}) & 0.136 & 0.348  \\ 
    Super learner (\texttt{sl3})                                          &  1.000     & 0.341  \\ 
\end{tabularx}
  \begin{tabularx}{\linewidth}{l*{3}{Y}}
   \rowcolor{Gray}     \multicolumn{3}{l}{\textbf{Binomial treatment model (Reference)}} \\
    Gradient boosting (\texttt{xgboost}) & 0.278 & 0.443  \\ 
	Elastic net regression, $\alpha=0.25$ (\texttt{glmnet}) & 0.132 & 0.441  \\ 
	Elastic net regression, $\alpha=0.50$ (\texttt{glmnet}) & 0.136 & 0.441  \\ 
	Elastic net regression, $\alpha=0.75$ (\texttt{glmnet}) & 0.137 & 0.441  \\ 
    Lasso regression, $\alpha=1$ (\texttt{glmnet}) & 0.143 & 0.441  \\ 
    Random forests, $\text{num.trees} = 100$ (\texttt{ranger}) & 0.002 & 0.449  \\ 
    Random forests, $\text{num.trees} = 500$ (\texttt{ranger}) & 0.172 & 0.447  \\ 
    Super learner (\texttt{sl3})                                          & 1.000      & 0.440  \\ 
   \end{tabularx}
  \begin{tabularx}{\linewidth}{l*{3}{Y}}
   \rowcolor{Gray}     \multicolumn{3}{l}{\textbf{Multinomial treatment model}} \\
    Gradient boosting (\texttt{xgboost}) & 0.342 & 1.577  \\ 
	Elastic net regression, $\alpha=0.25$ (\texttt{glmnet}) & 0.116 & 1.576  \\ 
	Elastic net regression, $\alpha=0.50$ (\texttt{glmnet}) & 0.110 & 1.576  \\ 
	Elastic net regression, $\alpha=0.75$ (\texttt{glmnet}) & 0.108 & 1.576  \\ 
    Lasso regression, $\alpha=1$ (\texttt{glmnet}) & 0.105 & 1.576  \\ 
    Random forests, $\text{num.trees} = 100$ (\texttt{ranger}) & 0.036 & 1.592  \\ 
    Random forests, $\text{num.trees} = 500$ (\texttt{ranger}) & 0.183 & 1.585  \\ 
    Super learner (\texttt{sl3})                                          & 1.000      & 1.569  \\ 
       	\bottomrule
   \end{tabularx}
  \begin{tablenotes}[flushleft]
\small
\item \emph{Notes:} 
\begin{itemize}
\item \emph{Weight} is the ensemble weight for each algorithm
\item \emph {NLL} is the average cross-validated error across $V=5$ folds in terms of negative log likelihood (NLL) for each algorithm and the super learner ensemble
\item \textsf{R} package used for implementing each algorithm in parentheses: the parameter for the \texttt{glmnet} models is the elastic net mixing parameter $\alpha$; for random forests, \text{num.trees} is the number of trees used to grow the forest
\item The binomial outcome model corresponds to the model for the combined outcome (diabetes diagnosis or death) and the binomial treatment model corresponds to the model for the Reference drug
\end{itemize}
\end{tablenotes}
\end{threeparttable}
\end{table}

\begin{table}[htbp]
\captionsetup{justification=justified,singlelinecheck=false}
 \centering
	\caption{Censoring rates all-cause, by event, and mean days to end of follow-up for the initial cohort of $n=64120$ patients over 36 months.} \label{tab:censoring}
 \begin{adjustbox}{max width=0.95\textwidth}
\begin{tabular}{lccccc}
	\toprule
	&  & \multicolumn{3}{c}{Censoring event} & \\
	\cline{3-5}
	 & All-cause & End of study & Loss of coverage & Turned 65 & Mean days to\\ 
 	\textbf{Antipsychotic} & censoring (\%) & (\%) & (\%) & (\%) & end of follow-up \\ 
	\midrule
	Reference & 29.33 & 20.82  & 6.98  & 1.53   & 985.43  \\
	A & 28.67 & 20.86    & 5.84   & 1.97  & 985.80 \\ 
	B  & 24.29 & 16.77  & 5.34   & 2.18  & 998.09 \\
	C  & 27.97 & 19.42  & 6.77   & 1.78 & 981.48 \\
	D & 27.43 & 19.82   & 5.77   & 1.85 & 988.10 \\
	E & 25.78 & 17.78    & 6.68   & 1.31  & 996.80 \\ \hline
	All & 27.36 & 19.31   & 6.27 & 1.79 & 988.00 \\
	\bottomrule
\end{tabular}
\end{adjustbox}
  \begin{tablenotes}[flushleft]
	\small
 \item \emph{Notes:} 
 \begin{itemize}
\item \emph{All-cause censoring} values are the sums of the respective values in the censoring event columns (does not include death or three-year followup end)
\item \emph{End of study} represents the percentage of patients who were censored due to the conclusion of the study period
\item \emph{Loss of  coverage} denotes the percentage of patients who were censored because they lost their insurance coverage during the study period
\item \emph{Turned 65} signifies the percentage of patients who reached the age of 65 during the study and were subsequently censored
\end{itemize}
\end{tablenotes}
\end{table} 

\begin{figure}[htbp]
	\centering
	\includegraphics[width=0.8\columnwidth]{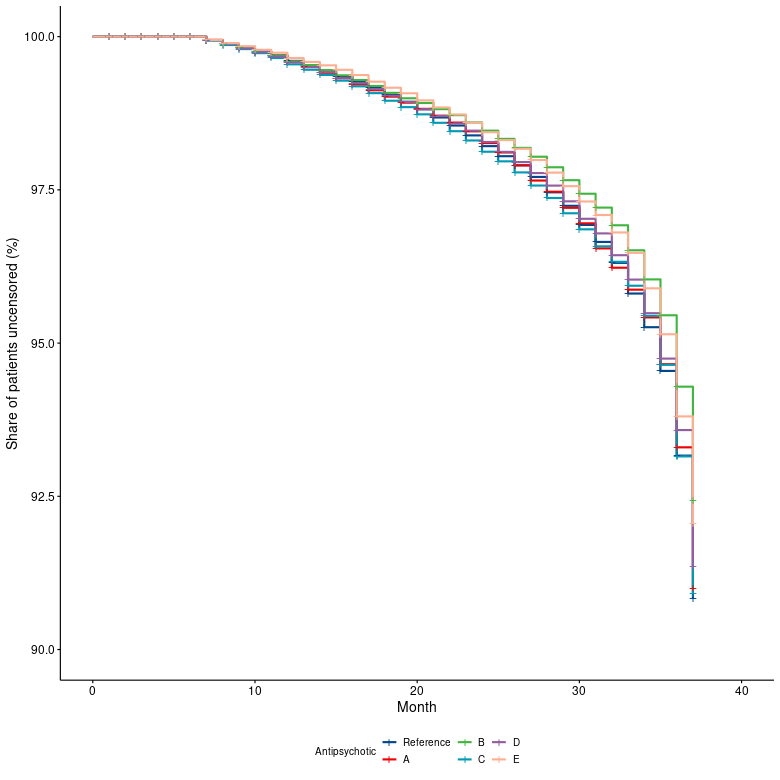}
	\caption{Kaplan-Meier curves depicting monthly patient retention (or lack of censoring) over a 36-month period for different antipsychotic treatments.} \label{censoring_KM_plot}
\end{figure}

\begin{table}[htbp]
\captionsetup{justification=justified,singlelinecheck=false}
\caption{Restricted Mean Survival Time (RMST) differences between the Reference drug and each comparator drug, among the initial cohort of $n=64120$ patients over 36 months.}\label{rmst}
\begin{tabular}{lcccc}
\hline
& \multicolumn{1}{c}{Model with covariates} & \multicolumn{1}{c}{Model without covariates} \\
\textbf{Antipsychotic}  & Est. (SE) & Est. (SE) \\
\hline
A & 0.016 (0.033) & 0.002 (0.016) \\
B & -0.042 (0.022) & 0.079 (0.011) \\
C & 0.015 (0.019) & -0.015 (0.010) \\
D & -0.037 (0.020) & 0.018 (0.010) \\
E & 0.056 (0.025) & 0.071 (0.013) \\
\hline
\end{tabular}
  \begin{tablenotes}[flushleft]
	\small
\item \emph{Note:} The standard errors reported in this table for the RMST are calculated using the Greenwood plug-in estimator for asymptotic variance. 
\end{tablenotes}
\end{table}

\begin{table}[htbp]
\captionsetup{justification=justified,singlelinecheck=false}
\caption{Cox proportional hazards models for assessing the impact of treatment assignment and covariates on the hazard of being censored, among the initial cohort of $n=64120$ patients over 36 months.}\label{cox_ph}
\begin{tabular}{lcc}
\hline
& \multicolumn{1}{c}{Model with covariates} & \multicolumn{1}{c}{Model without covariates} \\
\textbf{Antipsychotic} & Est. (SE) & Est. (SE) \\
\hline
A & -0.001 (0.029) & -0.005 (0.028) \\
B & 0.044 (0.022) & -0.153 (0.021) \\
C & 0.027 (0.018) & 0.026 (0.018) \\
D & 0.020 (0.019) & -0.033 (0.018) \\
E & -0.038 (0.026) & -0.137 (0.026) \\
\hline
\textbf{Model Statistics} & & \\
\hline
Concordance & 0.839 (0.002) & 0.517 (0.002)\\
Likelihood Ratio Test & 75216 ($p < 2e-16$) & 115.8 ($p < 2e-16$) \\
Wald Test & 50245 ($p < 2e-16$) & 113.4 ($p < 2e-16$) \\
Score (Logrank) Test & 85386 ($p < 2e-16$) & 113.6 ($p < 2e-16$) \\
Global Test for Proportional Hazards & 928 ($p < 2e-16$) & 12.5 ($p = 0.028$) \\
\hline
\end{tabular}
  \begin{tablenotes}[flushleft]
	\small
\item \emph{Notes:} 
\begin{itemize}
    \item \textit{Standard errors} associated with the coefficient estimates are calculated using the observed Fisher information matrix, which provides an estimate of the variability of the coefficient estimates.
    \item \textit{Concordance} is a measure of the discriminatory power of the model. A concordance of 0.5 suggests no discrimination, whereas a value close to 1 indicates good discriminatory ability.
    \item \textit{Likelihood ratio test} compares the likelihood of the data under the full model against the likelihood under a null model with fewer predictors. A significant test indicates that the predictors improve the model.
    \item \textit{Wald test} assesses the significance of individual predictors in the model by comparing the estimated parameter to its standard error.
    \item \textit{Score (logrank) test} is another test for the global null hypothesis, similar to the likelihood ratio test but based on different mathematical principles. It is generally used as a confirmatory test.
    \item \textit{Global test for proportional hazards} examines the null hypothesis that the hazard ratios are constant over time. A significant p-value suggests that the proportional hazards assumption may not hold.
\end{itemize}
\end{tablenotes}
\end{table}




\itemize